\documentclass[fleqn,usenatbib]{mnras}
\usepackage{textcomp}
\usepackage{amsmath}
\usepackage[T1]{fontenc}

\usepackage{graphicx}	
\usepackage{amsmath}	
\usepackage{amssymb}	
\usepackage{xcolor}
\usepackage{bm}
\usepackage{nicefrac}
\usepackage{ulem}

\newcommand{\be}{\begin{equation}}
\newcommand{\ee}{\end{equation}}

\newcommand{\ba}{\begin{eqnarray}}
\newcommand{\ea}{\end{eqnarray}}

\DeclareSymbolFont{matha}{OML}{txmi}{m}{it}
\DeclareMathSymbol{\varv}{\mathord}{matha}{118}

\title[Relativistic hydrodynamics study of HMMQ jets along the orbit]{Relativistic hydrodynamical simulations of the effects of the stellar wind and the orbit on high-mass microquasar jets}

\author[M.V. Barkov and V. Bosch-Ramon]{
Maxim V. Barkov,$^{1}$\thanks{E-mail: barkov@inasan.ru} and
V. Bosch-Ramon,$^{2}$\thanks{E-mail: vbosch@fqa.ub.edu}
\\
$^{1}$ Institute of Astronomy, Russian Academy of Sciences, Moscow, 119017 Russia\\
$^{2}$ Departament de F\'{i}sica Qu\`antica i Astrof\'{i}sica, Institut de Ci\`encies del Cosmos (ICC),\\ Universitat de Barcelona (IEEC-UB), Mart\'{i} i Franqu\`es 1, E08028 Barcelona, Spain.
}

\date{Accepted XXX. Received YYY; in original form ZZZ}

\pubyear{2021}

\begin{document} 
\label{firstpage}
\pagerange{\pageref{firstpage}--\pageref{lastpage}}
\maketitle

\begin{abstract}
High-mass microquasar jets, produced in an accreting compact object in orbit around a massive star, must cross a region filled with stellar wind. The combined effects of the wind and orbital motion can strongly affect the jet properties on binary scales and beyond. The study of such effects can shed light on how high-mass microquasar jets propagate and terminate in the interstellar medium. We study for the first time, using relativistic hydrodynamical simulations, the combined impact of the stellar wind and orbital motion on the properties of high-mass microquasar jets on binary scales and beyond. We have performed 3-dimensional relativistic hydrodynamic simulations, using the PLUTO code, of a microquasar scenario in which a strong weakly relativistic wind from a star interacts with a relativistic jet under the effect of the binary orbital motion. The parameters of the orbit are chosen such that the results can provide insight on the jet-wind interaction in compact systems like for instance Cyg~X-1 or Cyg~X-3. The wind and jet momentum rates are set to values that may be realistic for these sources and lead to moderate jet bending, which together with the close orbit and jet instabilities could trigger significant jet precession and disruption. For high-mass microquasars with orbit size $a\sim 0.1$~AU, and (relativistic) jet power $L_j\sim 10^{37}(\dot M_w/10^{-6}\,{\rm M}_\odot\,{\rm yr}^{-1})$~erg~s$^{-1}$, where $\dot M_w$ is the stellar wind mass rate, the combined effects of the stellar wind and orbital motion can induce relativistic jet disruption on scales $\sim 1$~AU.

  \end{abstract}
  

%

\section{Introduction}
\label{intro}

High-mass microquasars (HMMQ) are a type of X-ray binary that consists of a massive star, which produces a slow and dense wind, and a compact object (CO), which accretes matter from the star and launches very fast bipolar jets \citep{mir99,rib02}. Typically, compact objects in microquasars (MQ) are thought to be black holes and weakly magnetized neutron stars, although strongly magnetized neutron stars, and even white dwarfs, have been associated to the accretion-ejection phenomenon in stellar-mass objects (see, e.g., \citealt{van18} and \citealt{koe08}, respectively). When MQ were discovered,
it was the presence of extended radio jets that led to the use of the term {\it microquasar} for this kind of X-ray binaries, as they seemed to be much smaller galactic counterparts of active galactic nuclei (ANG -or quasars-) with jets \citep{gel89,fom91,mir92}. The association between MQ and AGN was strengthened even further once the relativistic nature of the jets of MQ was revealed \citep{mir94}, and gamma-rays were proposed as coming for microquasar candidates \citep[e.g.,][and references therein]{par05}. Later on, MQ were also associated to gamma-ray bursts (GRB), as the  engines of GRB are also based on the accretion-ejection phenomenon, and it was even proposed that MQ energetic radiation could have played a role in the Universe
evolution \citep[e.g.][]{mir11,mir11b}. However, despite high-energy processes in MQ share strong similarities with AGN and GRB, in the case of MQ the associated timescales are approximately in between those of the other two types of sources, and the emitting regions may be quite distinct \citep[e.g.,][]{bos06,bos08b}. Thus, the study of microquasars is not important by itself, but also in connection to the other accretion-ejection sources accounting for both their similarities and differences. 

\begin{figure}
\includegraphics[width=7.95cm]{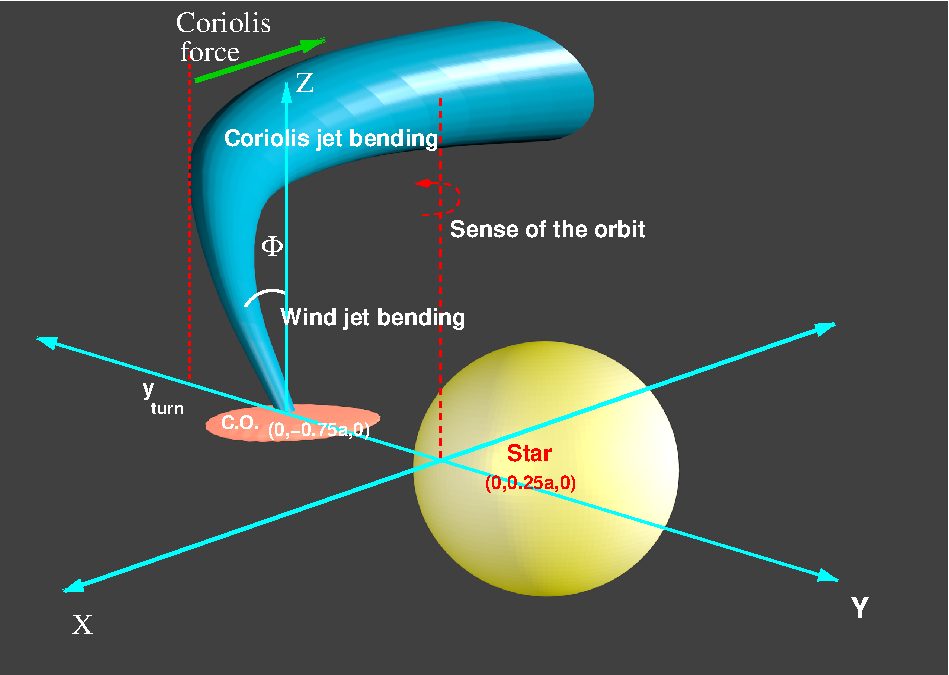}
\caption{Sketch of the simulated scenario (not to scale) including its main ingredients (star, stellar wind, jet) and physical processes and effects (wind-induced jet bending, Coriolis effect and beginning of the jet helical geometry); adapted from figure~1 in \citet{bos16}. The orientation of the axes in the figure corresponds to that at the beginning of the simulation (orbital phase 0).}
\label{f1}
\end{figure}

Among those differences between MQ, AGN and GRB, the study of which can be very informative, a major one is the presence of a
non-degenerate star in MQ. In the case of HMMQ in particular, the star is massive, hence having a very high luminosity and a
strong wind. The stellar wind is produced by line-driven radiation pressure that overcomes the Eddington luminosity in the
stellar atmosphere. Typical values for the mass-loss rates and wind velocities in OB stars are $\dot{M}_{\rm w}\sim 10^{-9}-10^{-5}\,$M$_\odot$~yr$^{-1}$ and $\varv_w\sim 2\times 10^8$~cm~s$^{-1}$, respectively \citep[e.g.][]{mui12,krt14}. The wind acceleration is thought to be unstable and to form clumps \citep[e.g.][]{run02,mof08}. These and other complexities of the wind structure (e.g. gravity darkening, \citealt{owo98}; presence of circumstellar disk, \citealt{oka01}; corotating Interaction Regions, \citealt{cra96}; etc.) characterize the
circumstellar environment, and the jets of HMMQ must thus cross the complex region filled by the wind of the star. This can lead to strong jet-wind interaction, which can severely affect the jet flow by producing shocks, deflection, instability growth and mass-load (see, e.g., \citealt{ore05,per08,per10,yoo15,zdz15}; and \citealt{per12} in the case of clumpy winds, \citealt{cha21} accounting for thermal cooling, and L\'opez-Miralles et al. -in prep.- including the jet magnetic field). 

\begin{figure}
 \centering
   \includegraphics[width=7.1cm]{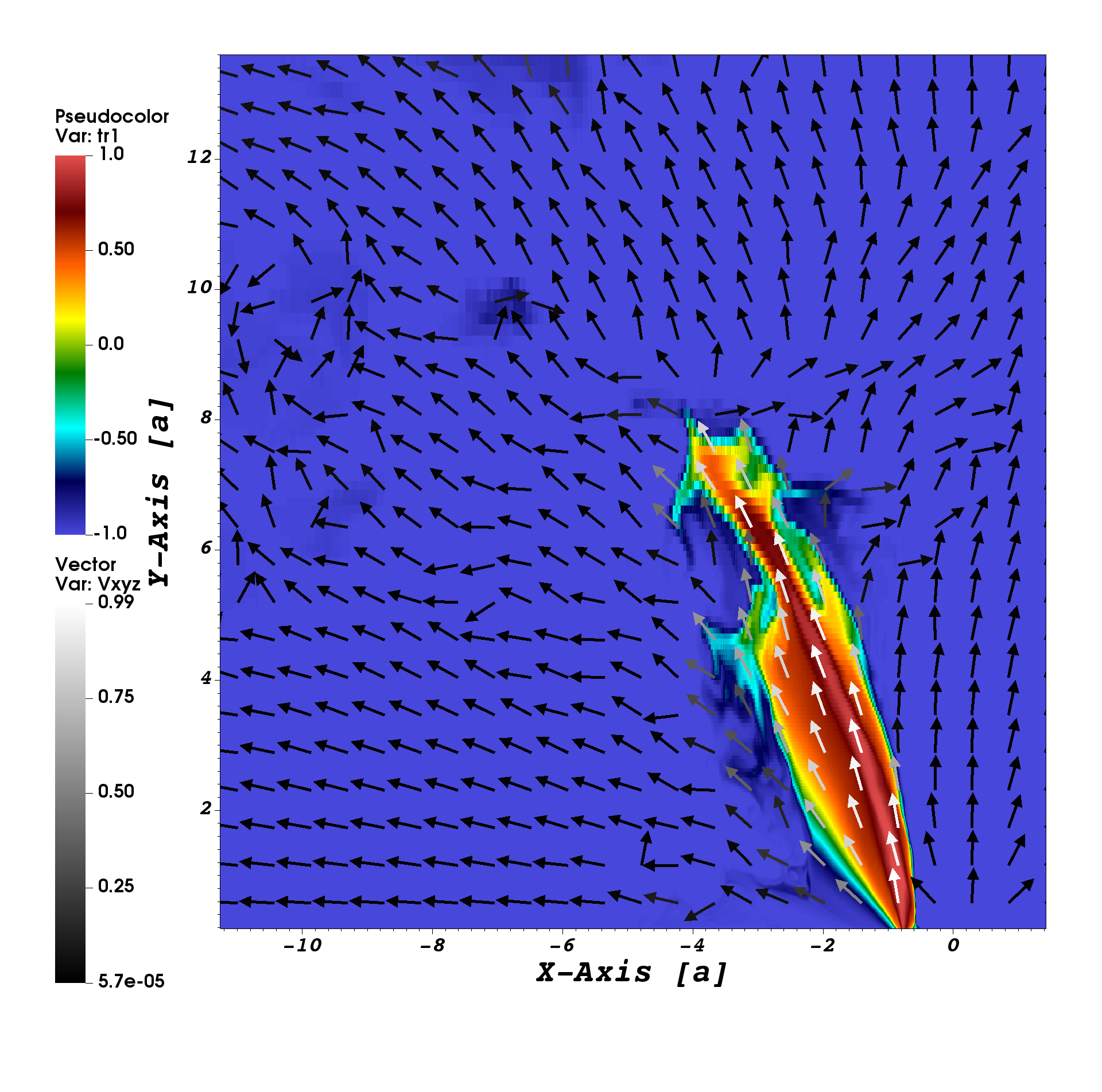}
   \includegraphics[width=7.1cm]{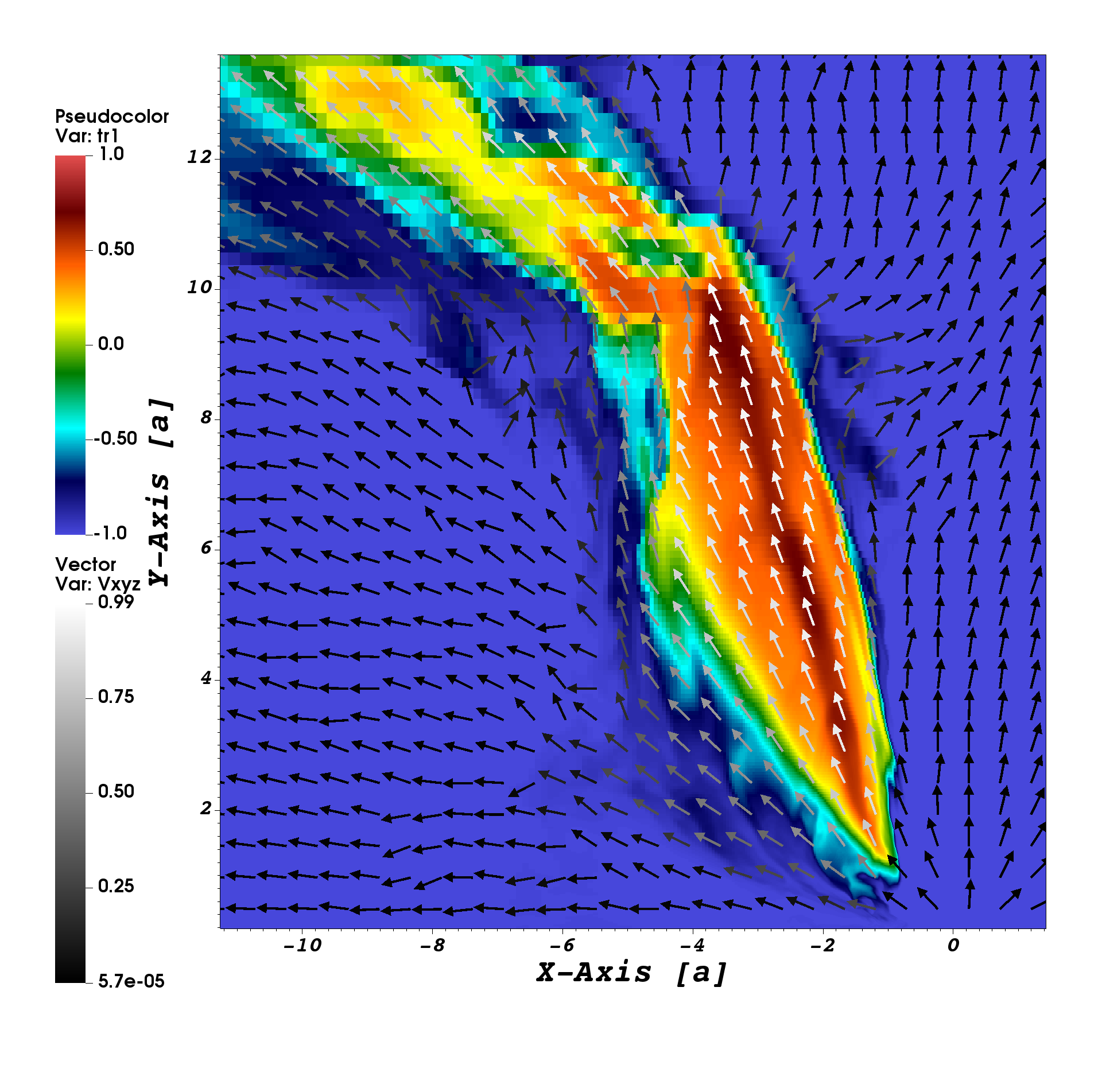}
   \includegraphics[width=7.1cm]{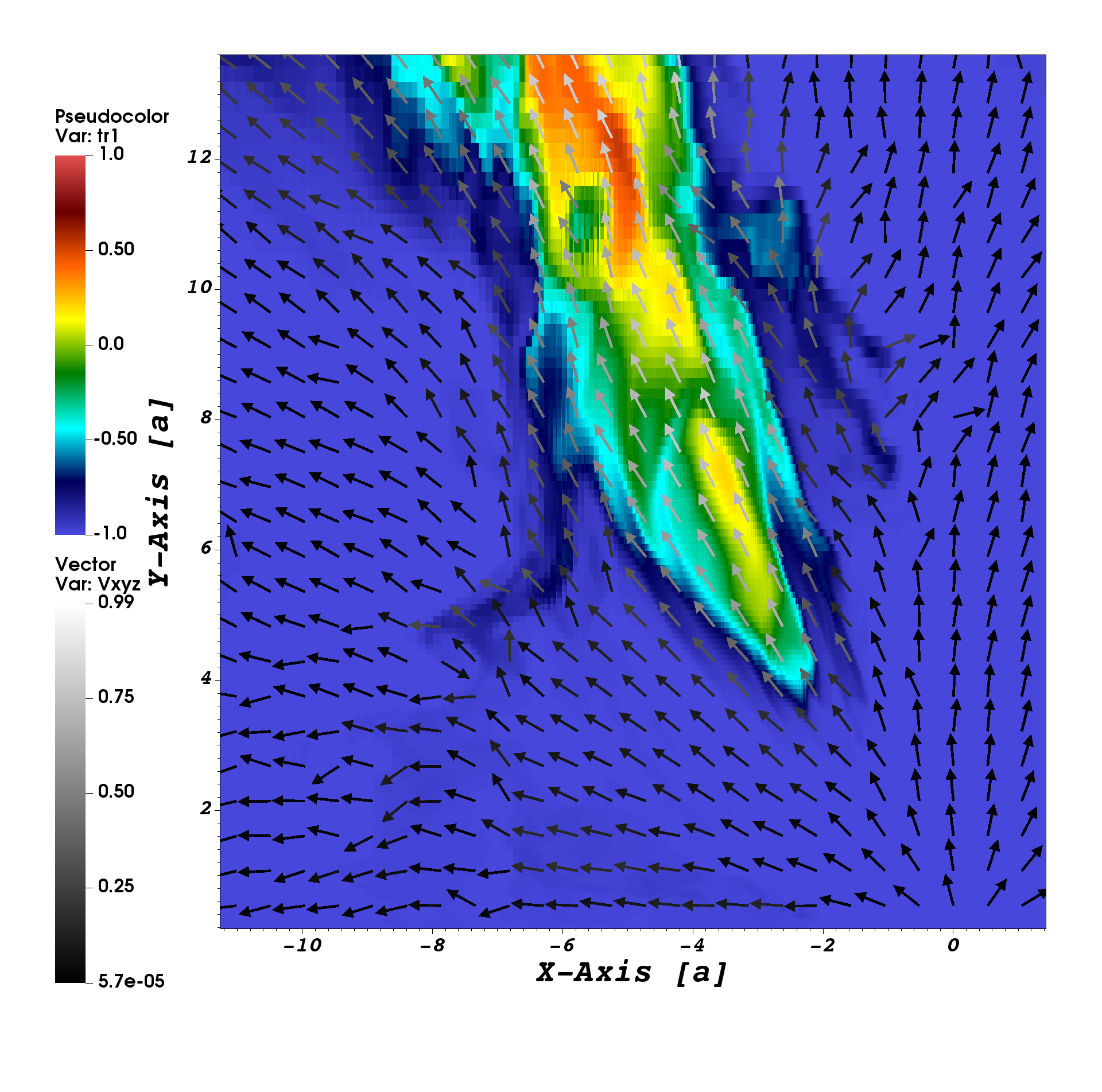}
   \caption{Colored tracer maps and velocity vector distributions for the relativistic jet, whose vertical axes are the $\hat z$-direction, and their horizontal axes have an angle in the orbit sense with respect to the $-\hat y$-direction of $\approx 215^\circ$ (top panel), $180^\circ$ (middle panel), and $153^\circ$ (bottom panel), capturing jet regions at different heights.}
   \label{fig:trsl}%
\end{figure}

\begin{figure}
 \centering
   \includegraphics[width=7.2cm]{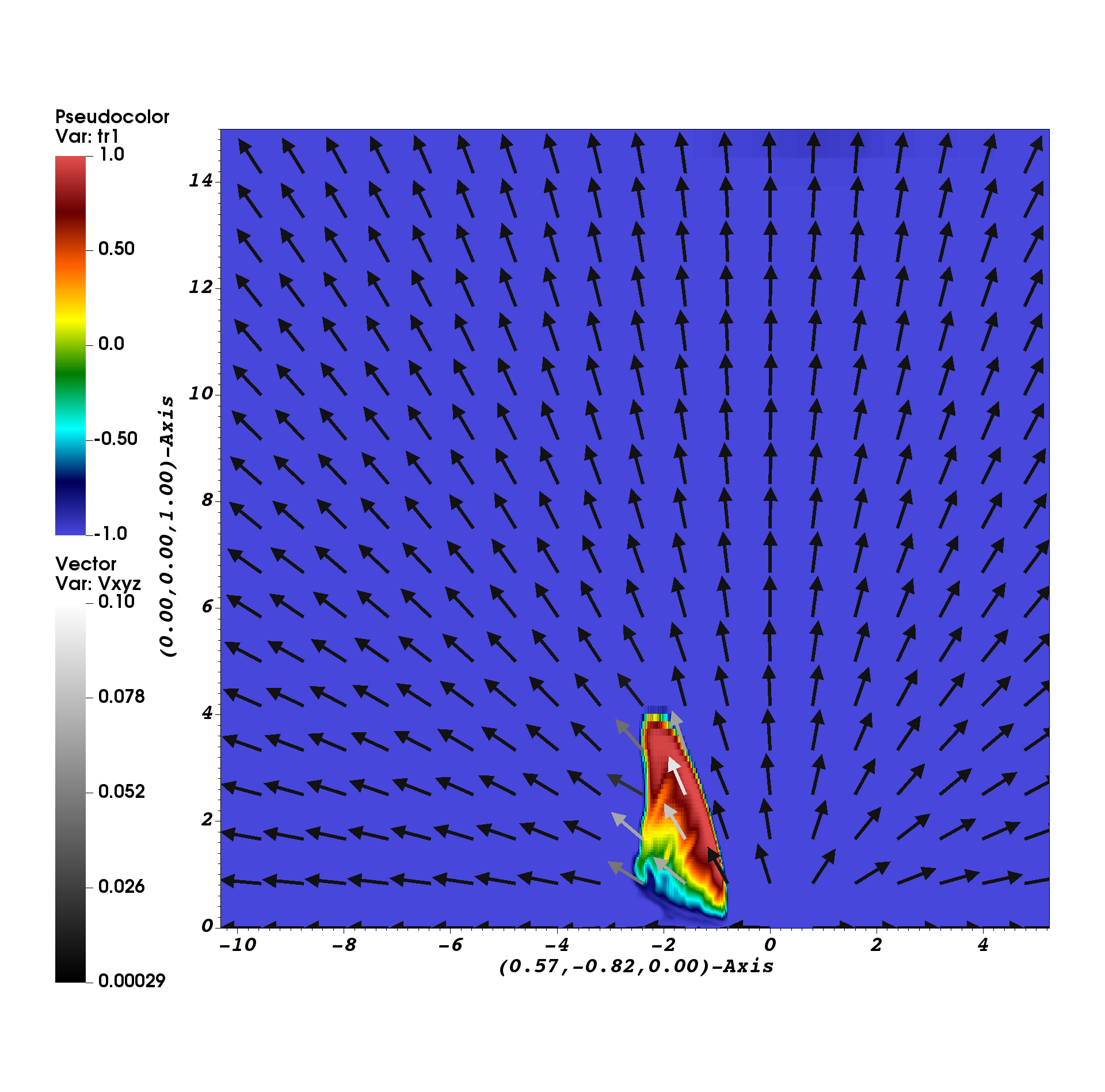}
   \includegraphics[width=7.2cm]{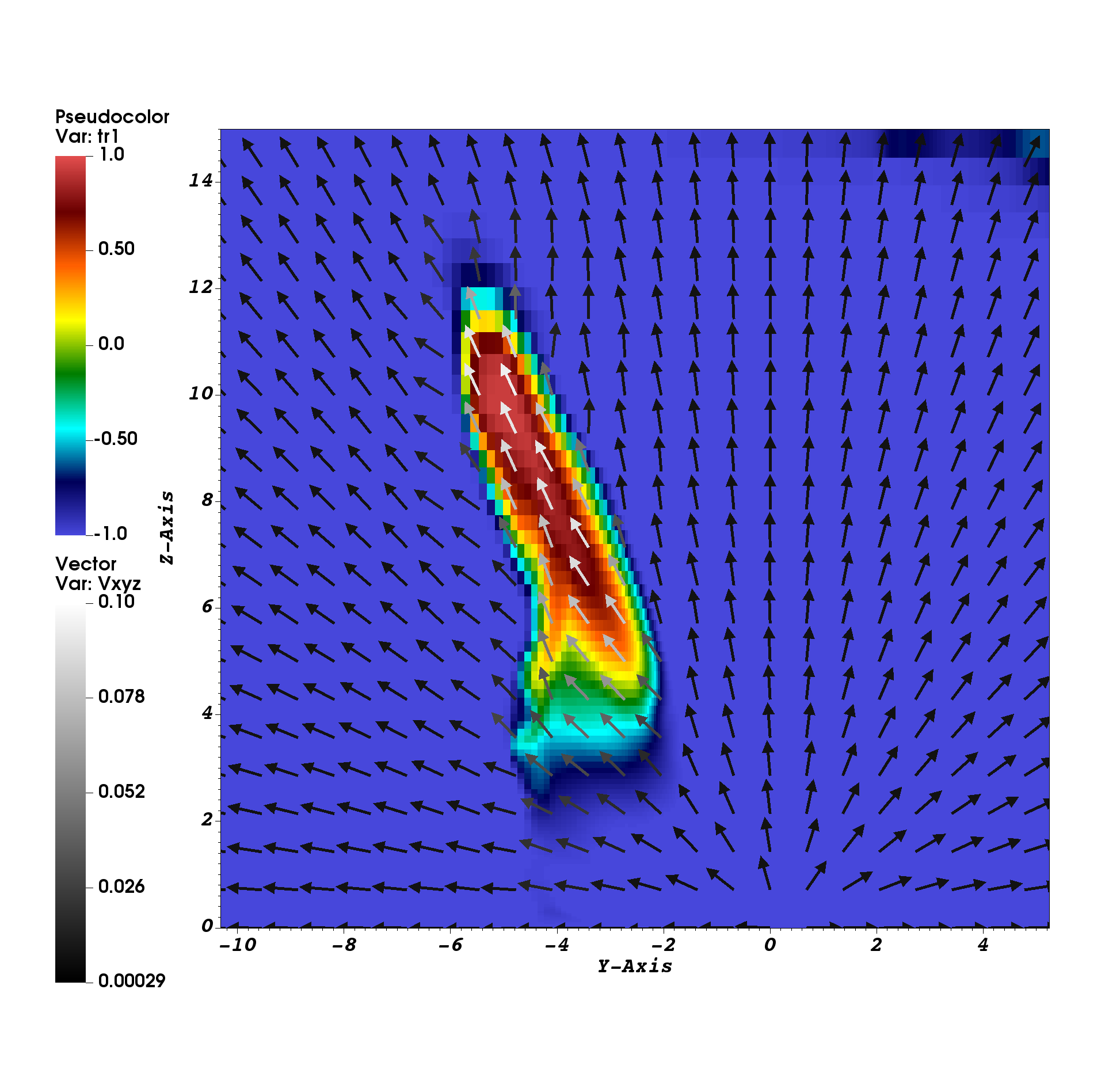}
   \includegraphics[width=7.2cm]{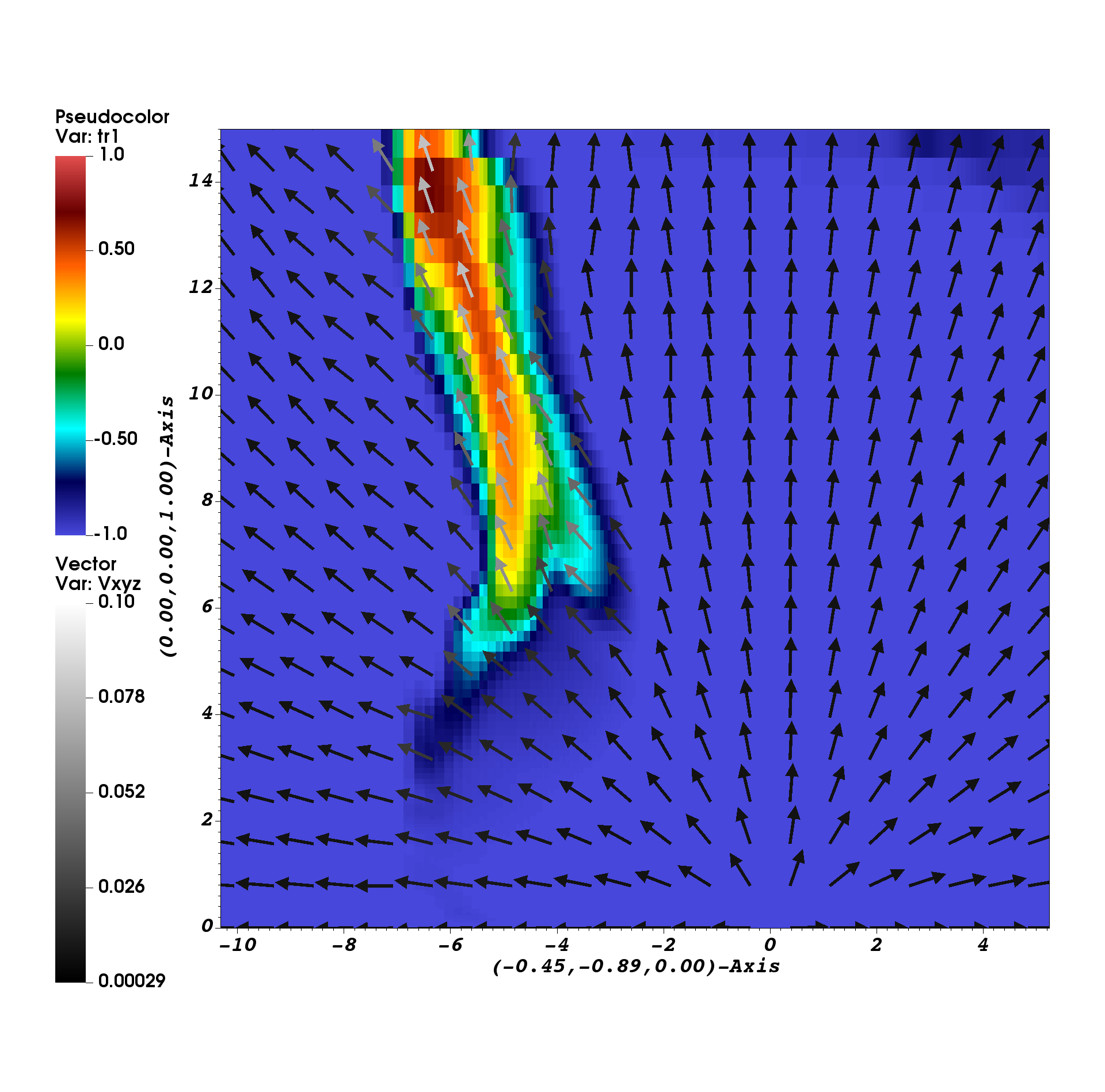}
   \caption{The same as shown in Fig.~\ref{fig:trsl} but for the weakly relativistic jet, and angles in the orbit sense with respect to the $-\hat y$-direction of $\approx 413^\circ$ (top panel), $384^\circ$ (central panel), and $336^\circ$ (bottom panel).}
   \label{fig:trsllrNR}%
\end{figure}

\begin{figure*}
 \centering
   \includegraphics[width=15cm]{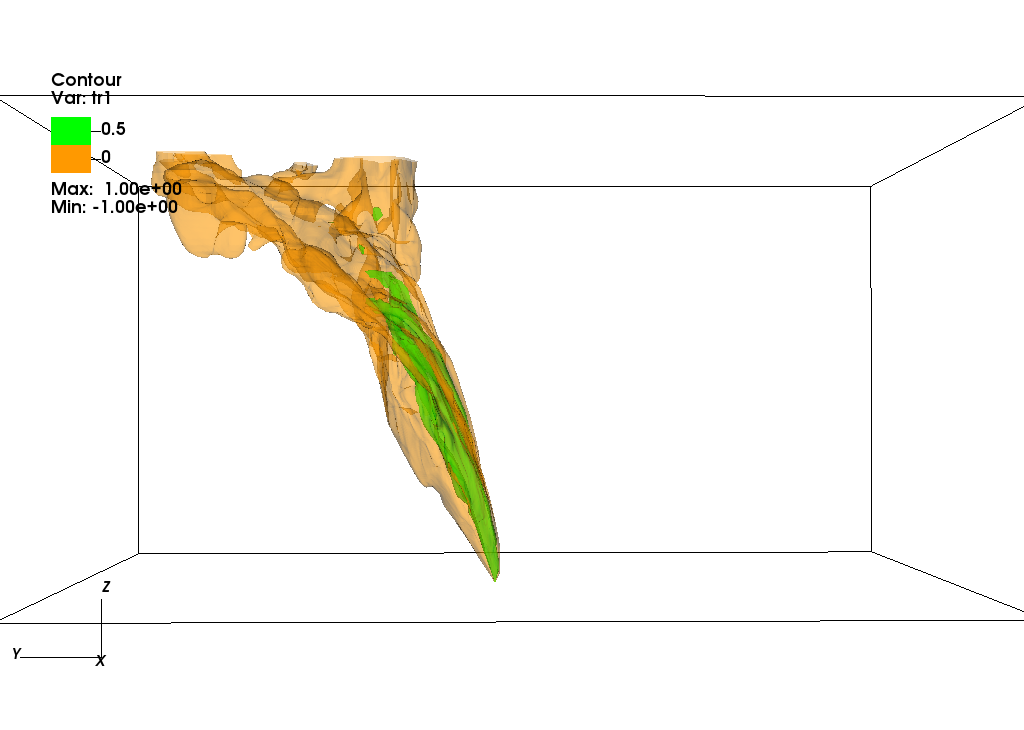}
   \includegraphics[width=15cm]{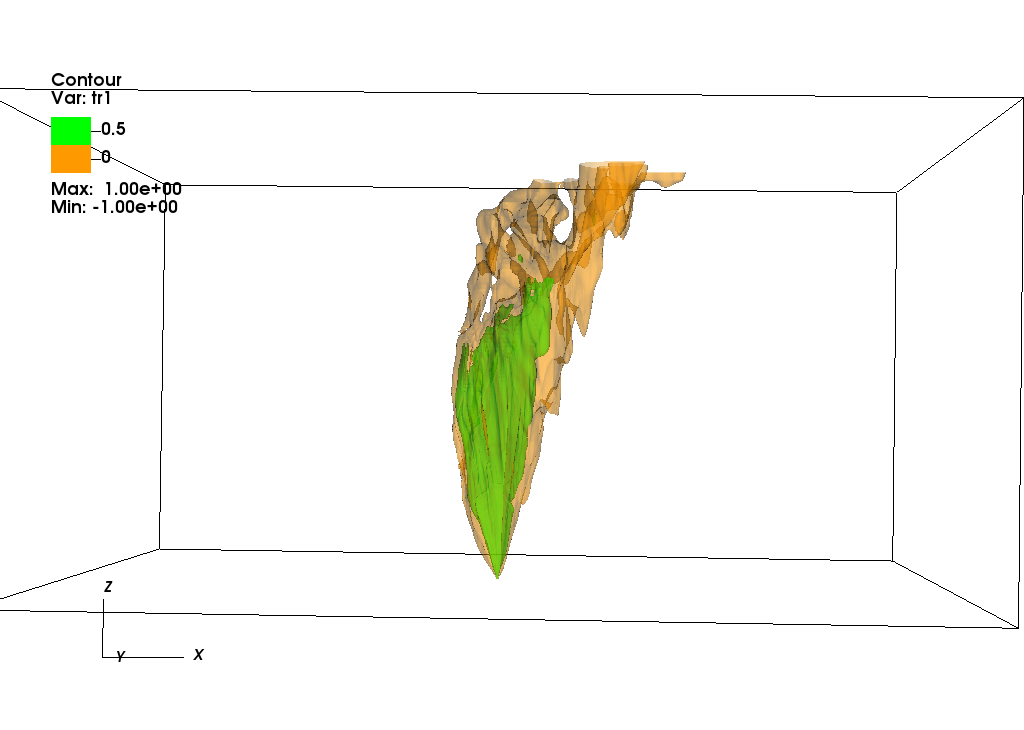}
   \caption{Tracer isosurfaces in 3D for the jet at phase $\approx 0.6$, determined by the cells in which the jet tracer is 0.5 (green) and 0 (brown), in the case of the relativistic jet. The choice of the coordinate system orientation (see bottom left inset) is determined by the jet being mostly inclined away from the star for heights with $z\gtrsim a$ (top panel), and strongly affected by orbital motion for $z\gtrsim 8\,a$ (bottom panel). The size of the canvas is that of the computational grid (from $-15\,a$ to $15\,a$ in both the $x$ and $y$ axes, and from 0 to $15\,a$ in the $z$ axis; see Sect.~\ref{codecg}). The simulation ran for a bit over half orbit.}
   \label{fig:tr}%
\end{figure*}

\begin{figure*}
 \centering
   \includegraphics[width=15cm]{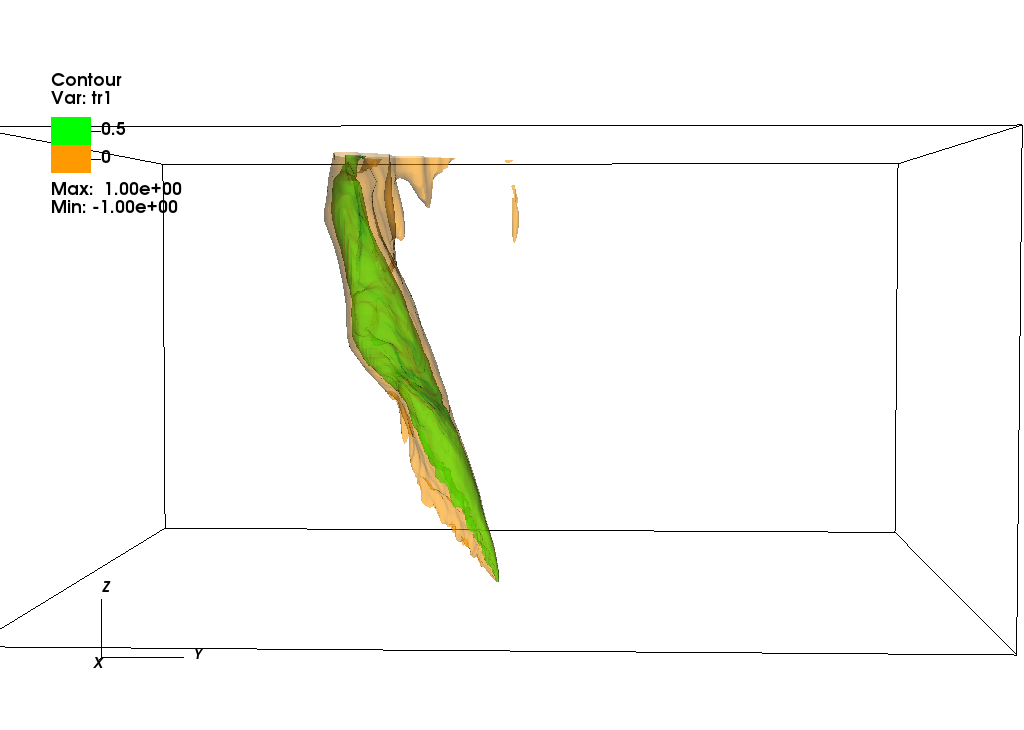}
   \includegraphics[width=15cm]{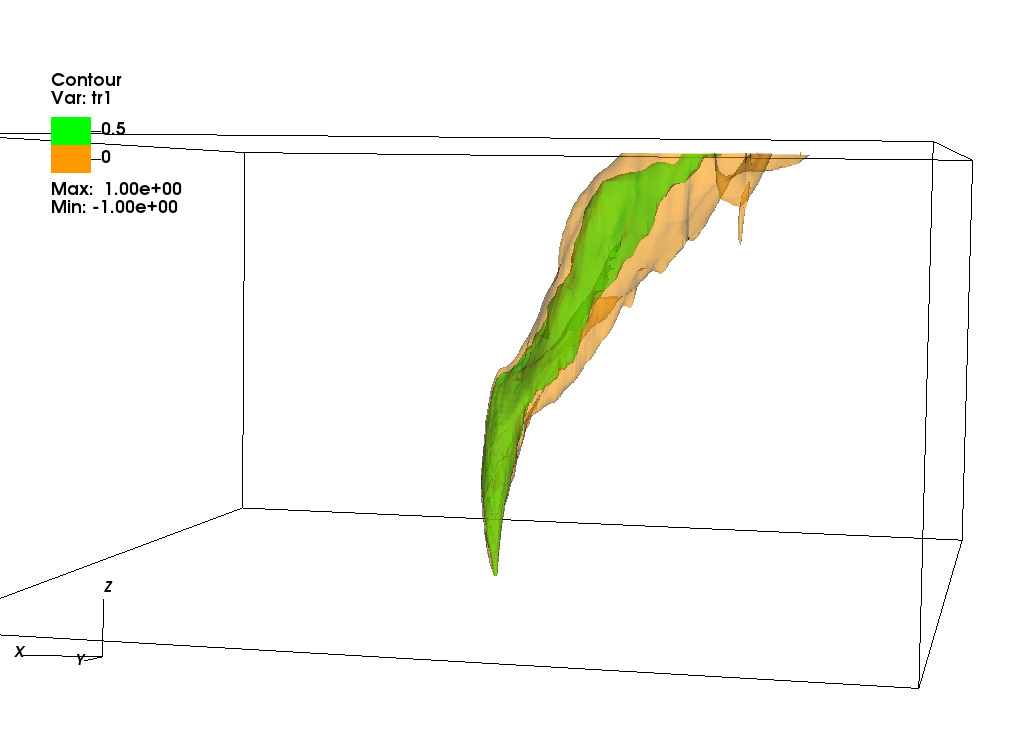}
   \caption{The same as shown in Fig.~\ref{fig:tr} but for the weakly relativistic jet (low resolution) at phase $\approx 0.15$; note that the orientations of the coordinate system (see bottom left inset) are different because the simulation run is longer in this case, a bit over one orbit).}
   \label{fig:trlrNR}%
\end{figure*}

\begin{figure*}
 \centering
   \includegraphics[width=7.95cm]{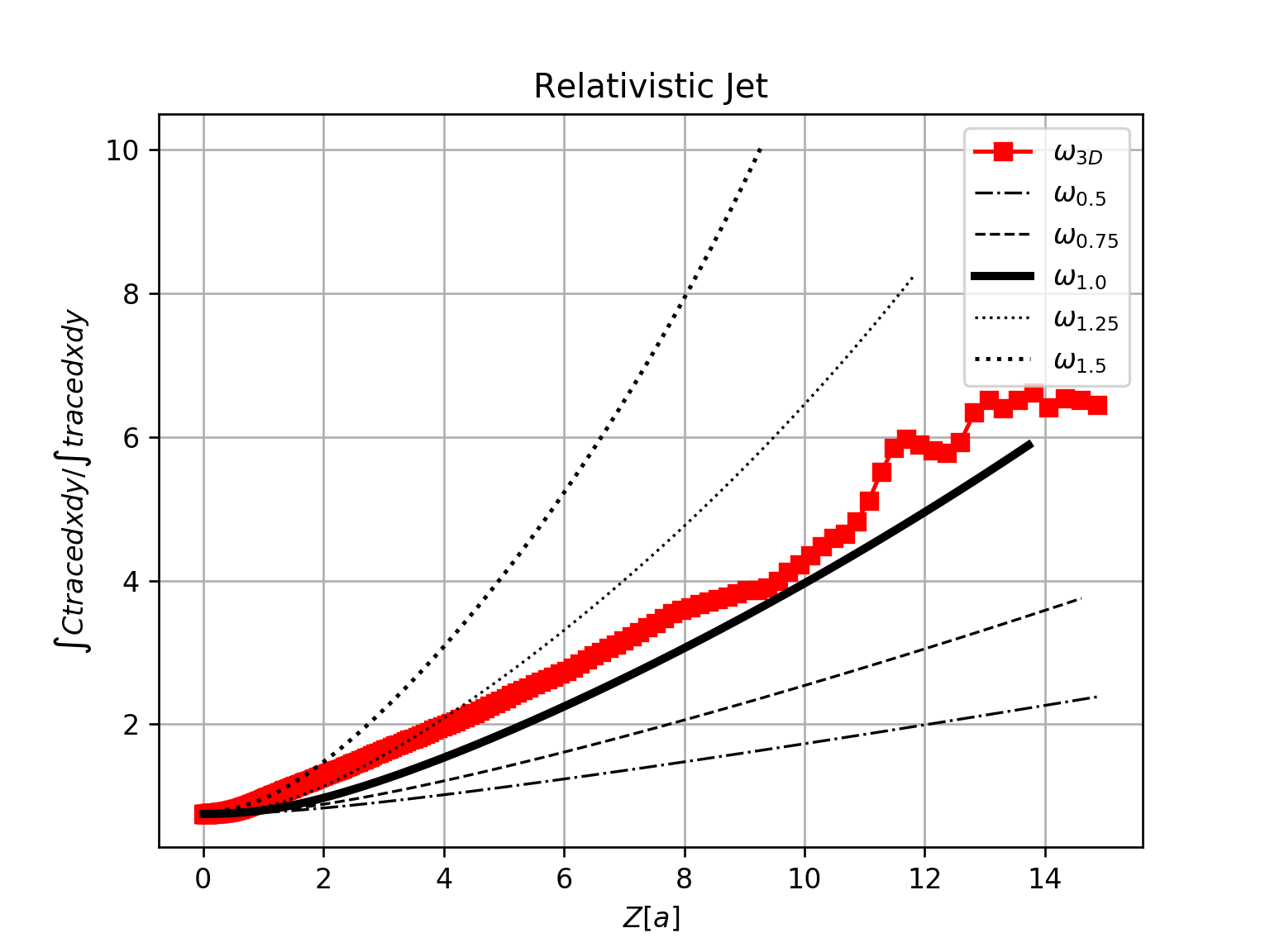}
   \includegraphics[width=7.95cm]{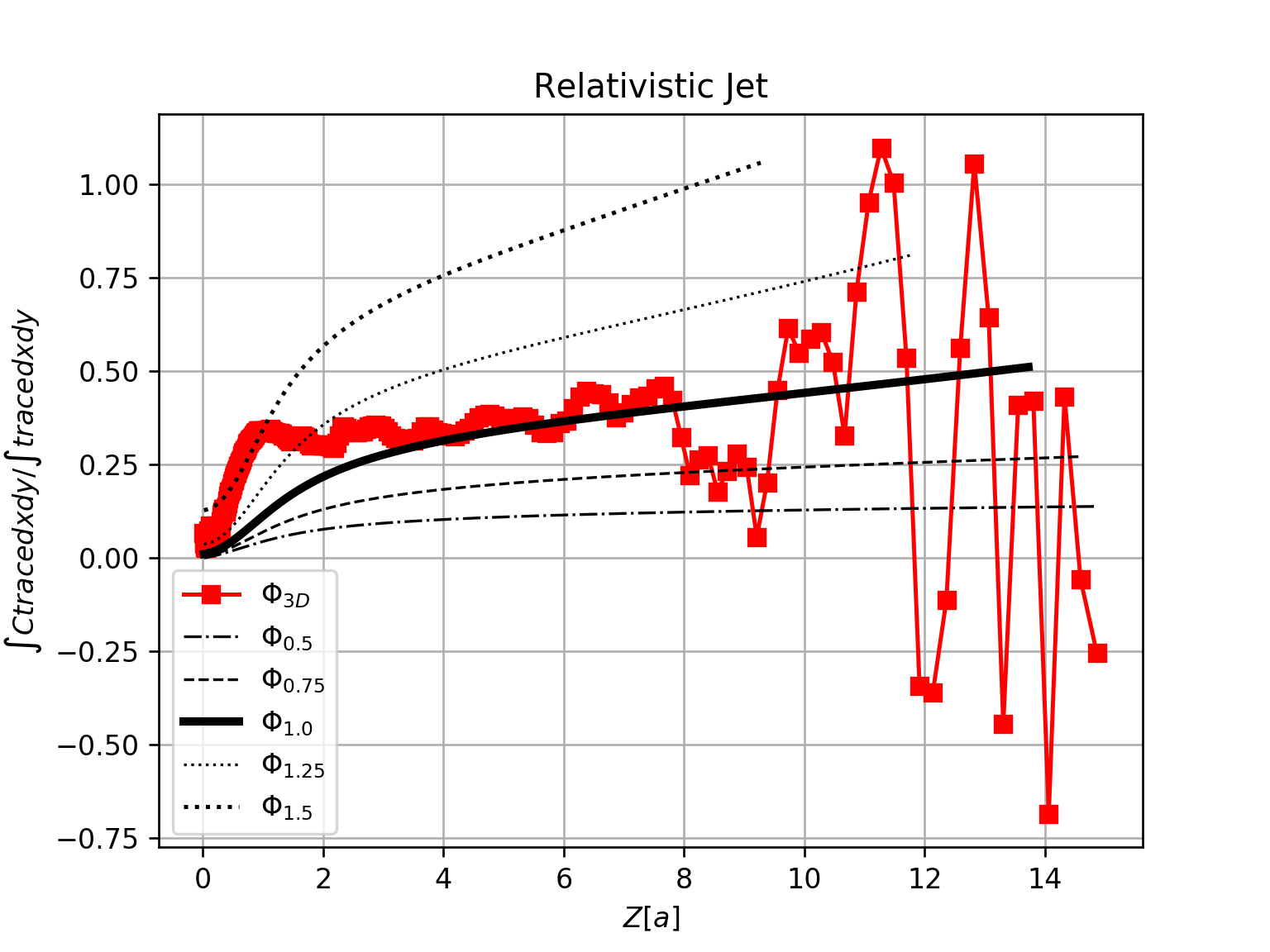}
   \includegraphics[width=7.95cm]{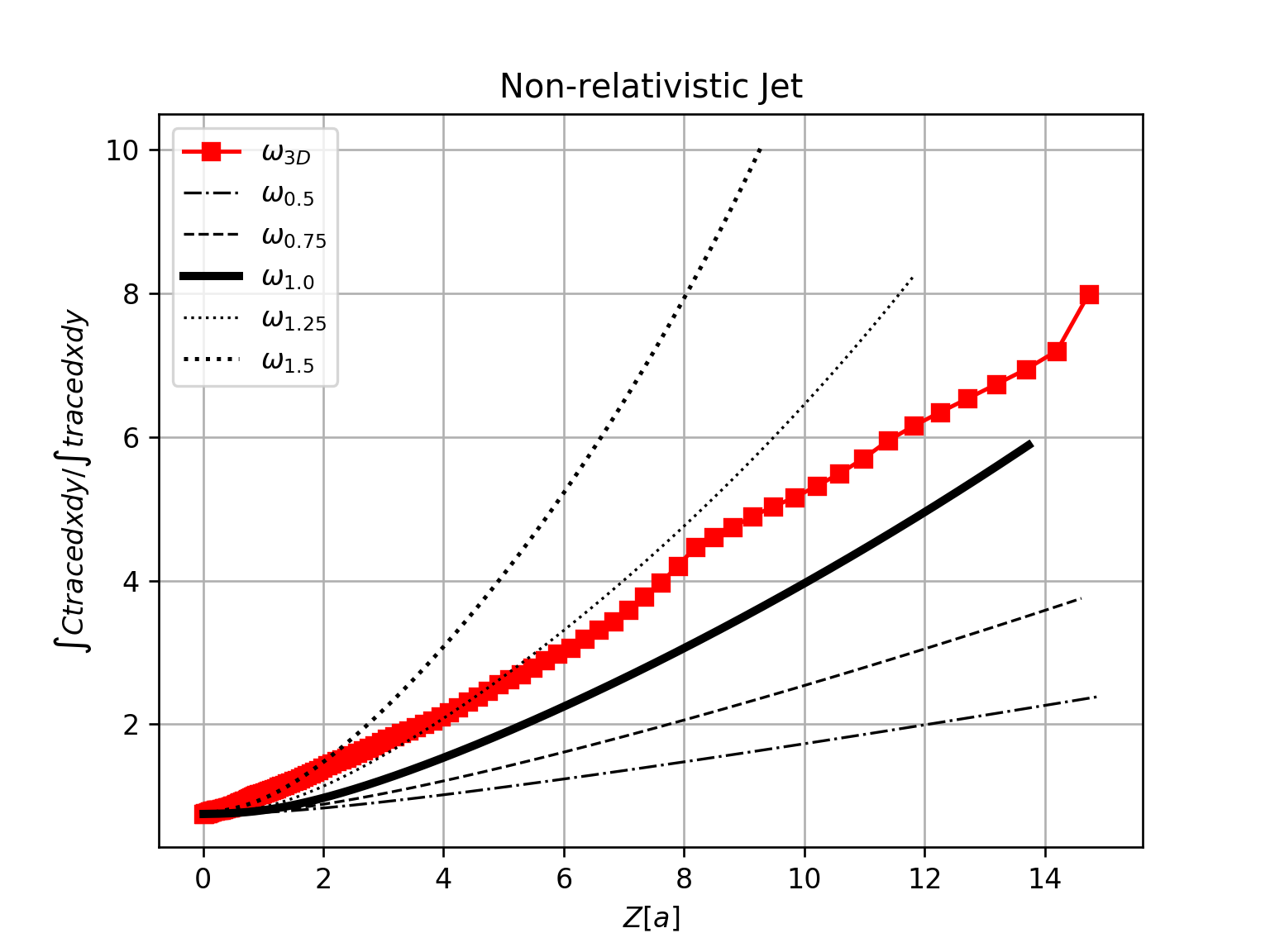}
   \includegraphics[width=7.95cm]{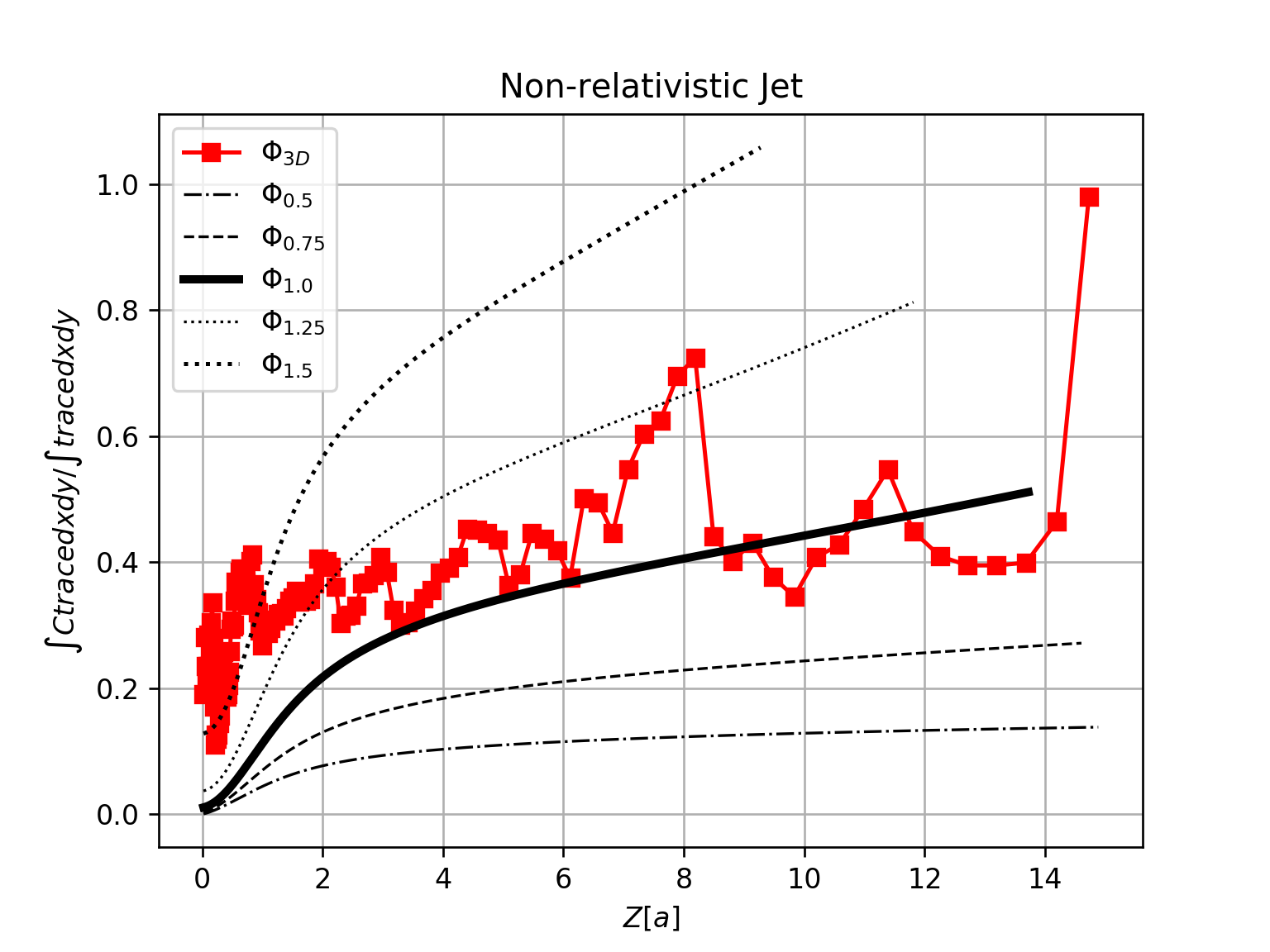}
   \caption{Comparison of the evolution with $z$ of the cylindrical radius {in $[a]$ units} ($\omega_c$; left panels) and the deflection angle in radians ($\Phi$; right panels) for a semi-analytical model with jet radius $\omega_k\propto z^k$ and $k=0.5$, 0.75, 1, 1.25, and 1.5 (see text -Sect.~\ref{res}-; black lines in the figure), and for the hydrodynamical solution (see text -Sect.~\ref{res}-; red solid lines with squares in the figure) in the relativistic (two top panels) and the weakly relativistic case (two bottom panels).}
   \label{fig:analsim}%
\end{figure*}

\begin{figure*}
 \centering
   \includegraphics[width=7.8cm]{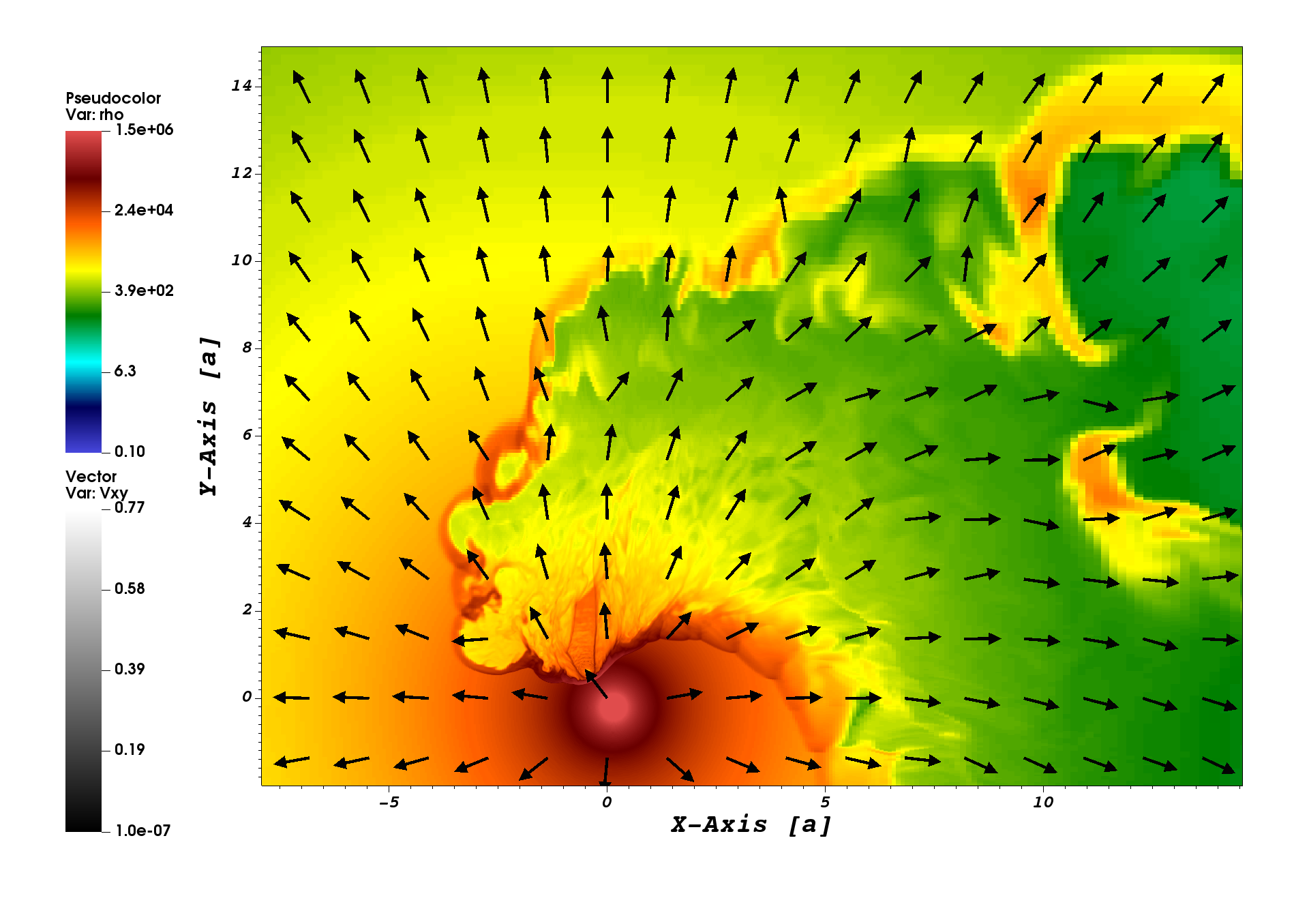}
   \includegraphics[width=7.8cm]{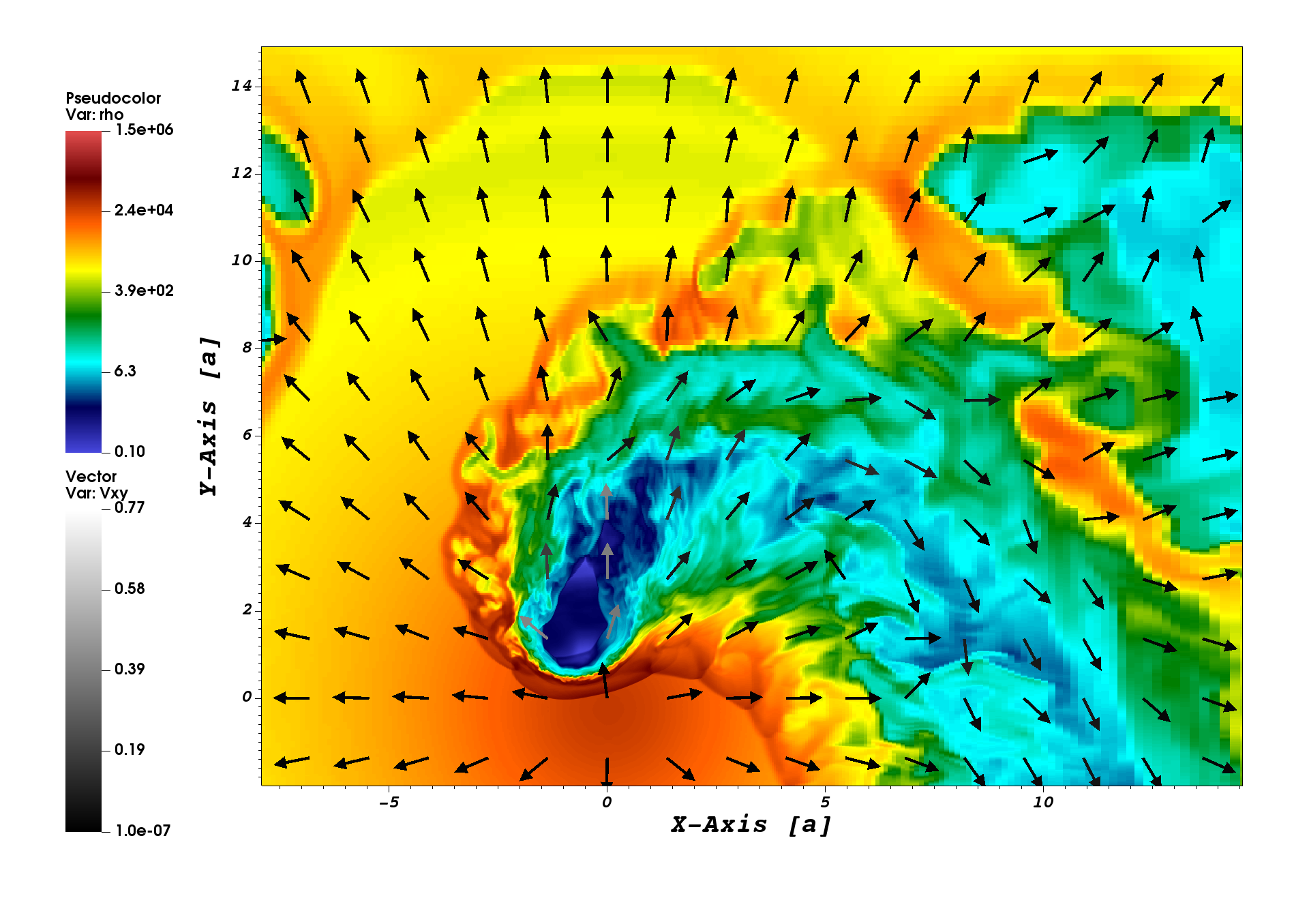}
   \includegraphics[width=7.8cm]{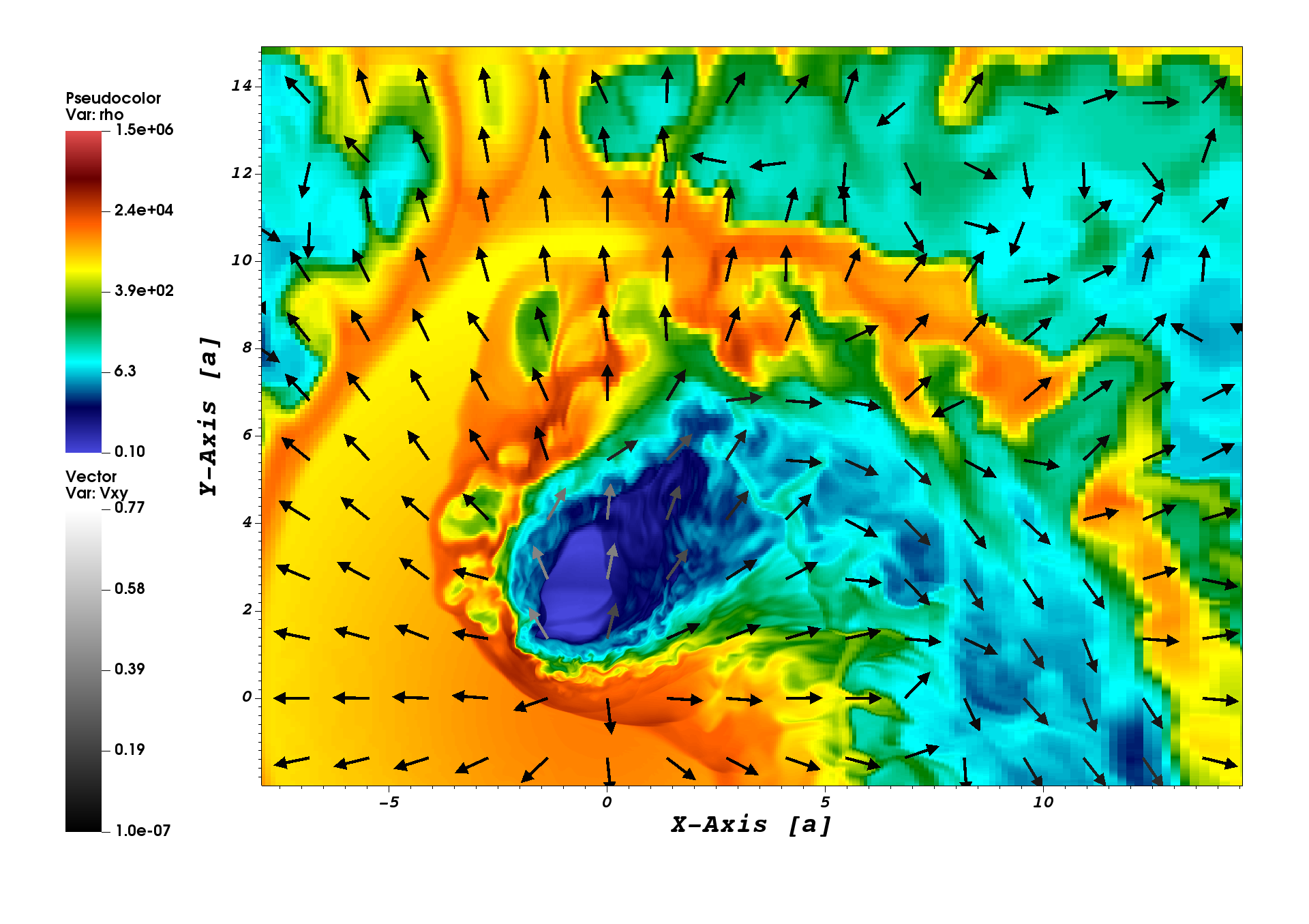}
   \includegraphics[width=7.8cm]{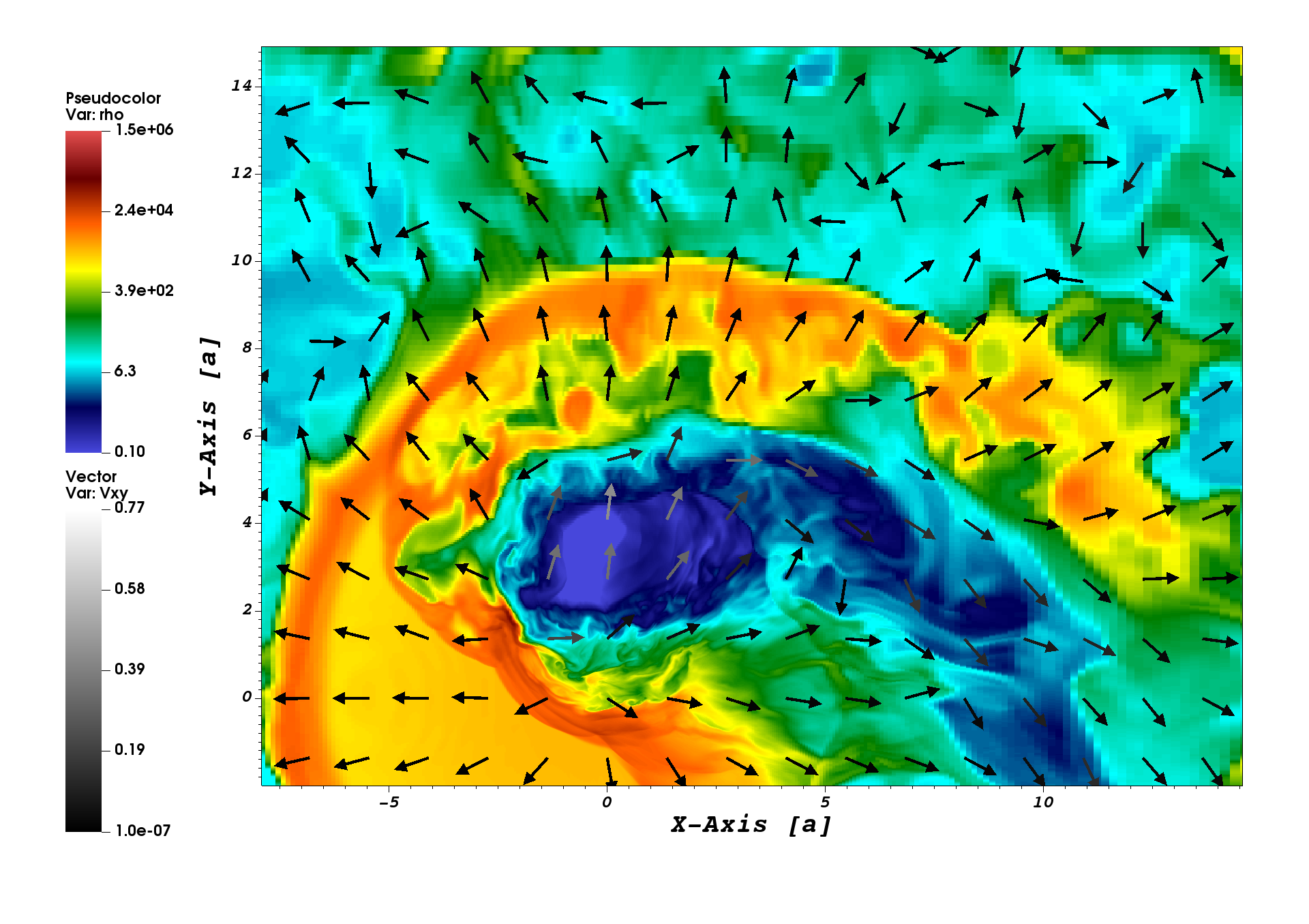}
   \includegraphics[width=7.8cm]{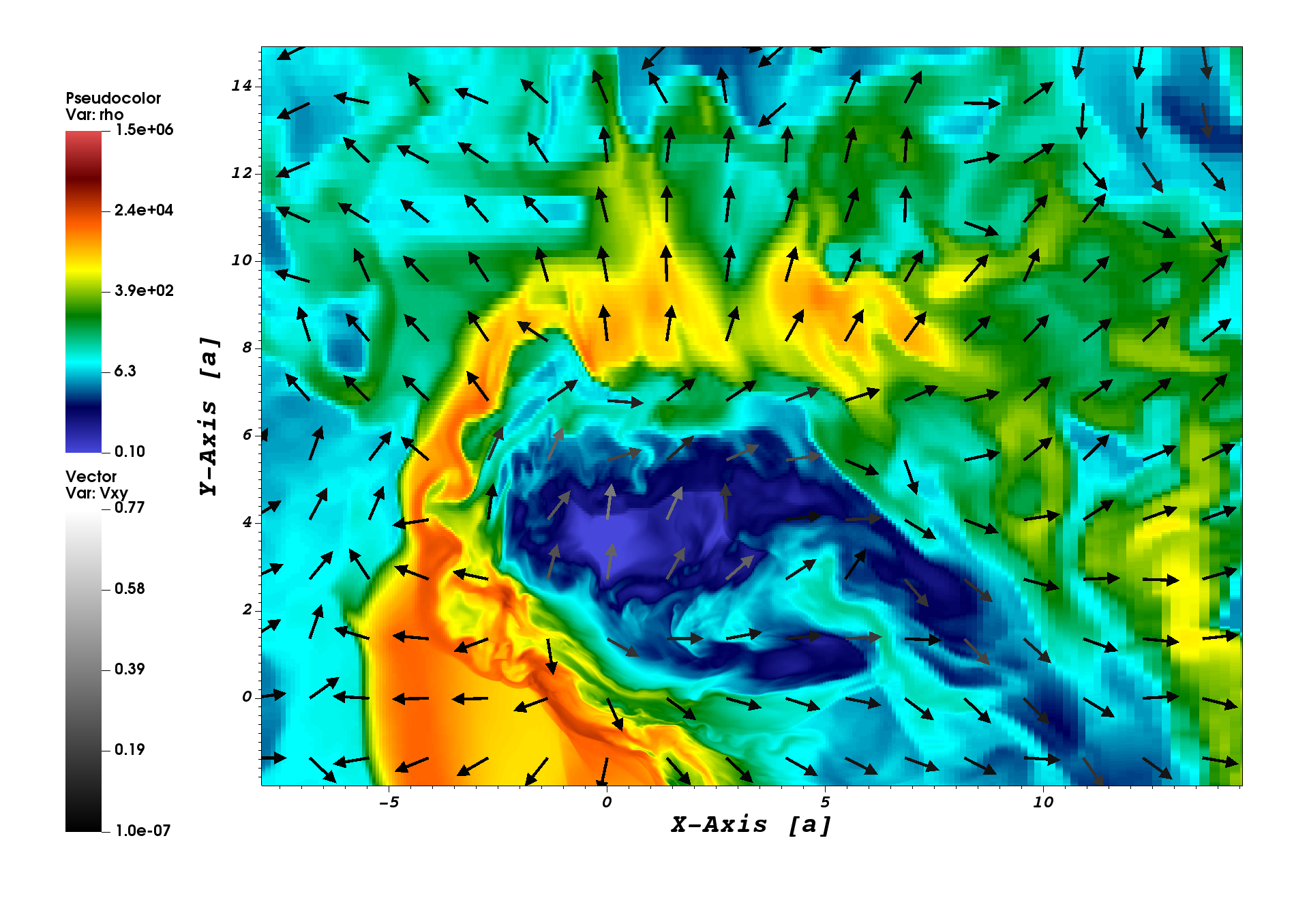}
   \includegraphics[width=7.8cm]{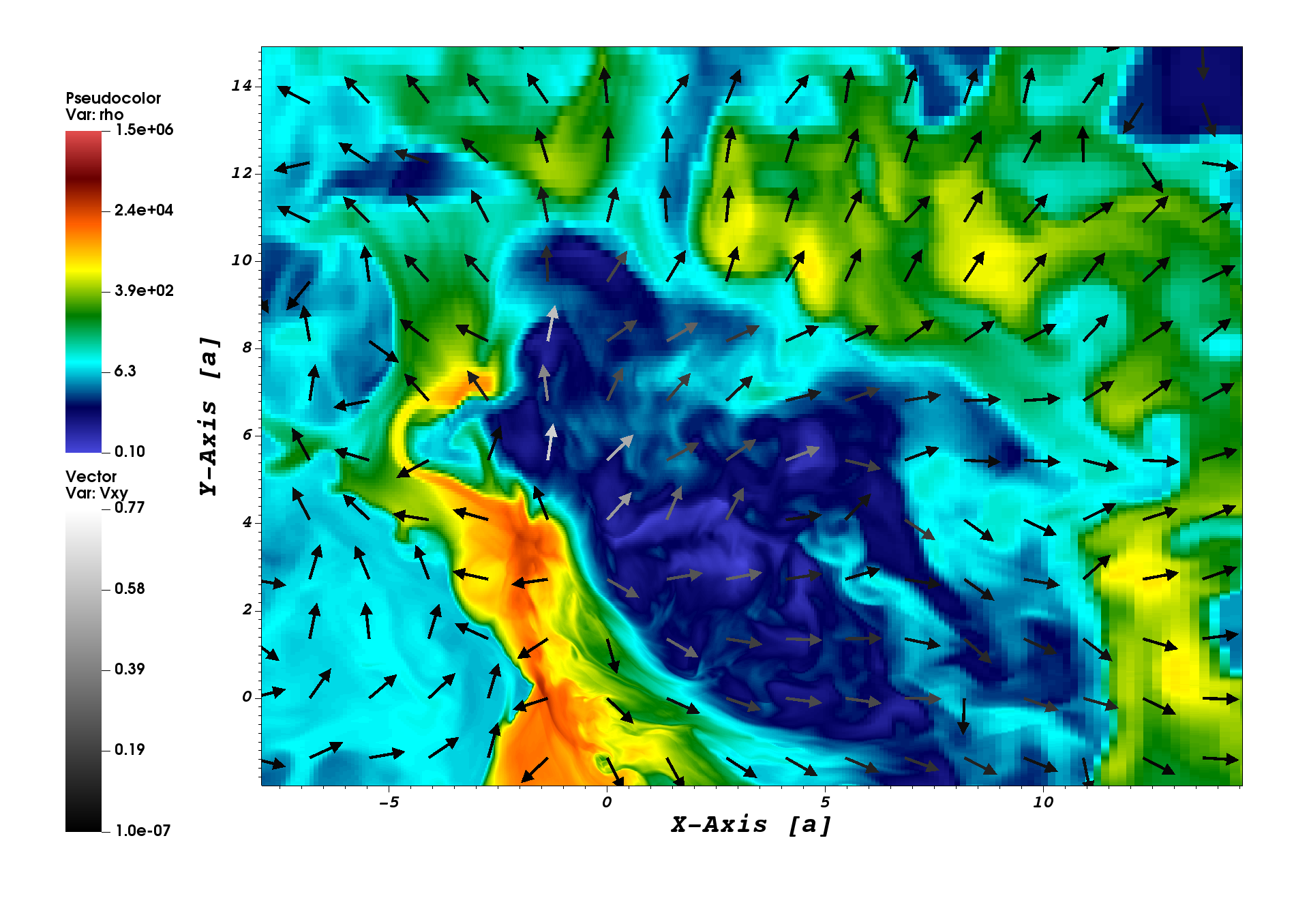}
   \includegraphics[width=7.8cm]{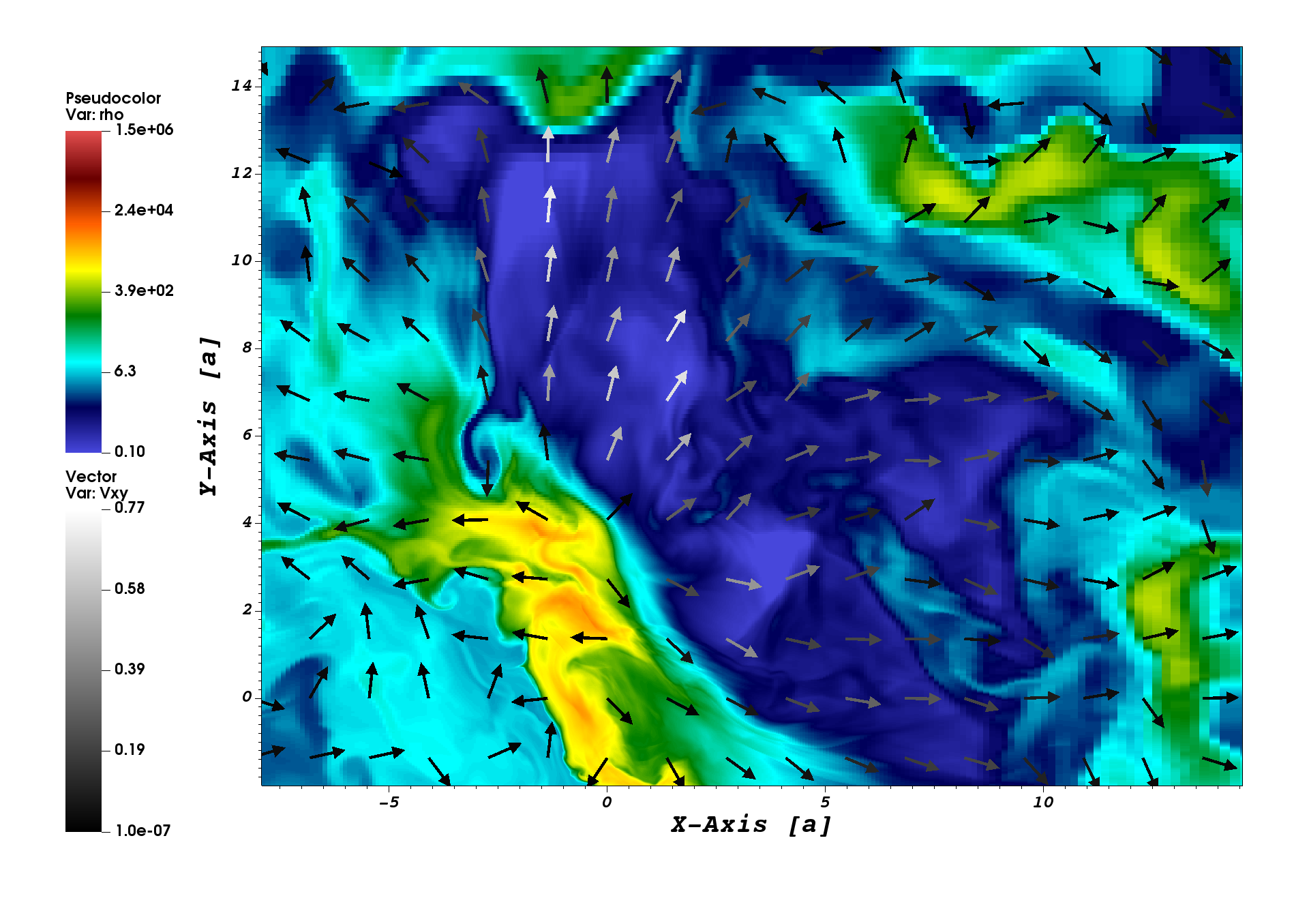}
   \includegraphics[width=7.8cm]{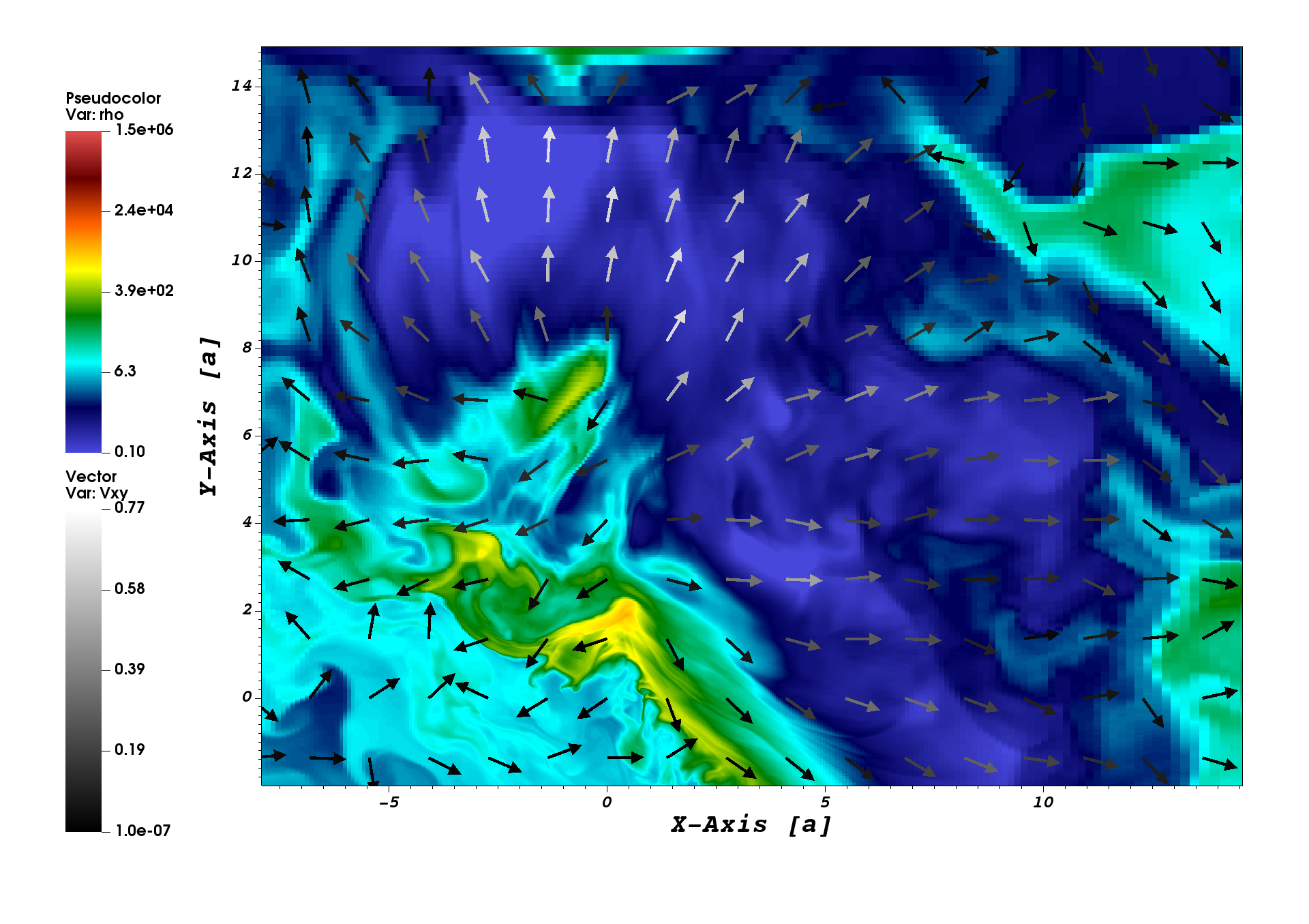}
   \caption{Colored density maps and velocity vector distributions for the relativistic jet. The panels, from left to right and from top to bottom, correspond to 2D cuts parallel to the $xy$-plane at heights $z = 0.1\,a$, $2\,a$, $4\,a$, $6\,a$, $8\,a$, $10\,a$, $12\,a$, and $14\,a$ (the locations of the star and the CO can be identified in the map with $z=0.1\,a$ -top left panel-).}
   \label{fig:rhoVxy}%
\end{figure*}

\begin{figure*}
 \centering
   \includegraphics[width=8.0cm]{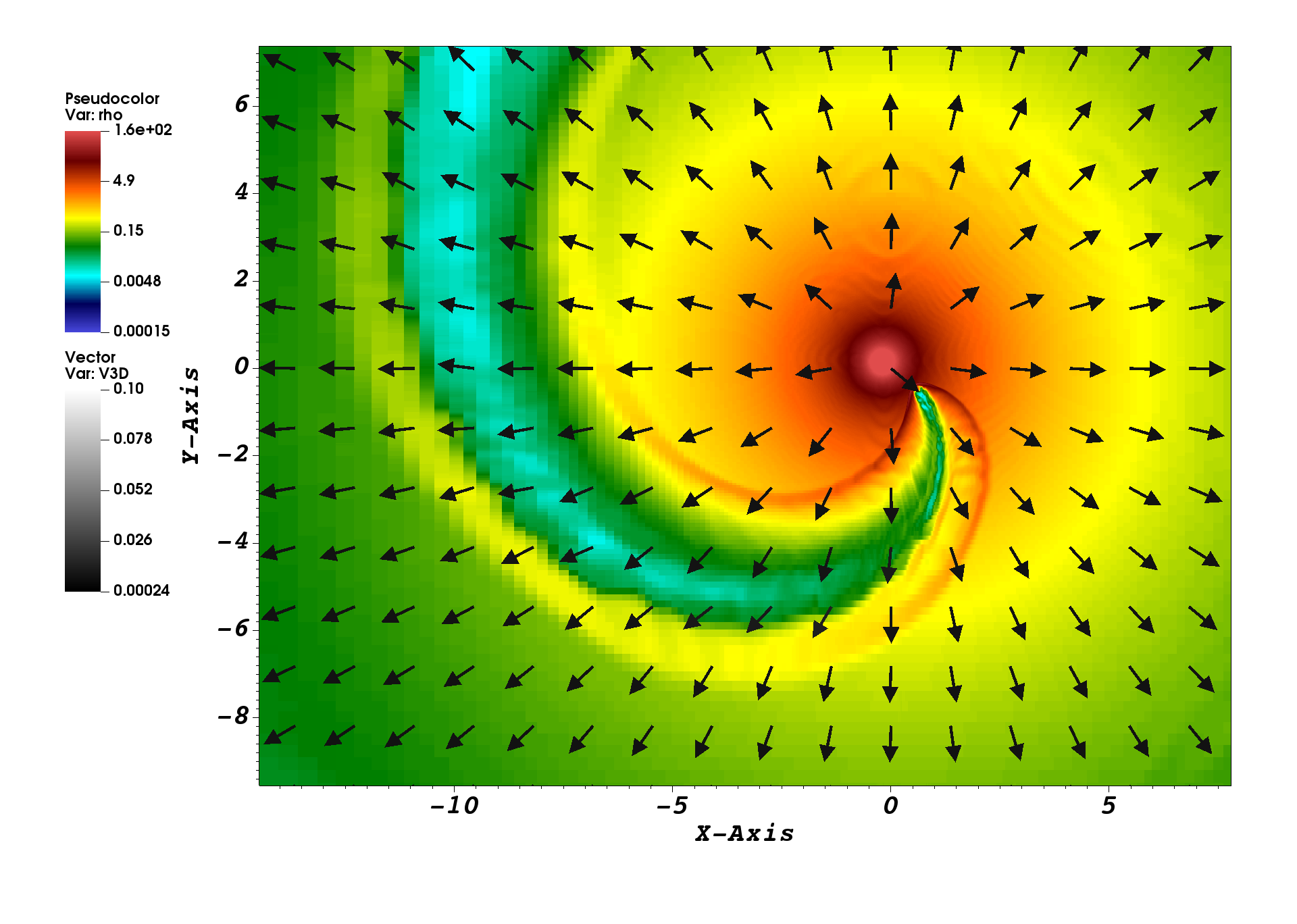}
   \includegraphics[width=8.0cm]{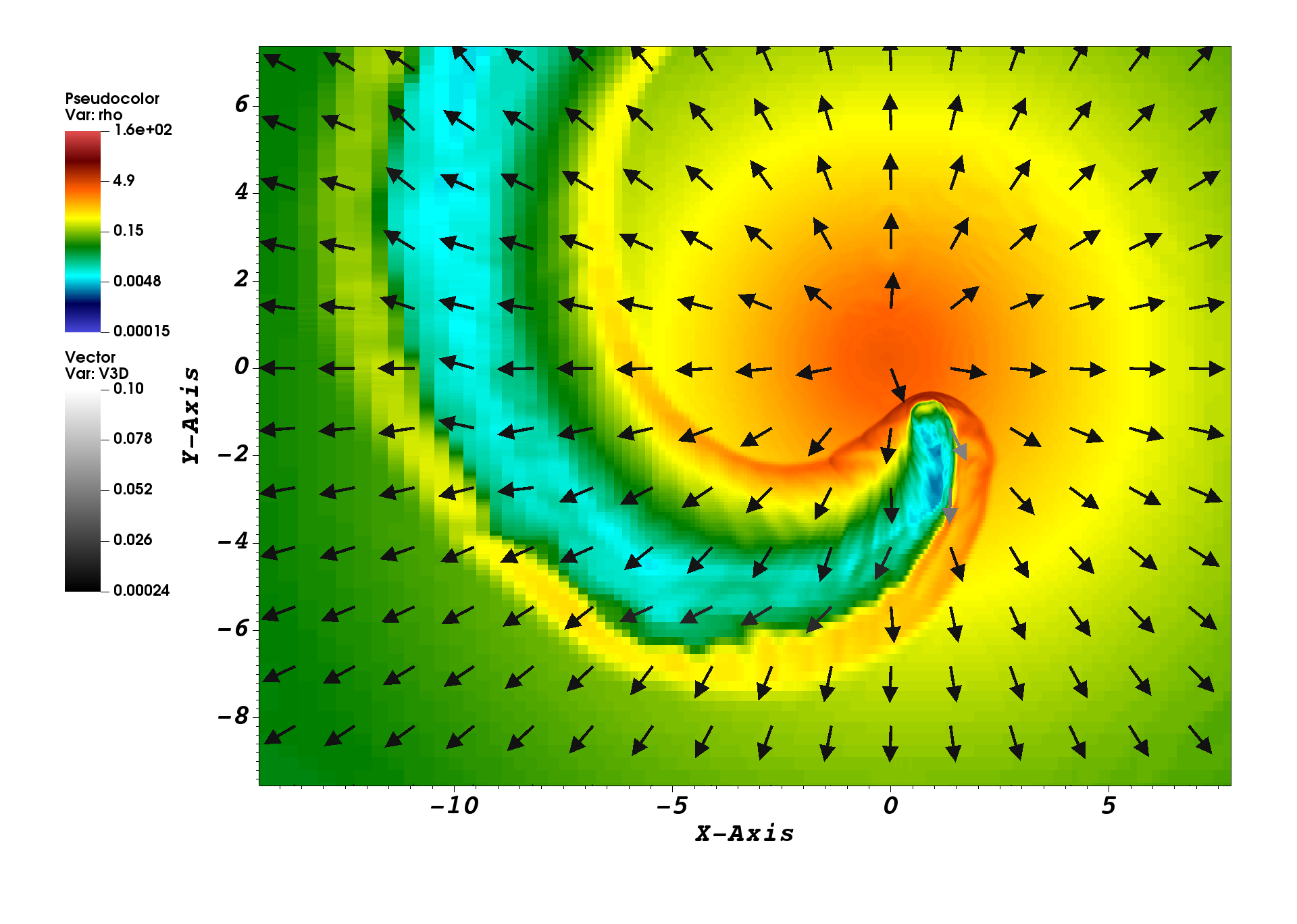}
   \includegraphics[width=8.0cm]{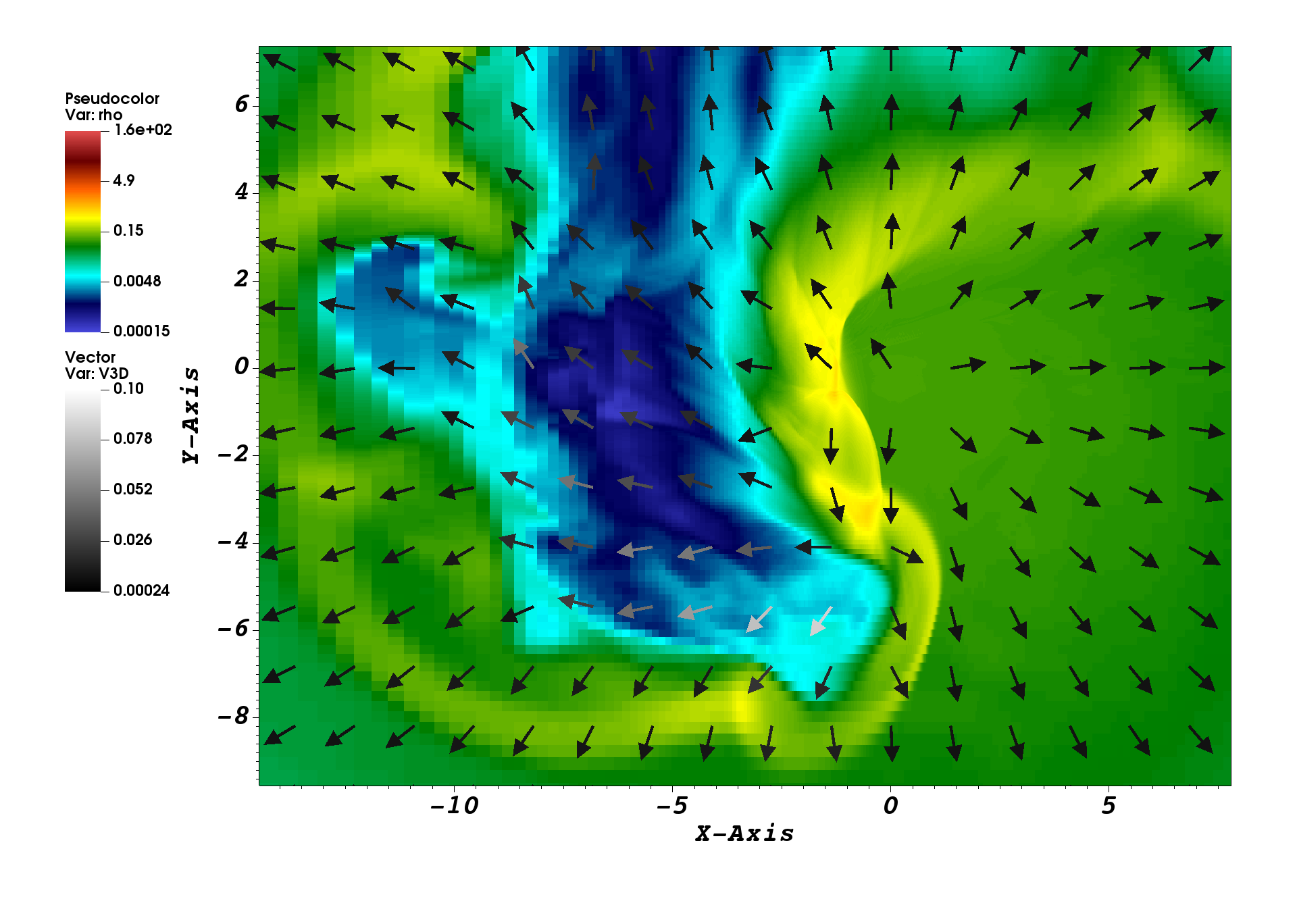}
   \includegraphics[width=8.0cm]{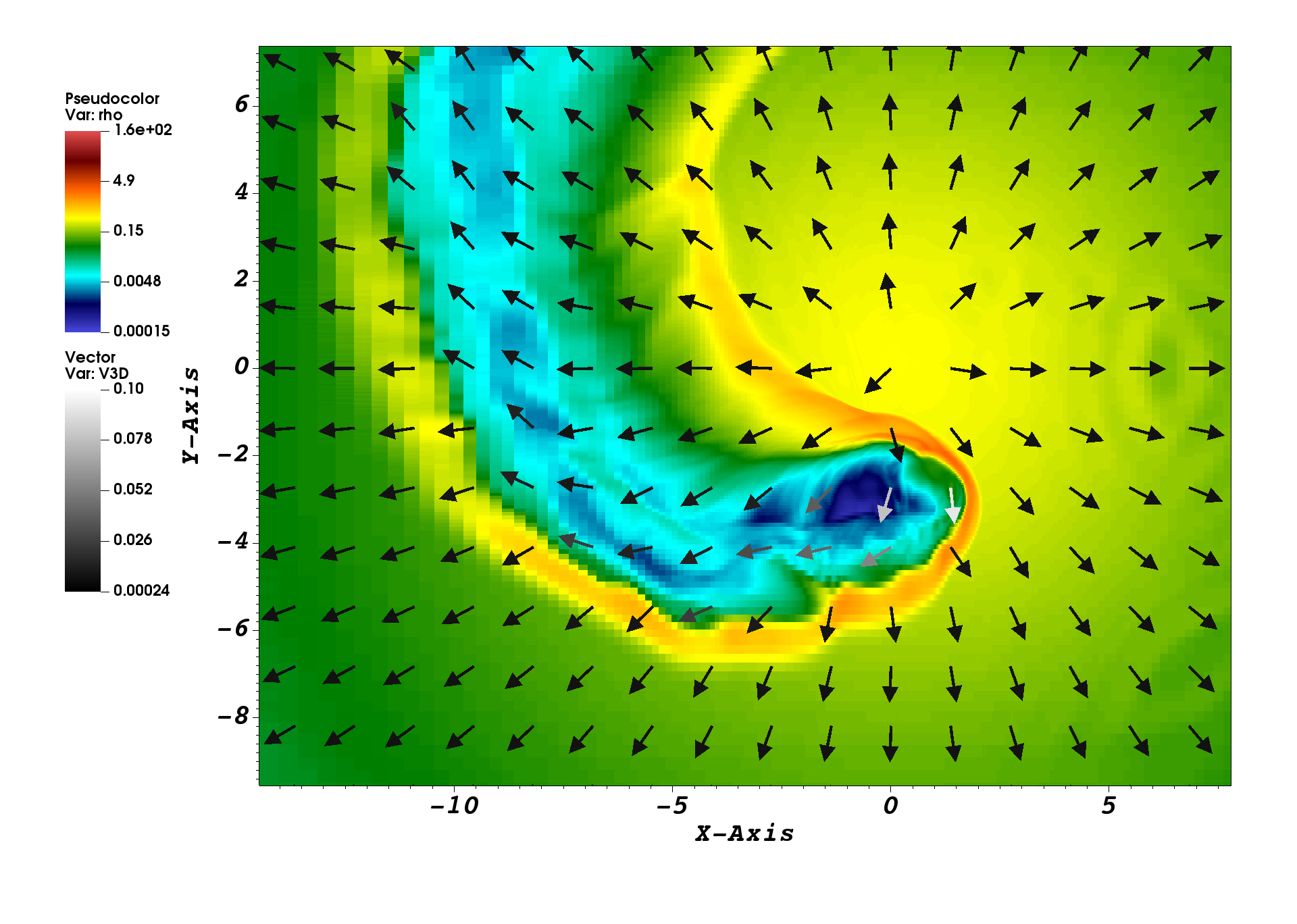}
   \includegraphics[width=8.0cm]{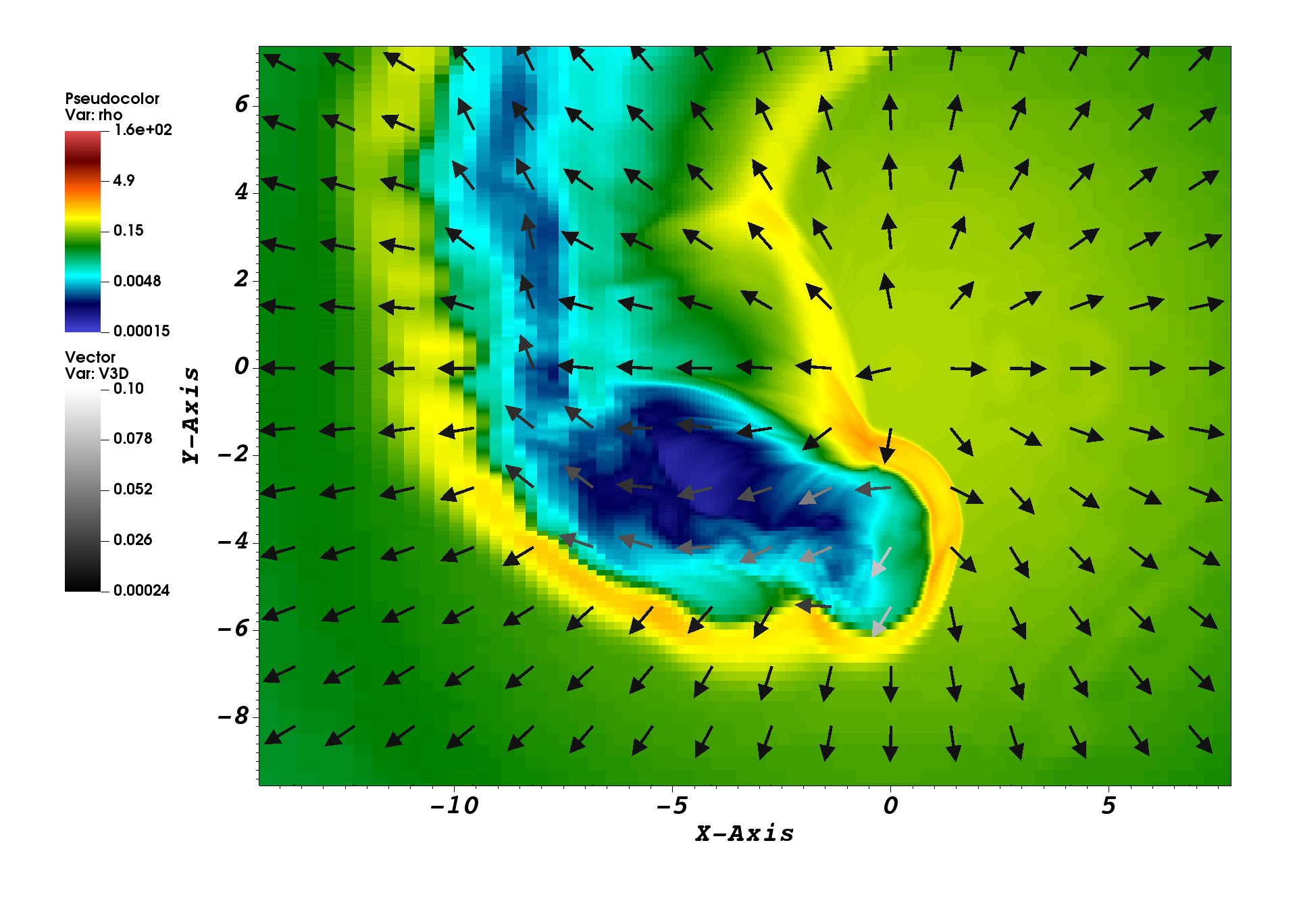}
   \includegraphics[width=8.0cm]{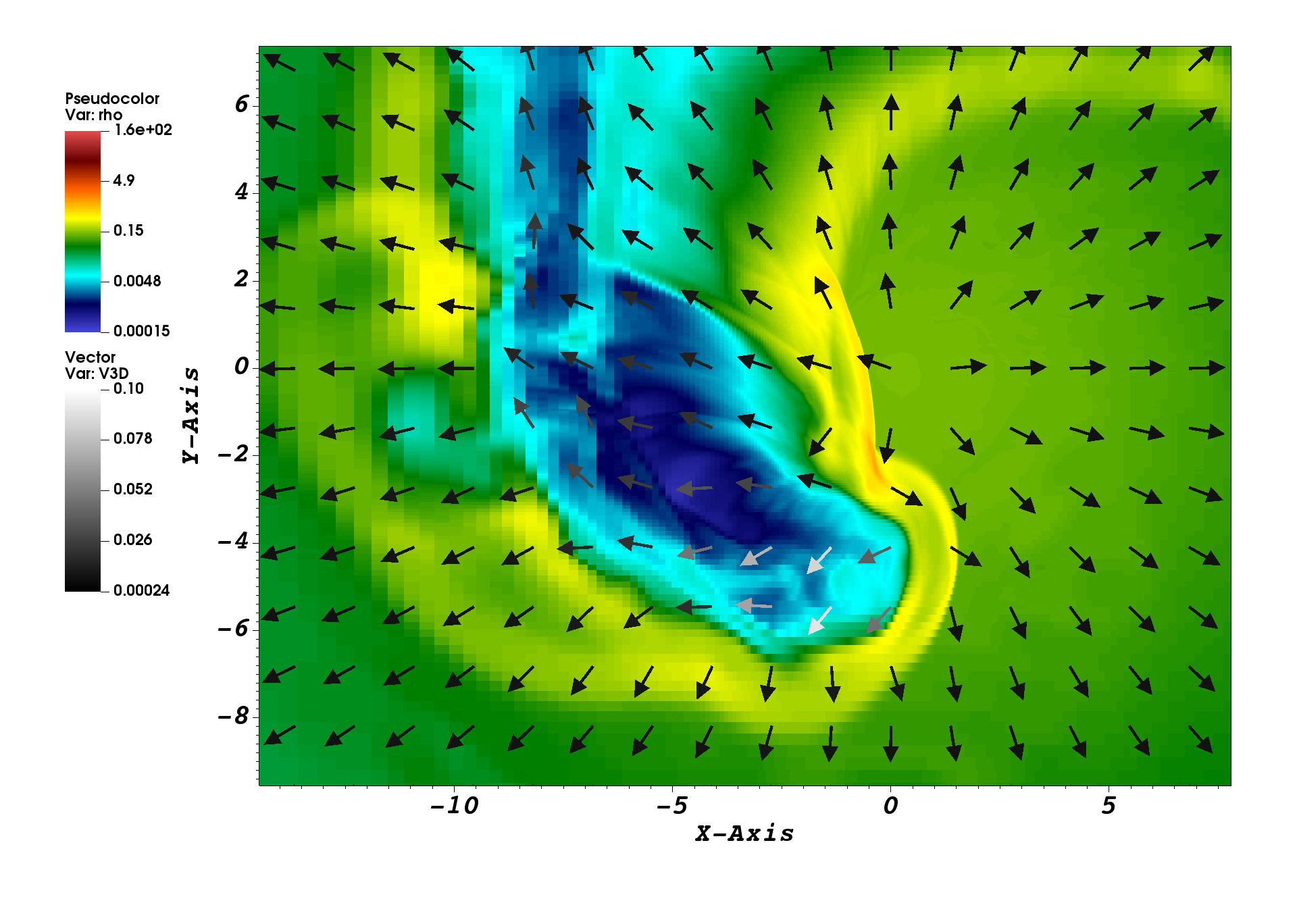}
   \includegraphics[width=8.0cm]{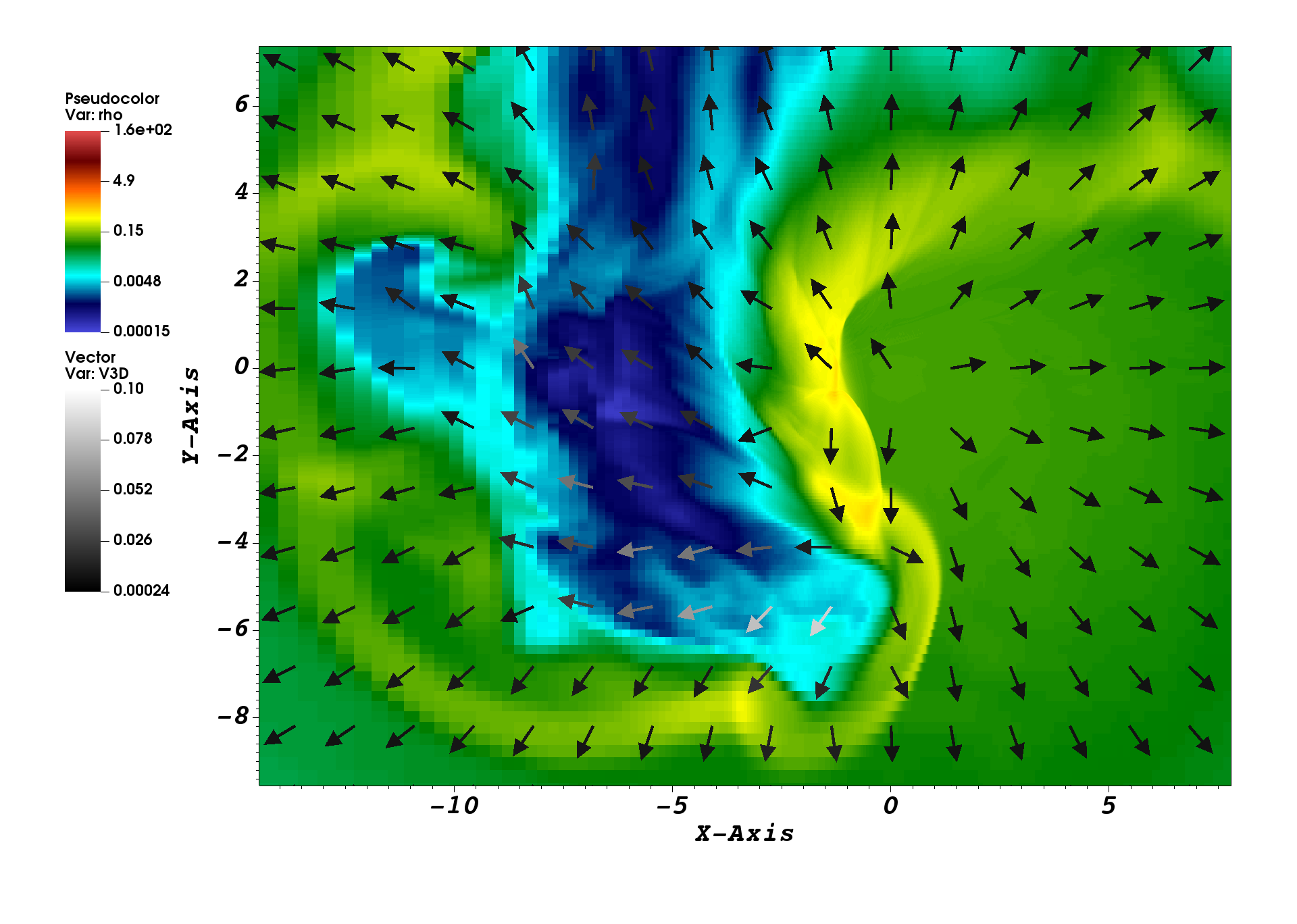}
   \includegraphics[width=8.0cm]{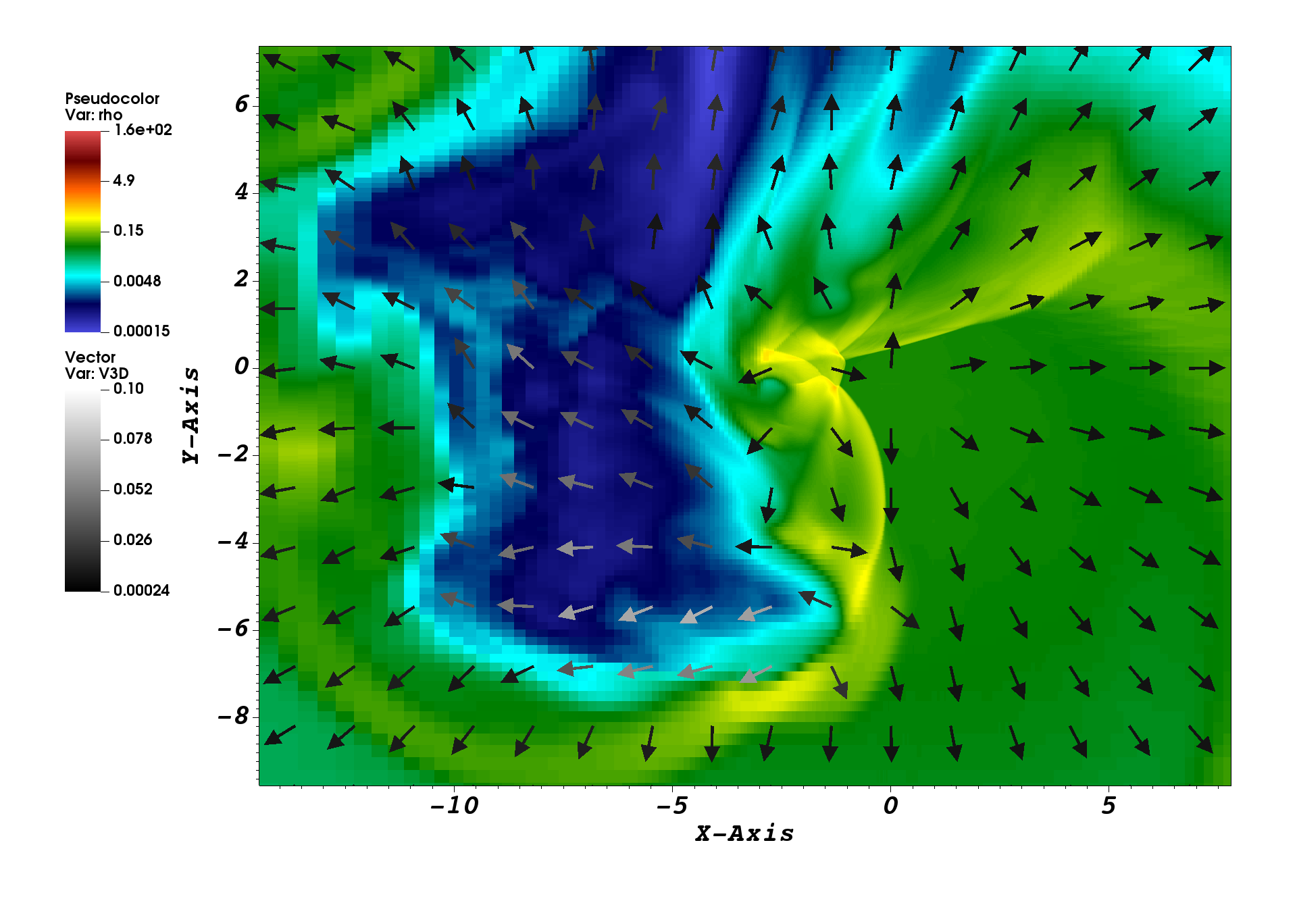}
   \caption{The same as shown in Fig.~\ref{fig:rhoVxy} but for the weakly relativistic jet (low resolution).}
   \label{fig:rhoVxylrNR}%
\end{figure*}

\begin{figure*}
 \centering
   \includegraphics[width=8.0cm]{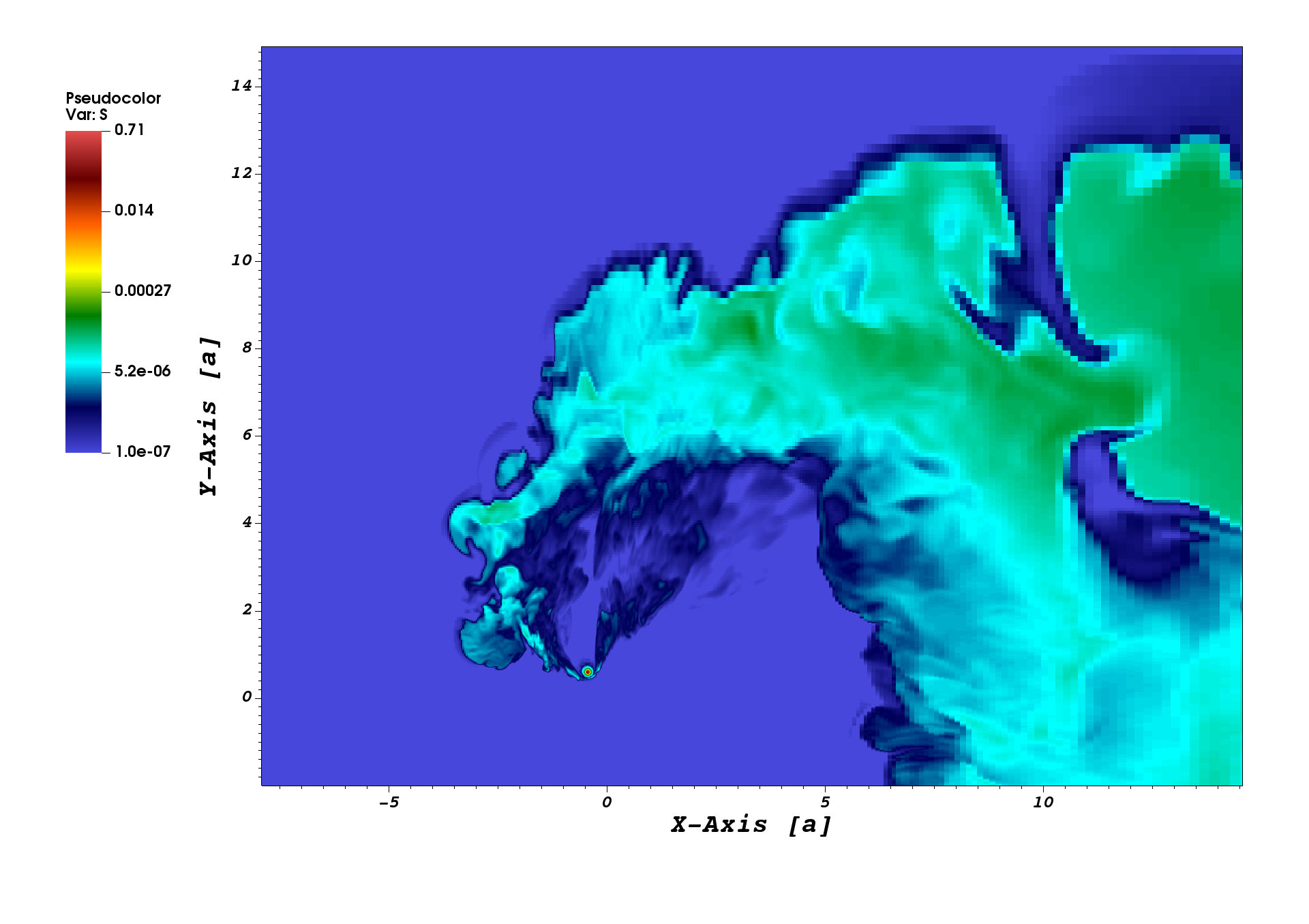}
   \includegraphics[width=8.0cm]{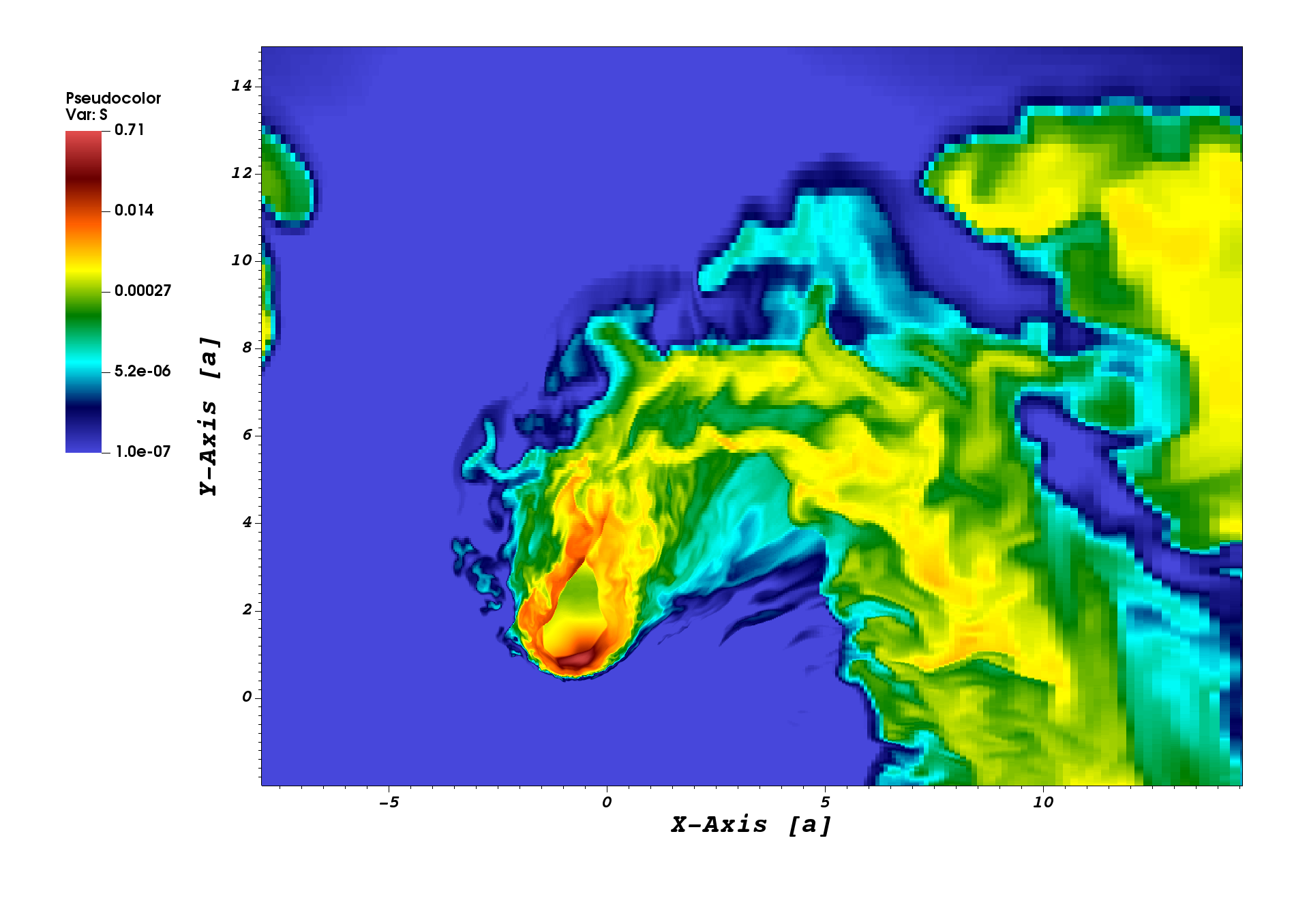}
   \includegraphics[width=8.0cm]{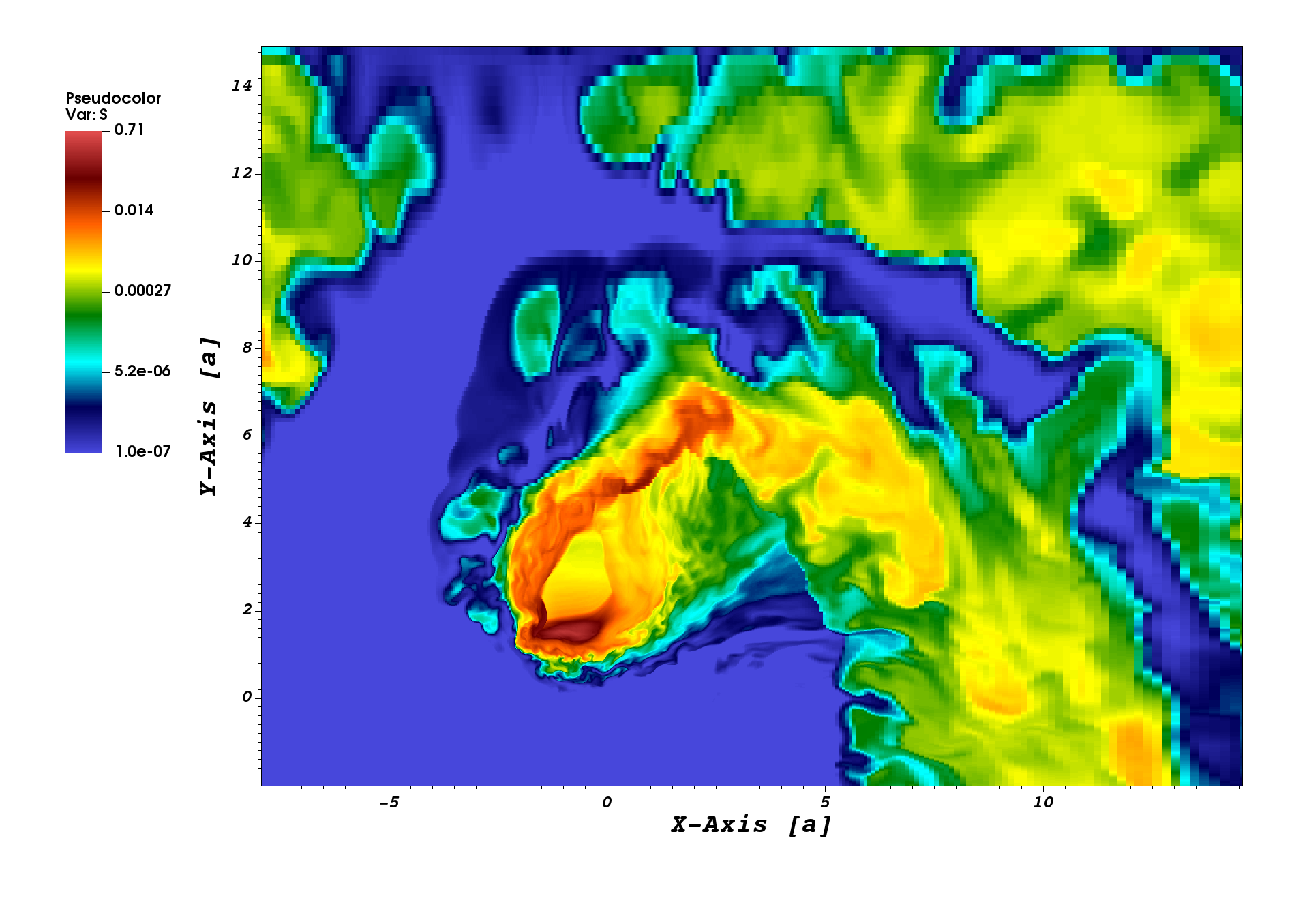}
   \includegraphics[width=8.0cm]{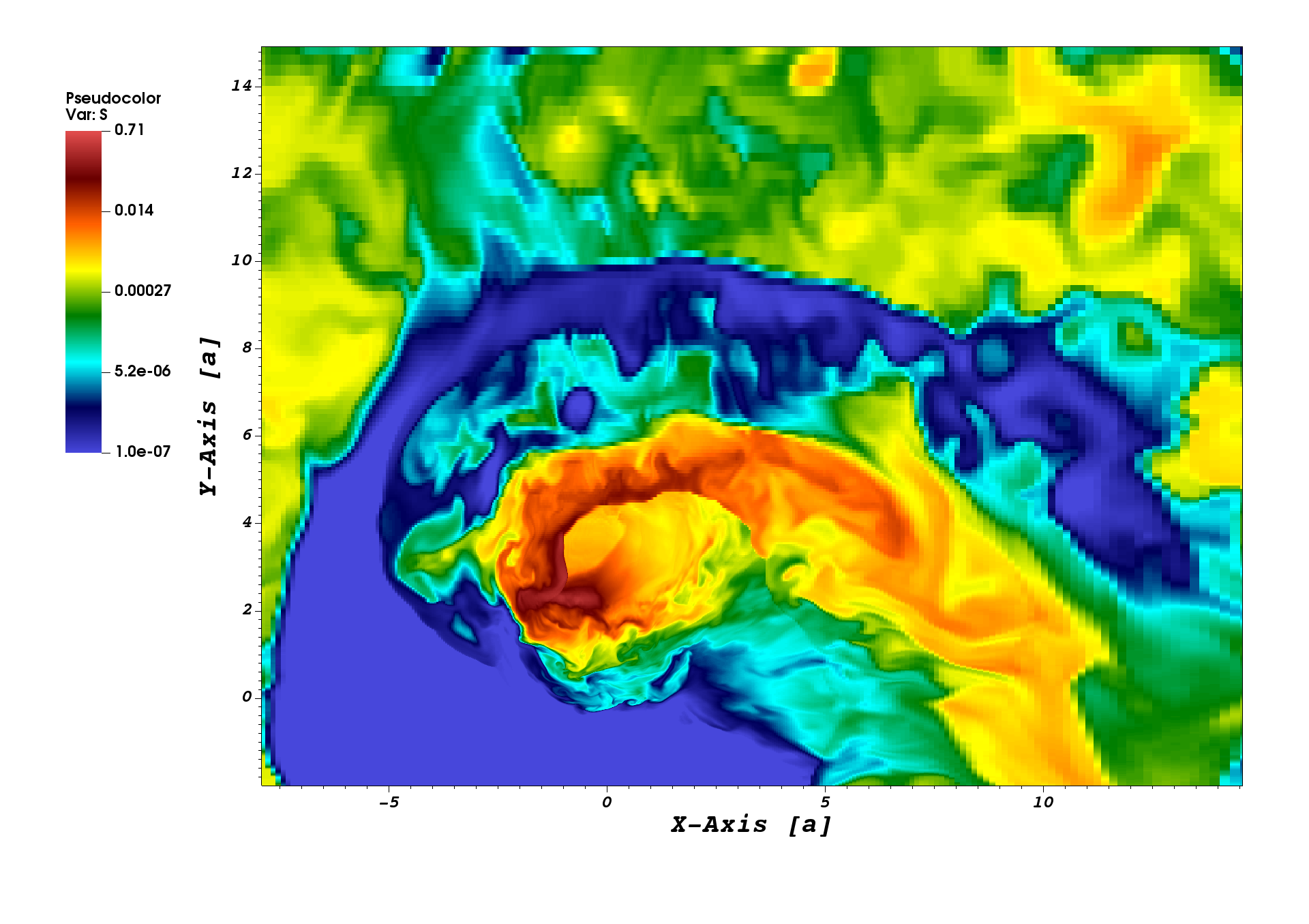}
   \includegraphics[width=8.0cm]{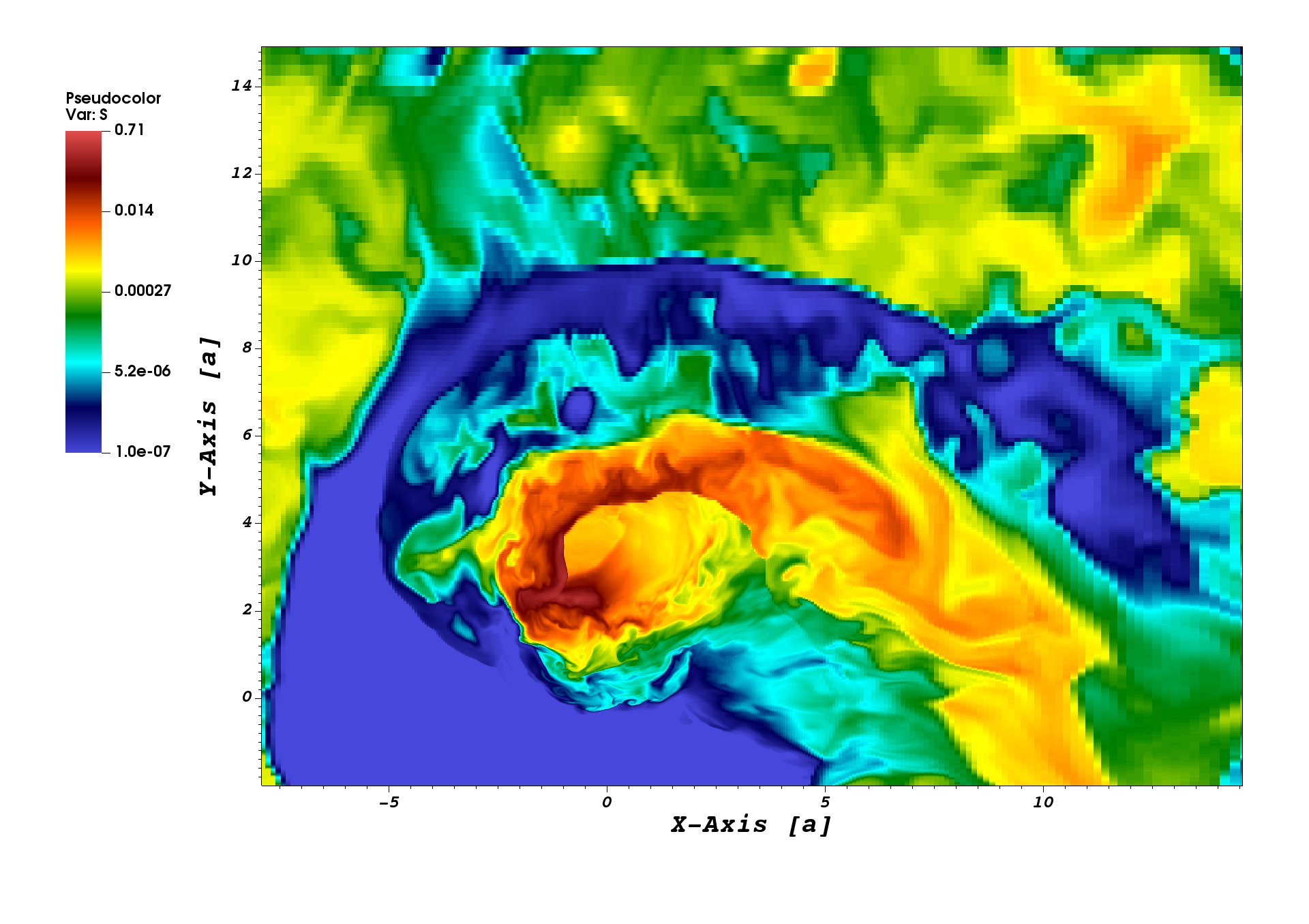}
   \includegraphics[width=8.0cm]{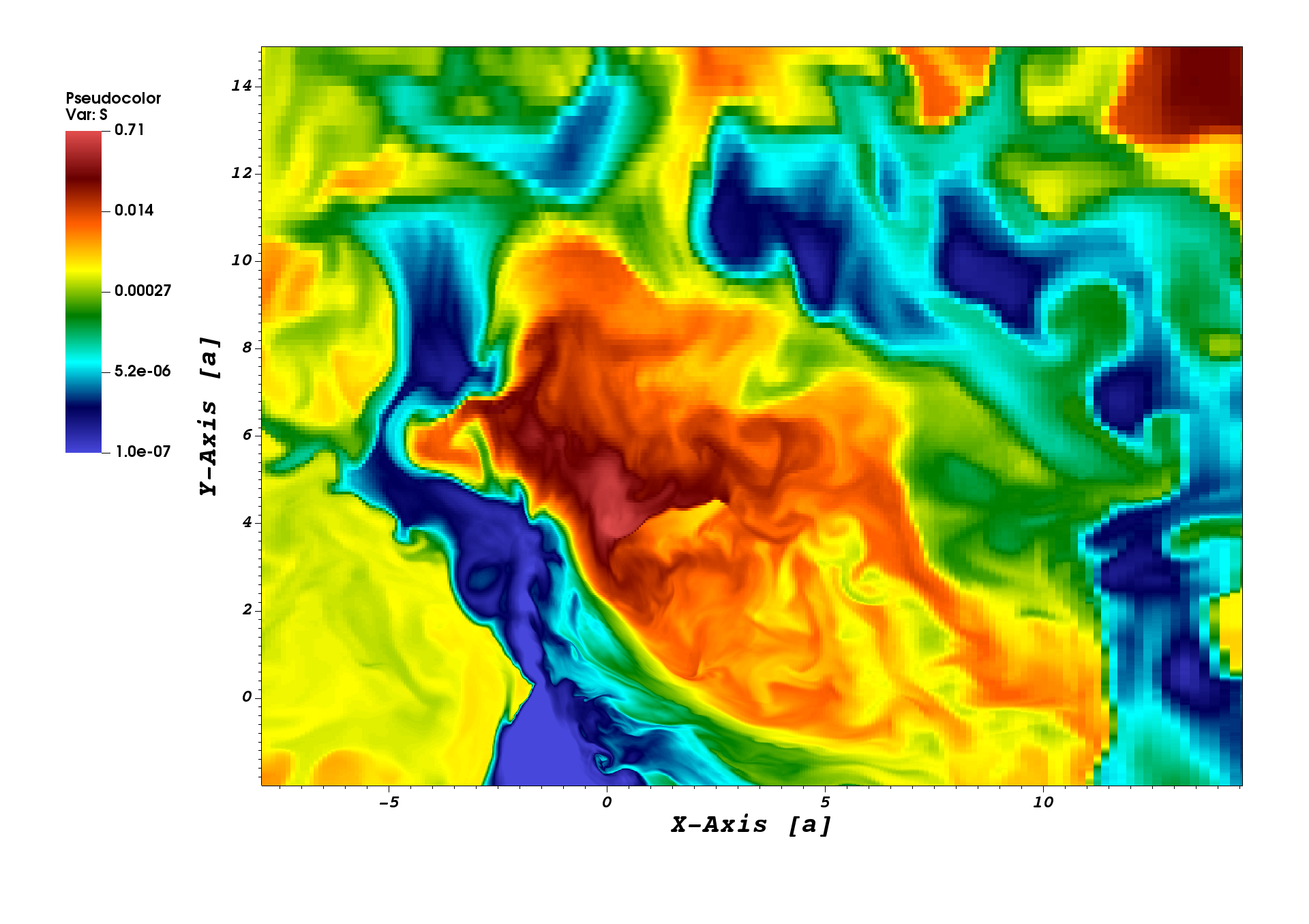}
   \includegraphics[width=8.0cm]{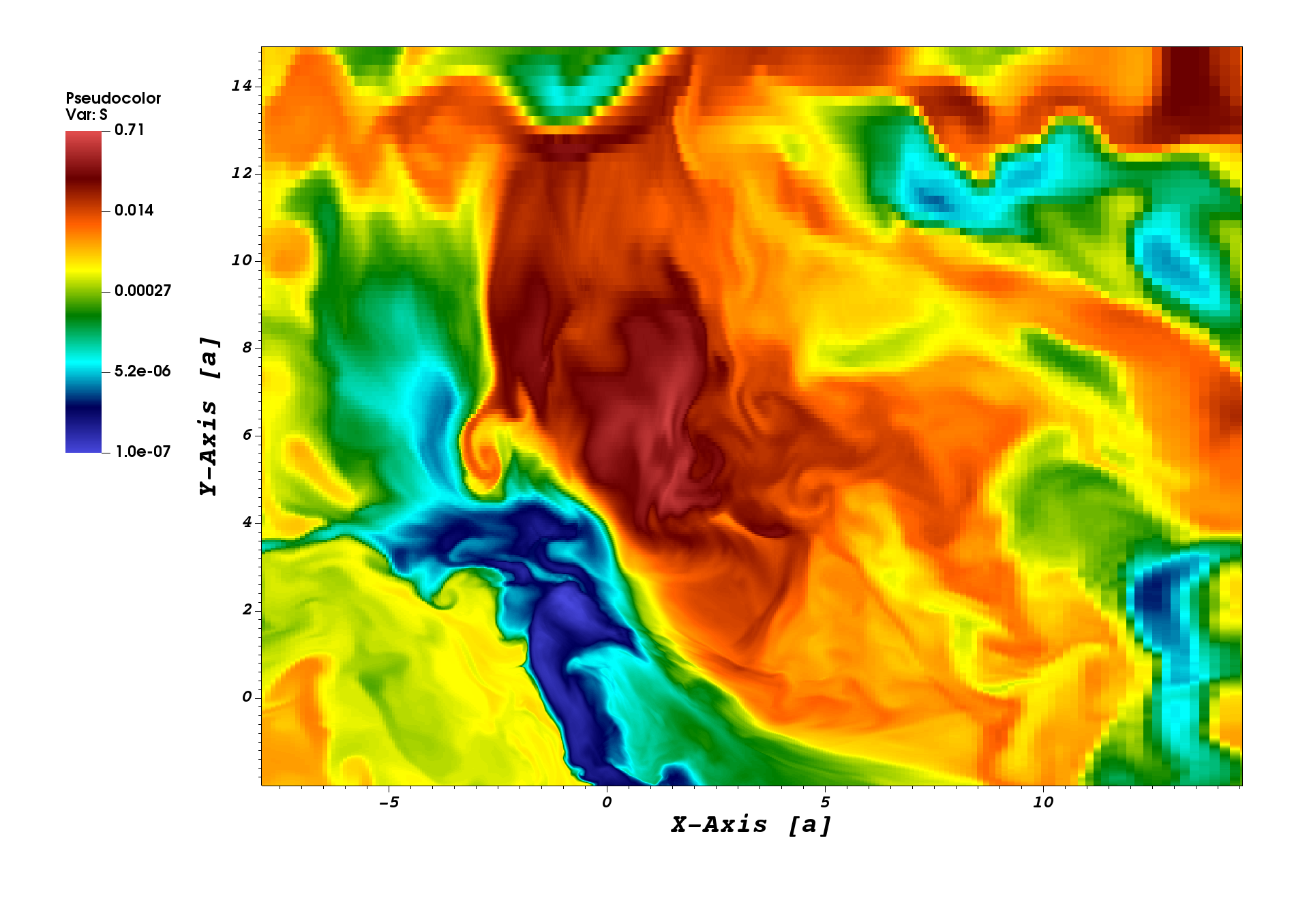}
   \includegraphics[width=8.0cm]{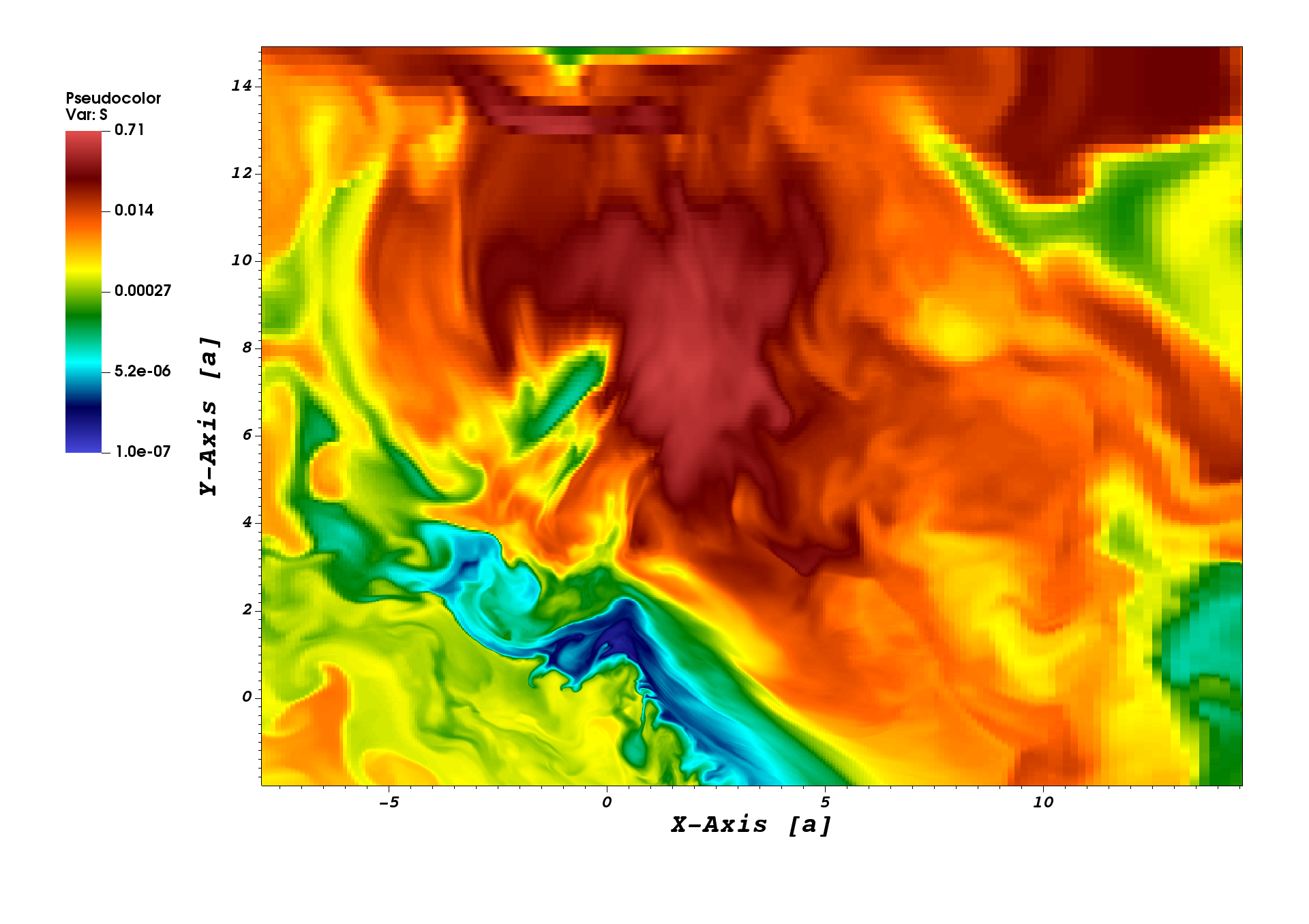}
   \caption{The same as shown in Fig.~\ref{fig:rhoVxy} but for the measure of flow disorder $S=P/\rho^{4/3}$ and without velocity vectors (see text -Sect.~\ref{res}-).}
   \label{fig:S}%
\end{figure*}

\begin{figure*}
 \centering
   \includegraphics[width=8.0cm]{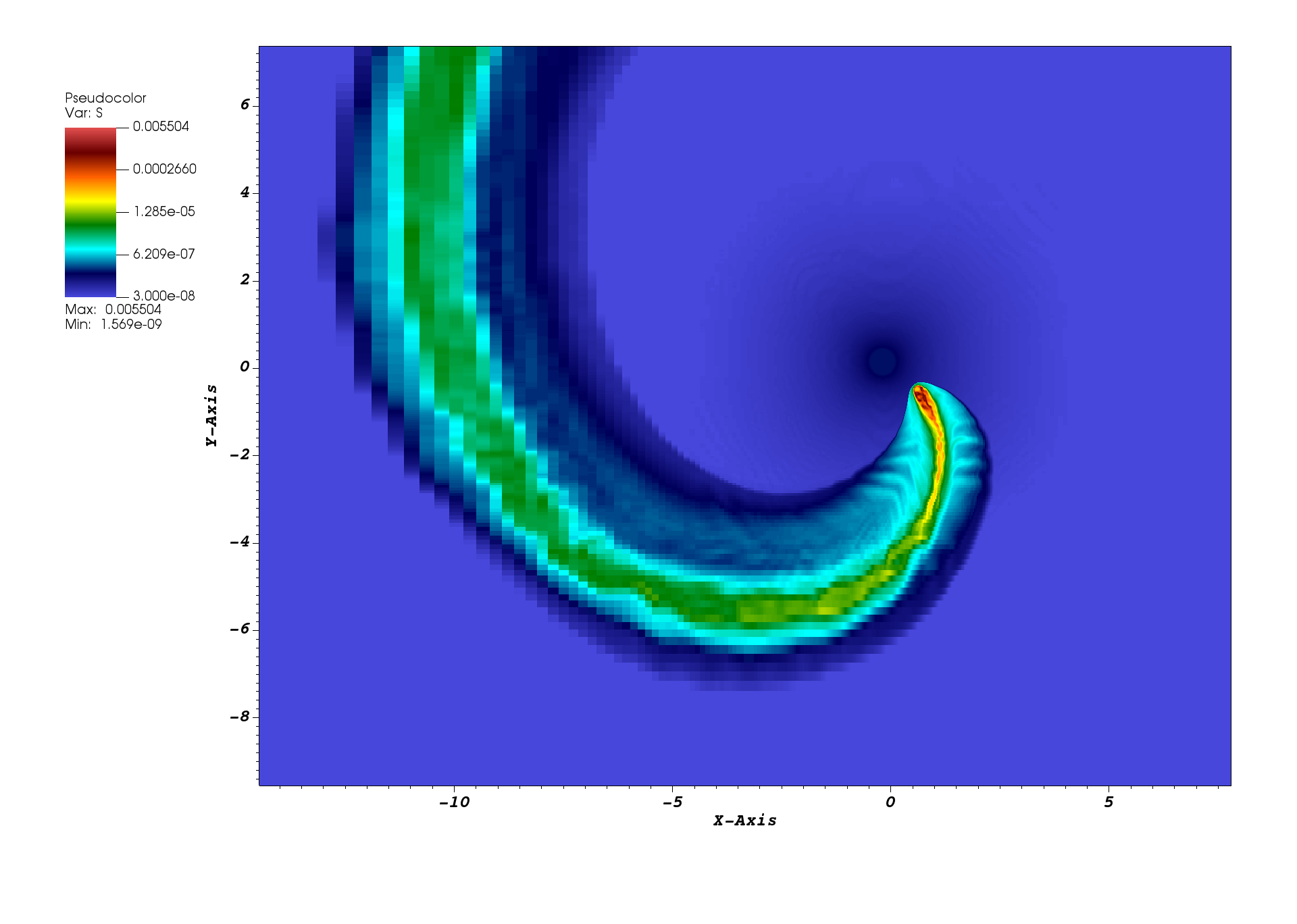}
   \includegraphics[width=8.0cm]{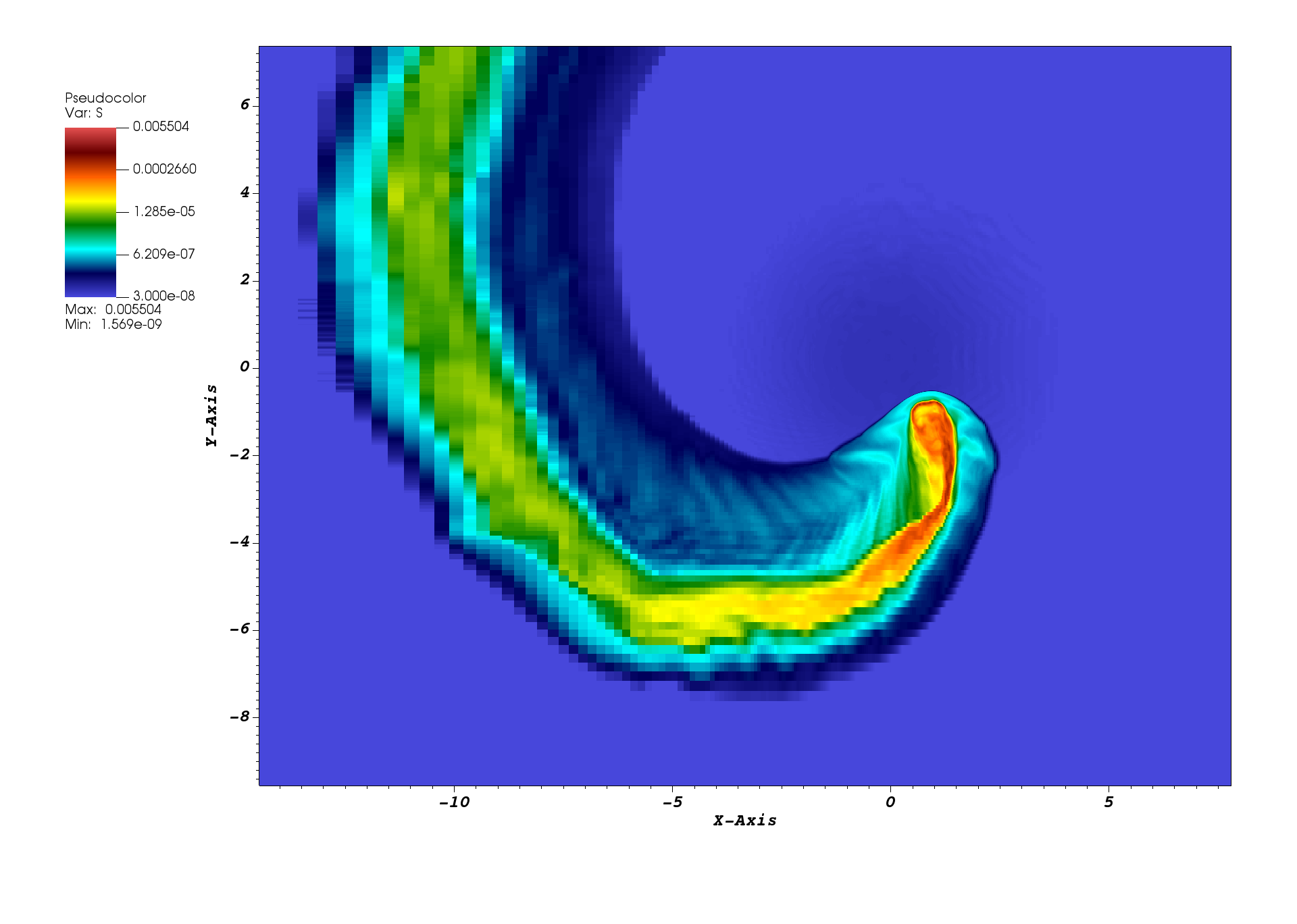}
   \includegraphics[width=8.0cm]{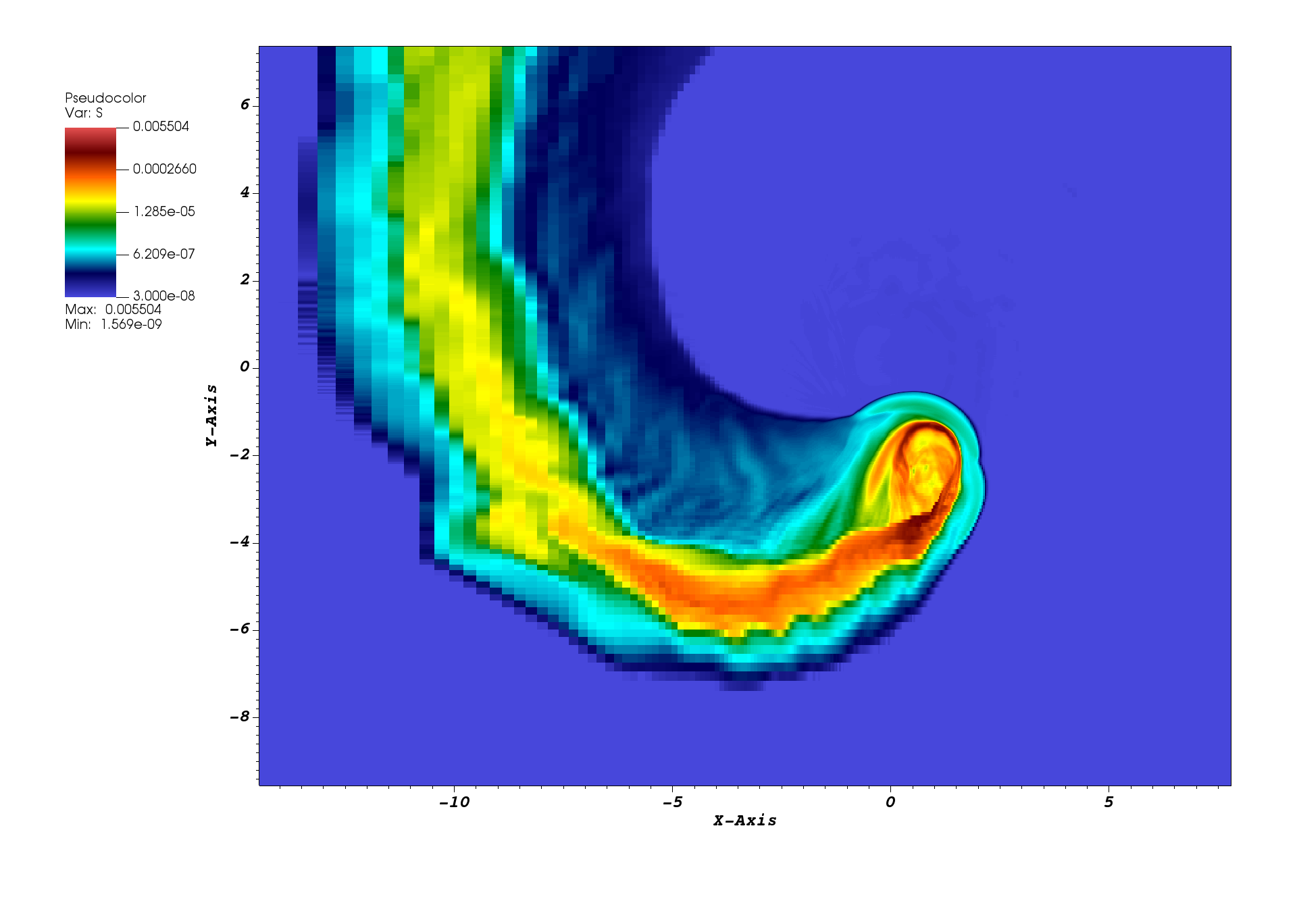}
   \includegraphics[width=8.0cm]{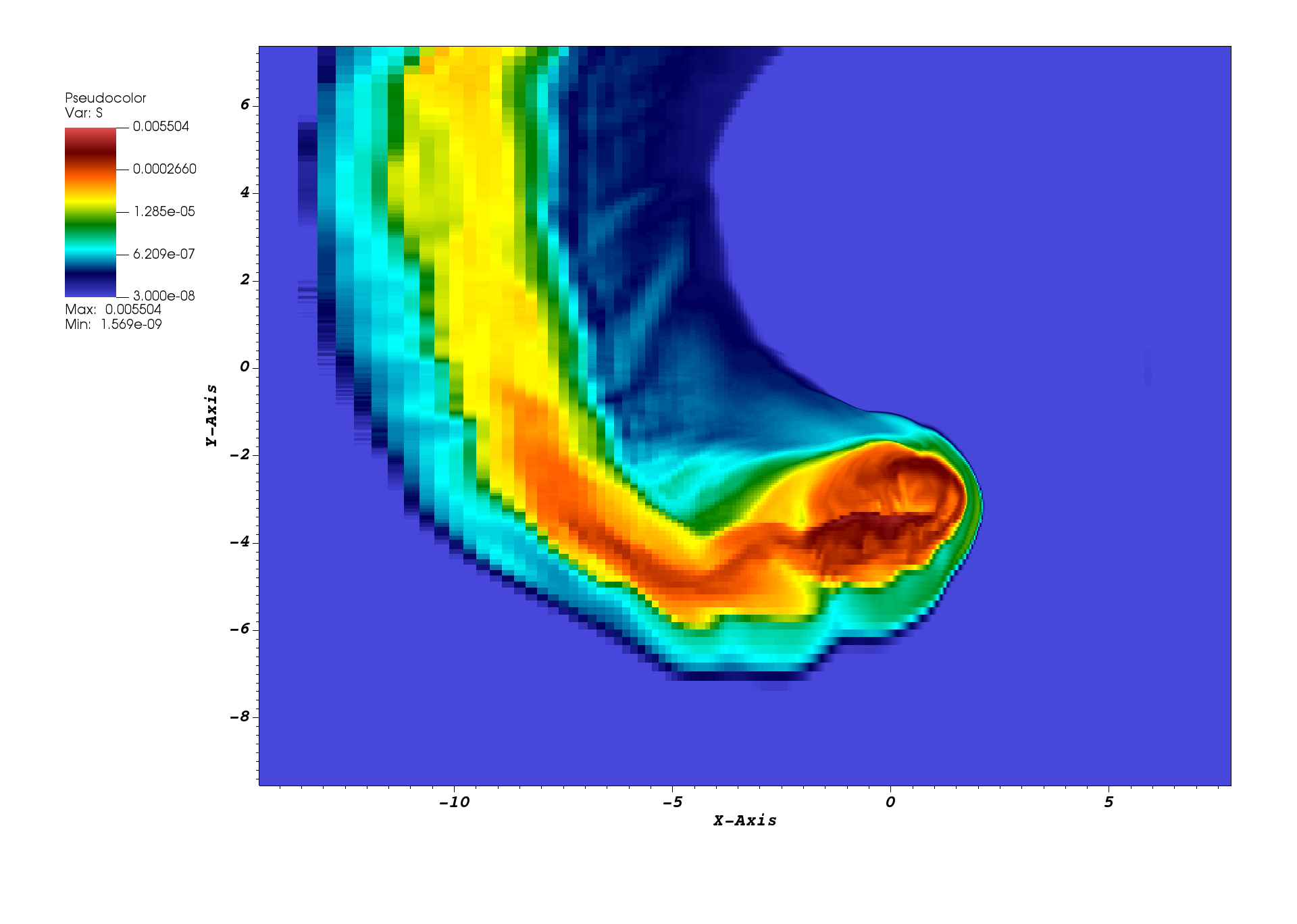}
   \includegraphics[width=8.0cm]{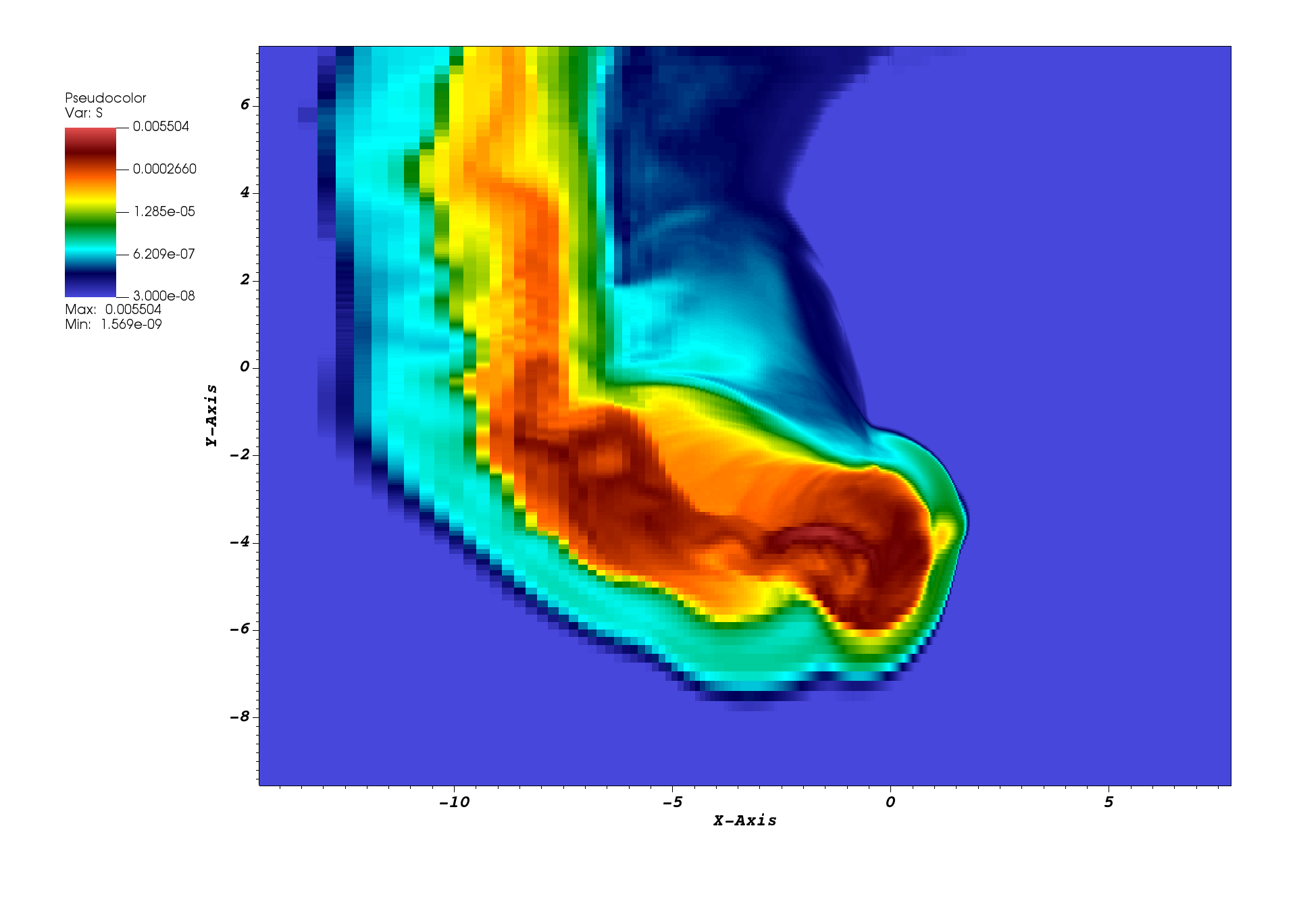}
   \includegraphics[width=8.0cm]{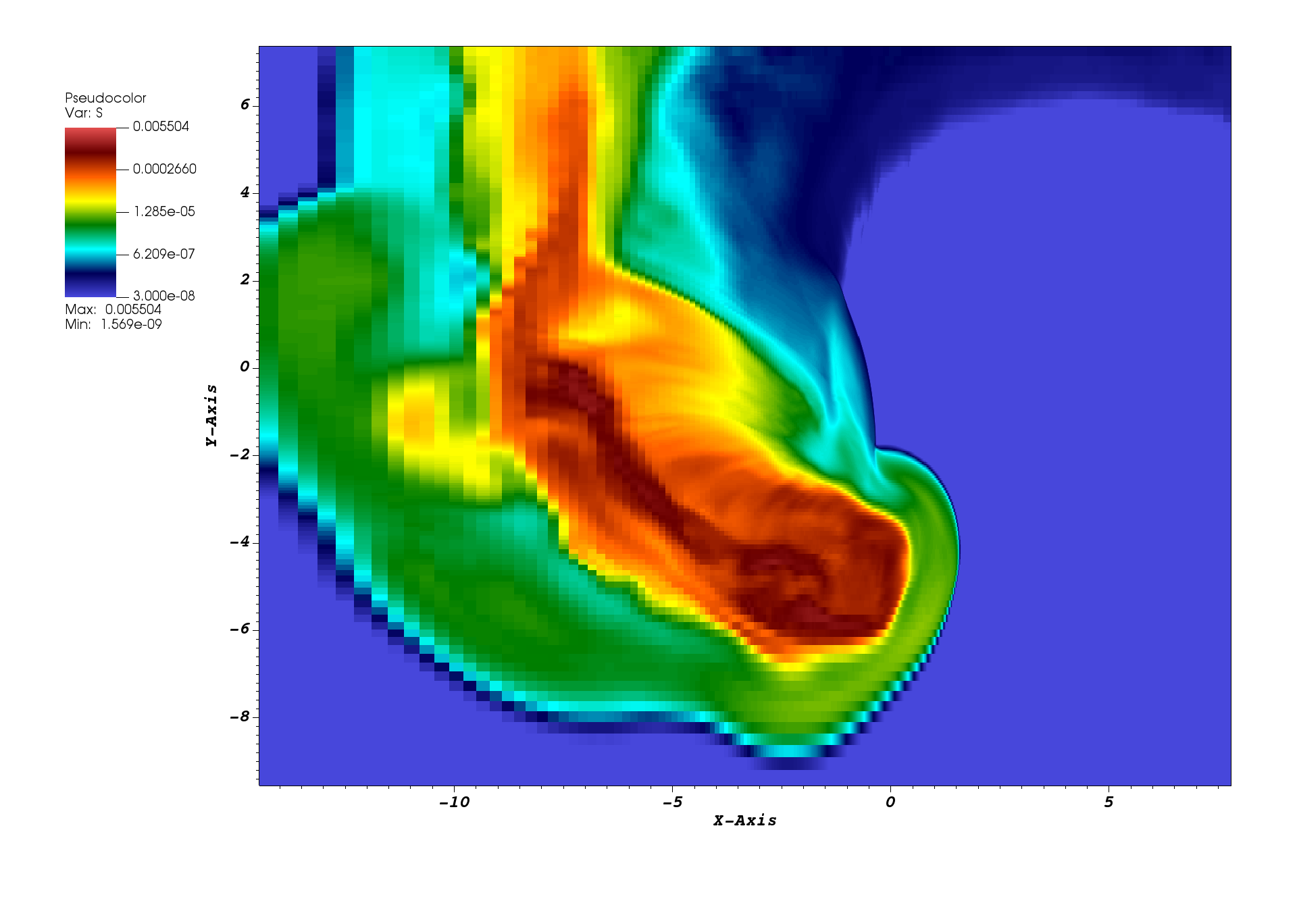}
   \includegraphics[width=8.0cm]{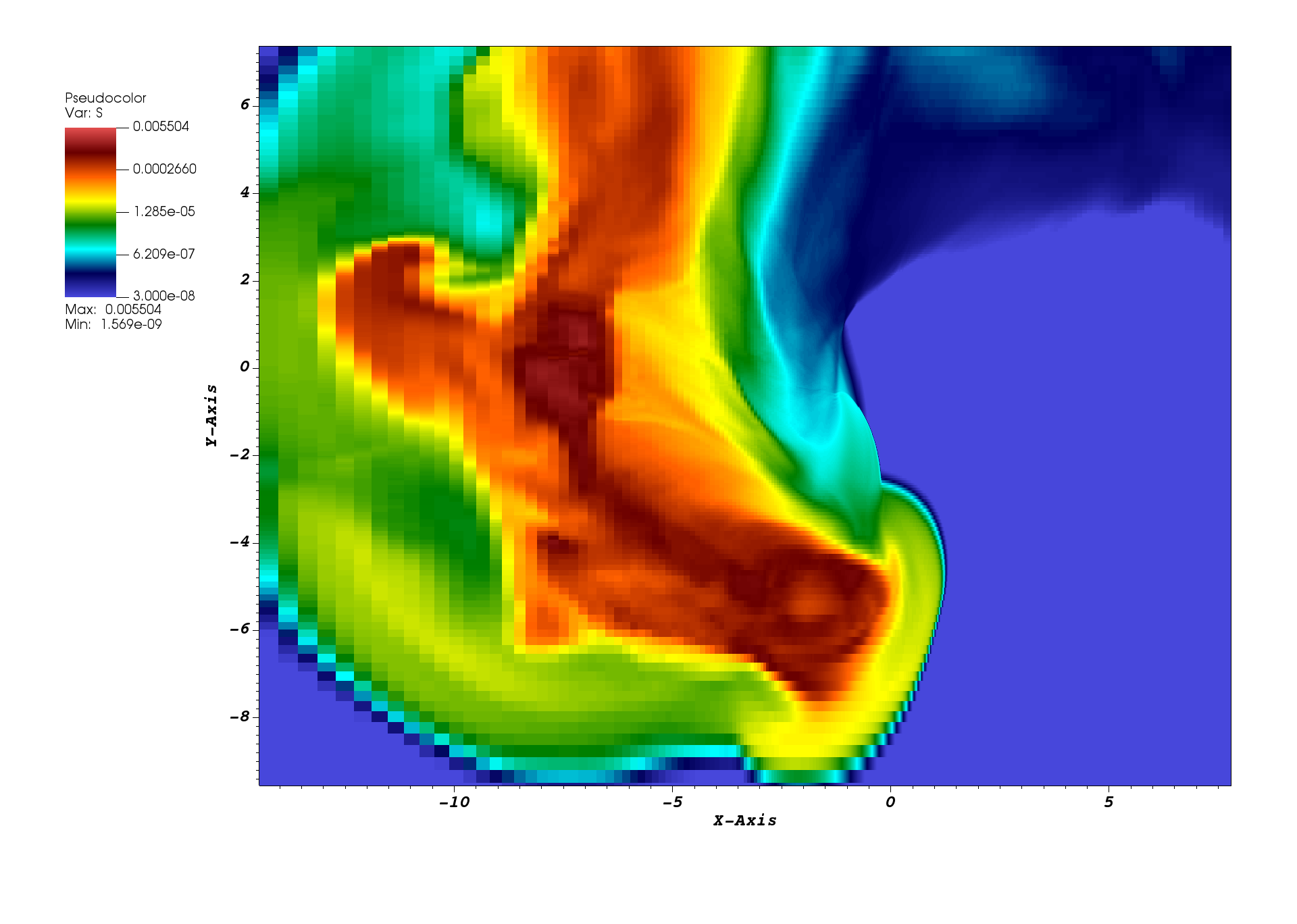}
   \includegraphics[width=8.0cm]{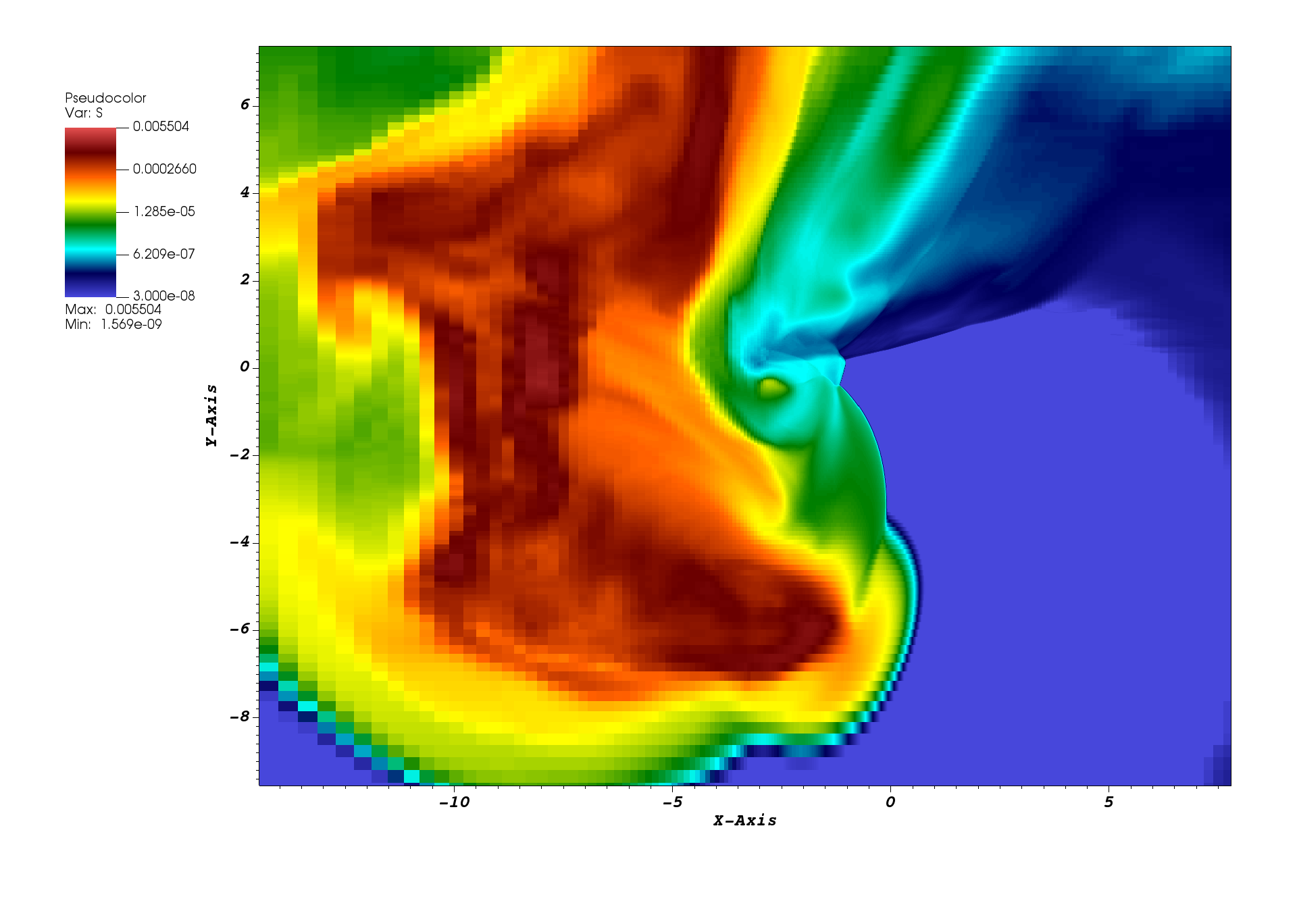}
   \caption{The same as shown in Fig.~\ref{fig:S} but for the weakly relativistic jet (low resolution).}
   \label{fig:SlrNR}%
\end{figure*}
 
\begin{figure}
 \centering
   \includegraphics[width=8.4cm]{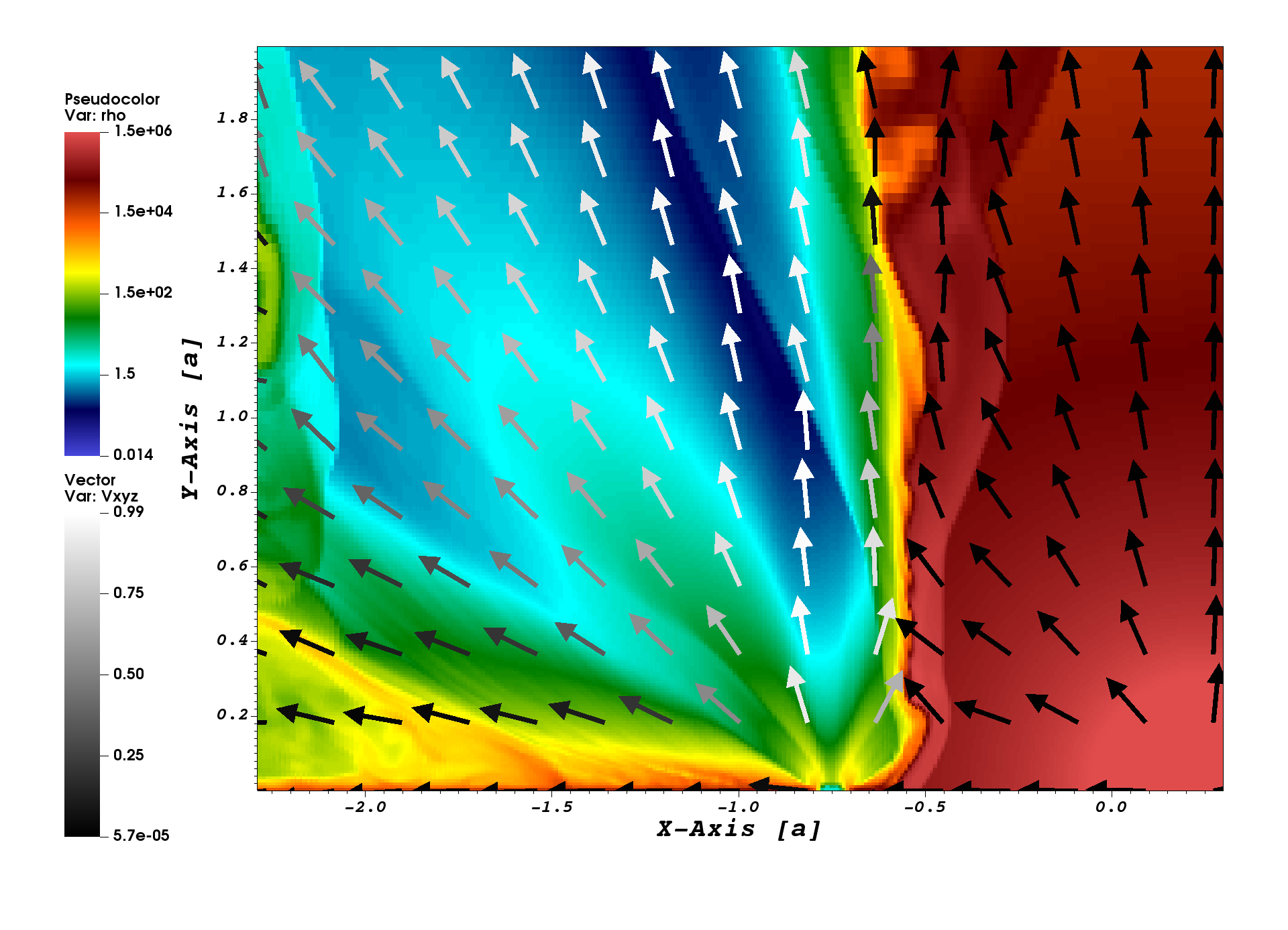}
   \includegraphics[width=8.4cm]{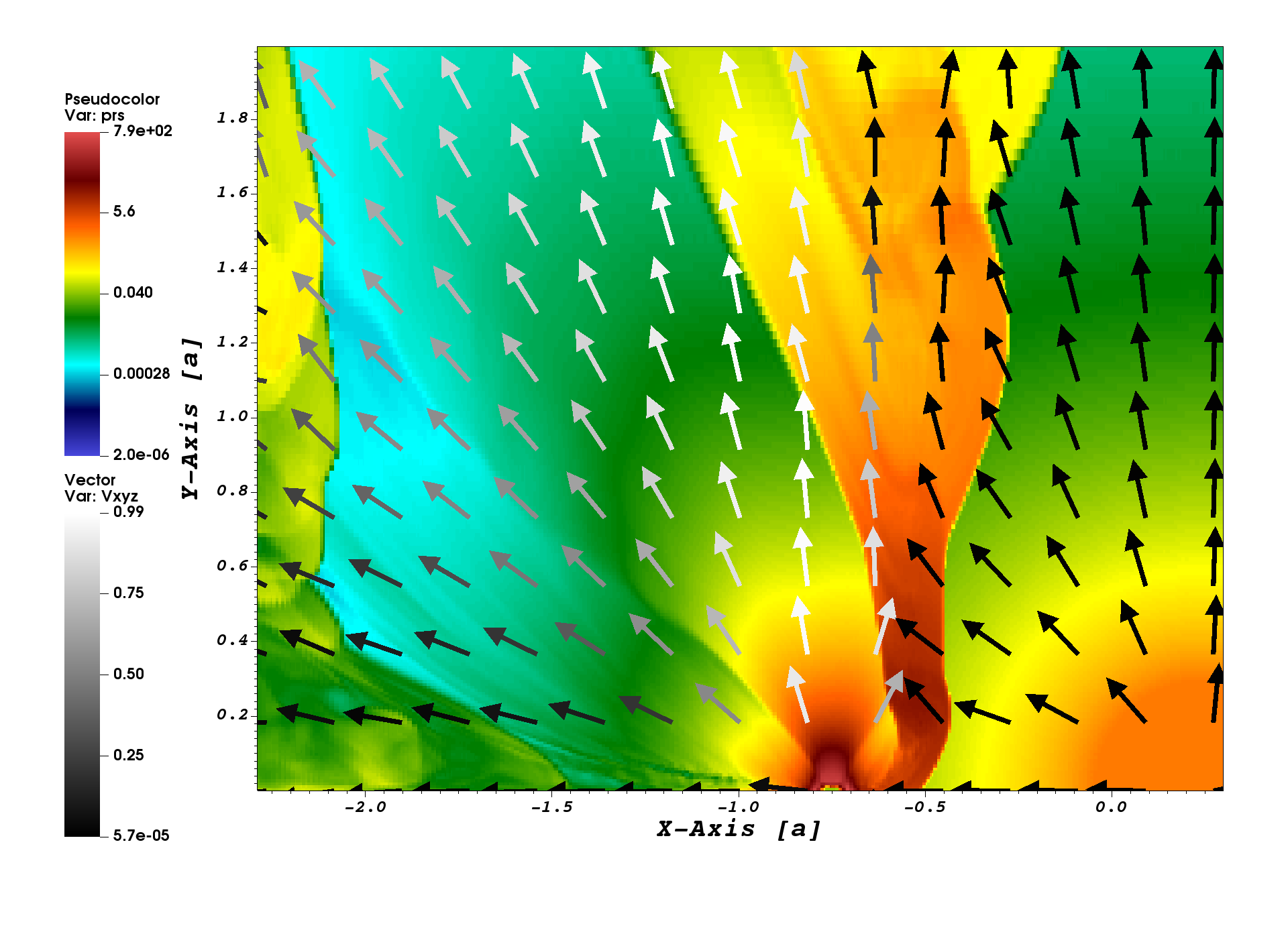}
   \includegraphics[width=8.4cm]{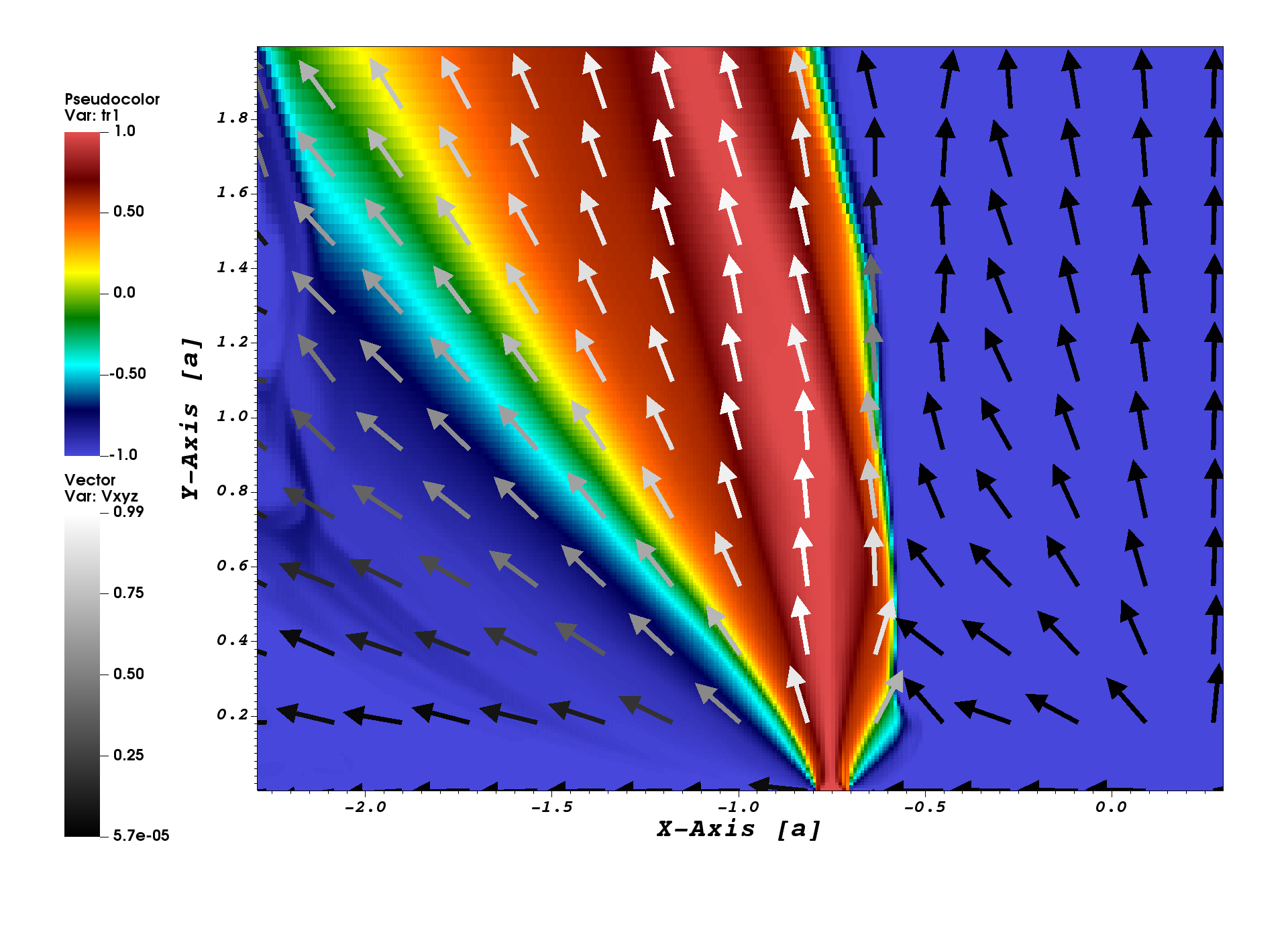}
   \caption{Zoomed-in, colored density (top panel), pressure (middle panel), and tracer (bottom panel) maps, and velocity vector distributions, in the case of the relativistic jet, for 2D cuts whose vertical axes are the $\hat z$-direction, and the horizontal axes go through the CO and the star, at $(-0.75\,a,0)$ and $(0.25\,a,0)$ in the plot, respectively.}
   \label{fig:HRRzoom}%
\end{figure} 

\begin{figure}
 \centering
   \includegraphics[width=8.4cm]{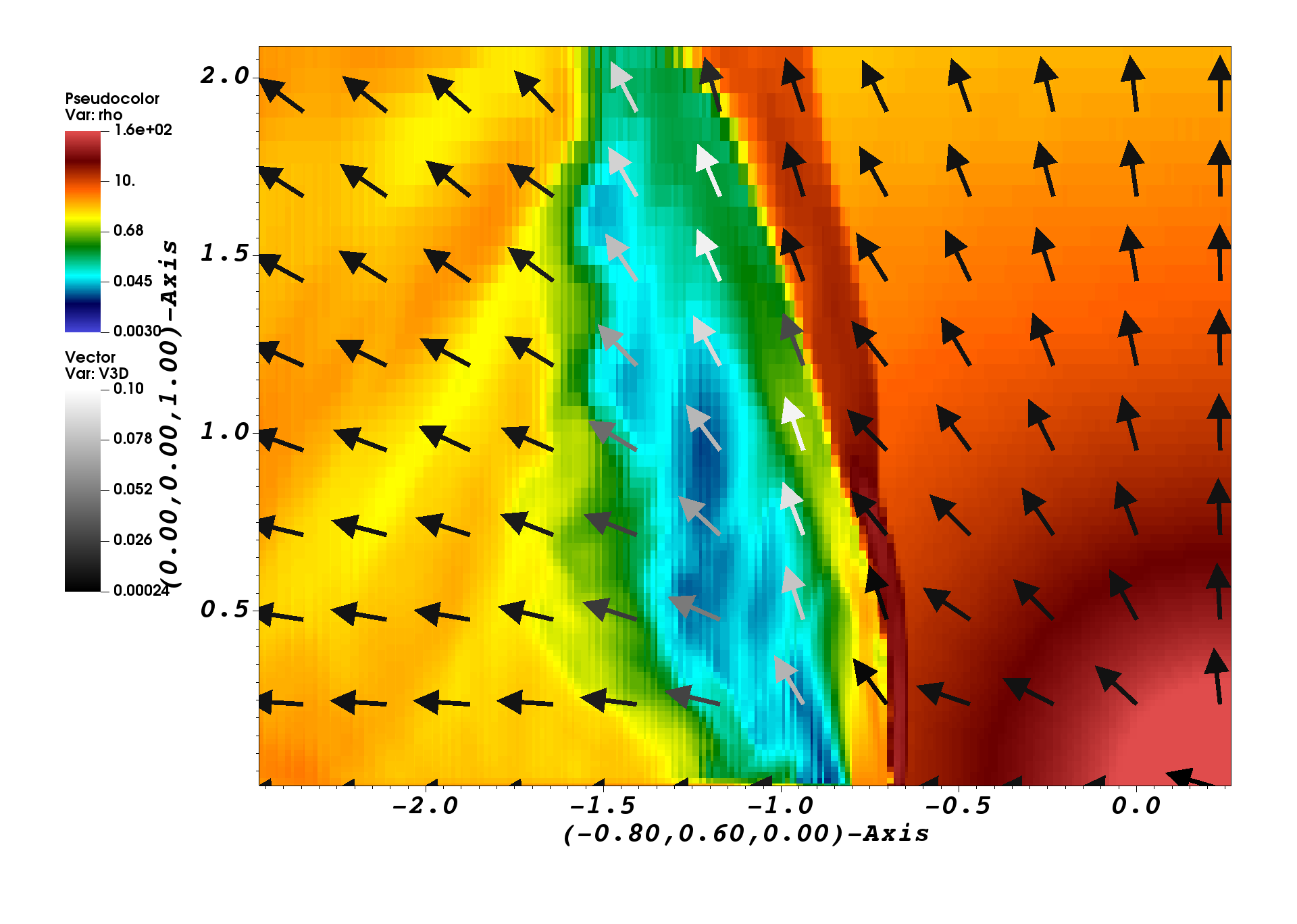}
   \includegraphics[width=8.4cm]{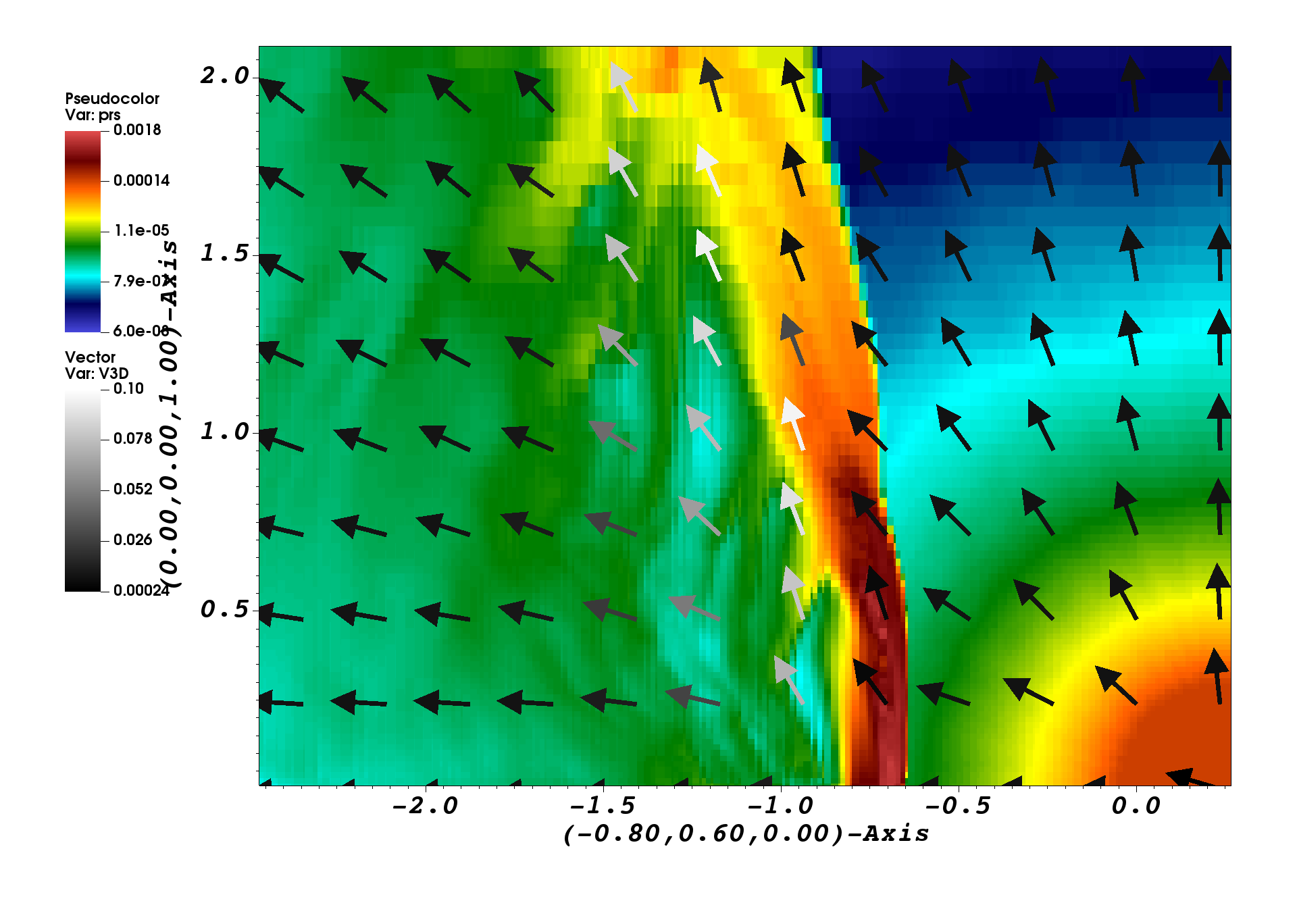}
   \includegraphics[width=8.4cm]{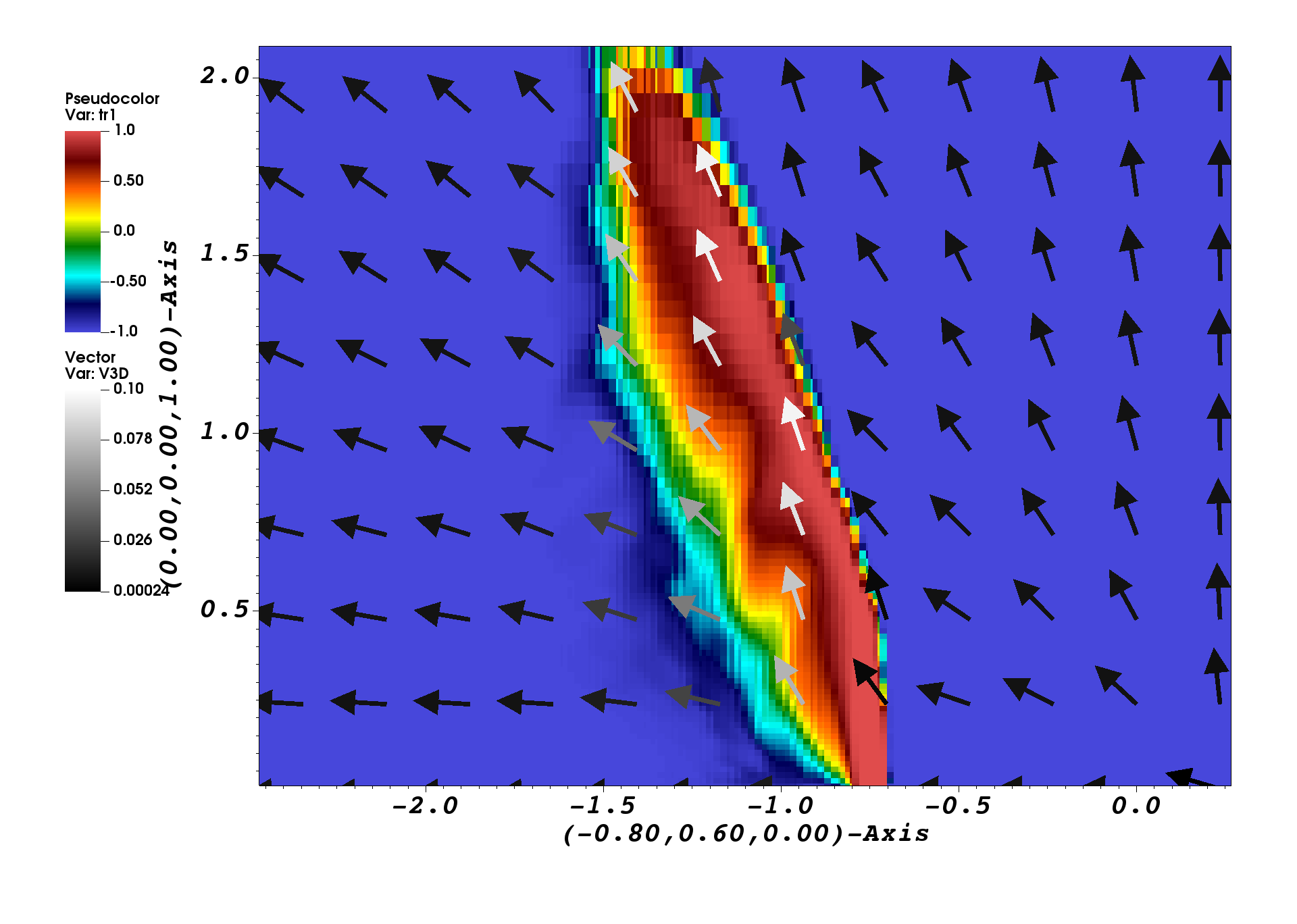}
   \caption{The same as shown in Fig.~\ref{fig:HRRzoom} but for the weakly relativistic jet (low resolution).}
   \label{fig:LRNRzoom}%
\end{figure}
   
\begin{figure*}
 \centering
   \includegraphics[width=8.6cm]{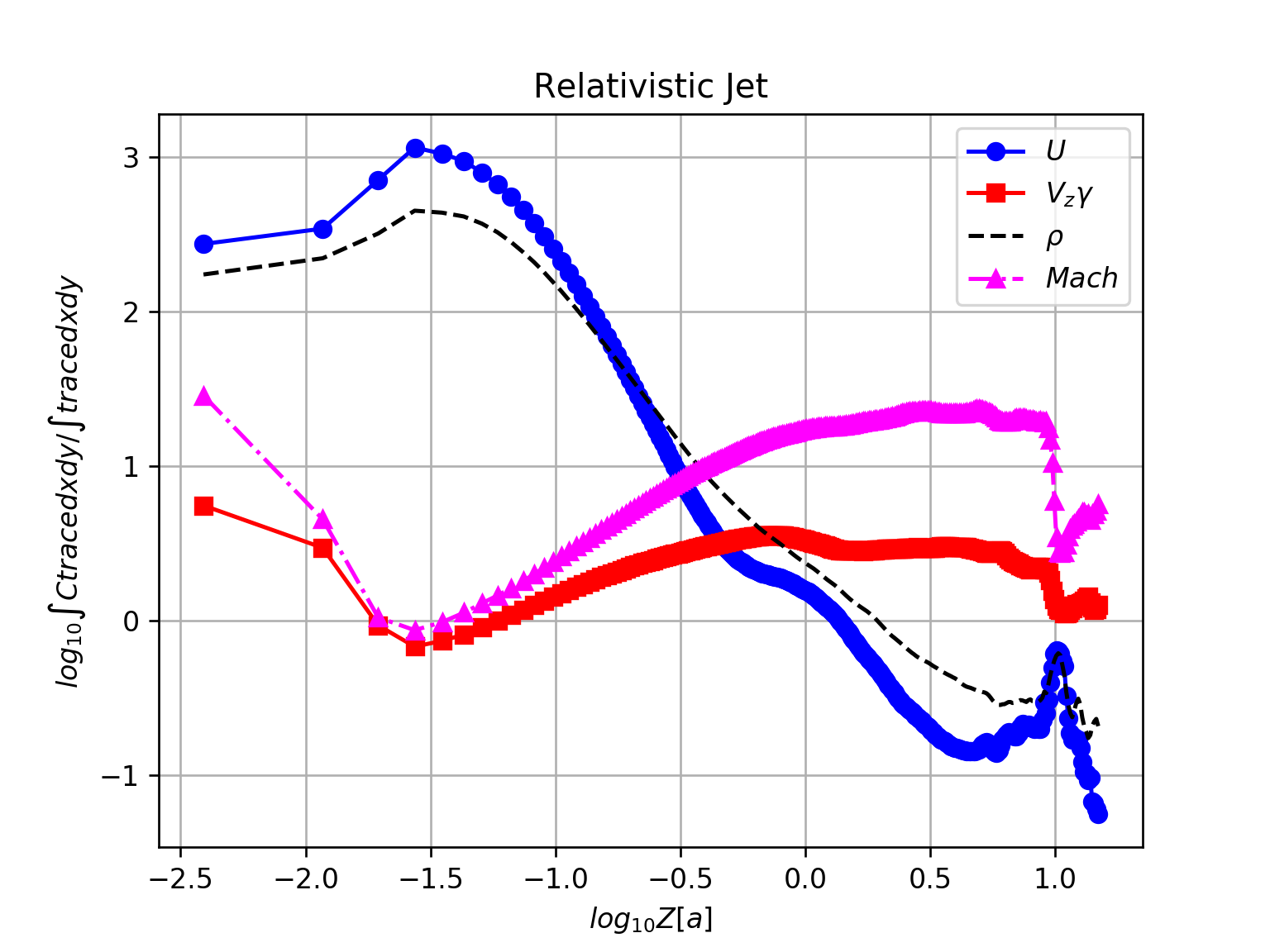}
   \includegraphics[width=8.6cm]{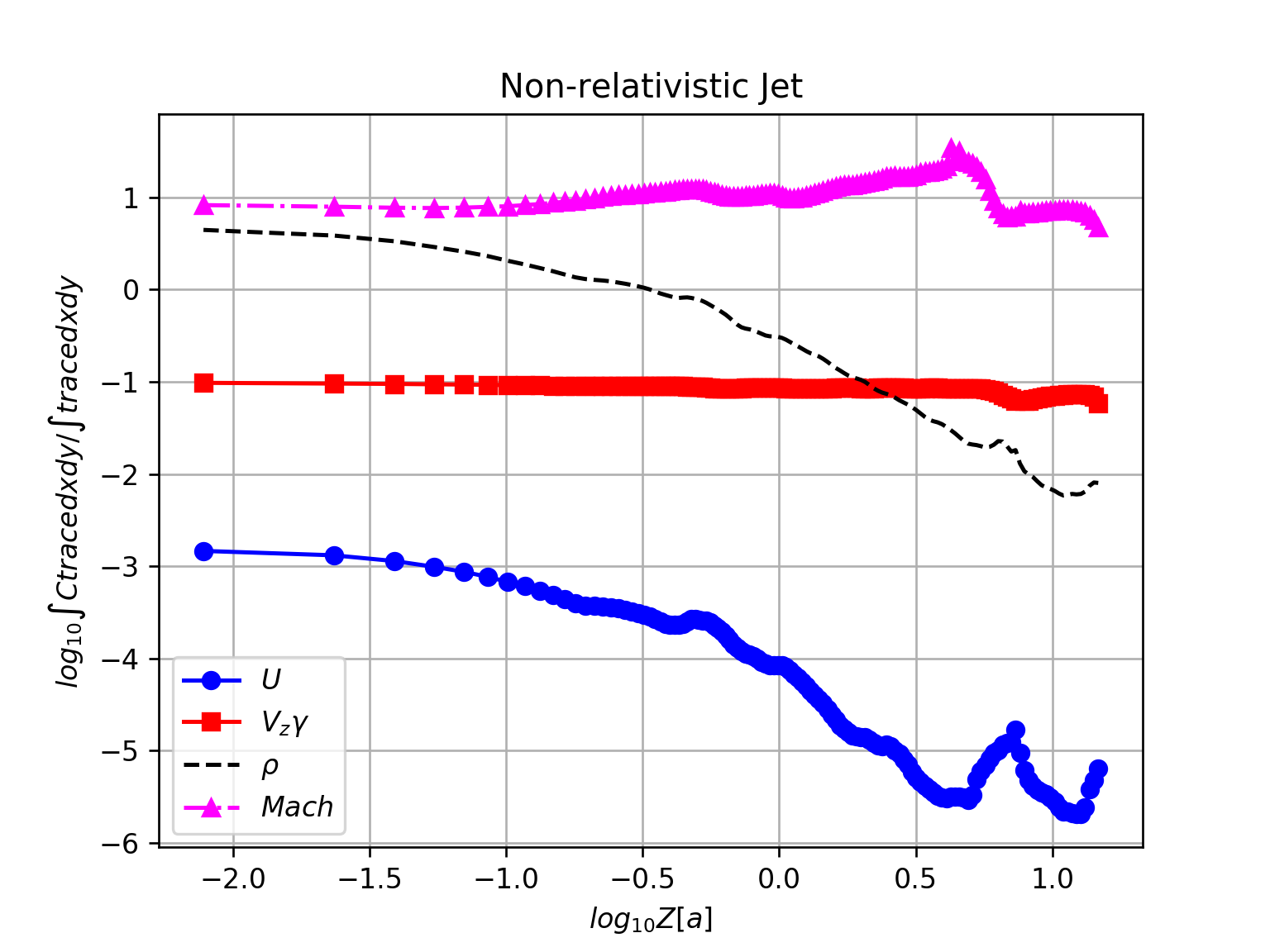}
   \caption{Logarithmic representations of the specific internal energy, density, Mach number, and the $z$-component of the 4-velocity (see text -Sect.~\ref{res}-), with $z$, for the relativistic (left) and the weakly relativistic jet (right). The y-axis units are arbitrary.}
   \label{fig:proflog}%
\end{figure*}

In addition to the wind momentum rate intercepted by the jet in the direction away from the star, and its role bending, destabilizing and mass-loading the jet, orbital motion can also dynamically influence the jet-wind interacting structure on scales similar to, or larger than, those of the binary. Analytical and semi-analytical treatments have shown that this influence leads to a helical geometry, which can further enhance instability development in the jet \citep{per12,bos13,bos16}. This influence can also affect the high-energy emission of HMMQ, which are confirmed gamma-ray sources, as shown by \cite{mol18} and \cite{mol19}. The strength of the wind and the orbit influences on the jet is expected to be directly related to the intercepted wind-to-jet momentum rate ratio: First, the more wind momentum rate is intercepted by the jet, the more inclined the jet becomes away from the star. Then, the larger the jet inclination, the stronger the orbital Coriolis force can push the jet flow in the direction opposite to the orbit sense. As argued in particular by \cite{bos16}, under such conditions the jet should likely disrupt after one or a few turns of the helical structure, not very far from the binary. A similar phenomenon has been shown to likely occur in colliding winds in high-mass binaries hosting a non-accreting pulsar \citep[see, e.g.,][for 2D and 3D simulations, respectively]{bbkp12,bbp15}. In the case of HMMQ, however, relativistic hydrodynamical simulations have not explored yet the regions in the jet affected by orbital motion. Observationally, jet bending and precession has been found in HMMQ (Cyg~X-3, SS~433, and possibly in Cyg~X-1), although the origin of these jet features is presently unclear, or ascribed to disk precession in the case of SS~433 \citep[see the discussion in][and references therein]{bos16}. Despite the different origin, nevertheless, the consequences of precession in the clearest case of precessing jet, SS~433, are expected to be similar to those of an orbital origin for the precession: jet eventual
disruption and mixing with the medium \citep[see, e.g.,][]{mon14,mon15,mil19}. Moreover, the evolution of precessing jets in HMMQ should have an impact as well on the jet termination region \citep[see, e.g.,][]{vel00,zav08,bos11b}.

In this work, we study in detail the jet behavior under the influence of the stellar wind and orbital motion. With this aim, we carried out for the first time 3-dimensional (3D) relativistic hydrodynamic simulations of the interaction of a jet with a stellar wind including the effect of orbital motion, on scales that include the helical jet formation region. The article is organized as follows: In Sects.~\ref{phys} and \ref{sim} we describe the physical scenario and the simulations carried out,
respectively. Then, in Sect.~\ref{res} the results are presented. Finally, a summary of the conclusions and a discussion are provided in Sect.~\ref{disc}. The convention $Q_b=Q/10^b$ is adopted all through the paper, with $Q$ in cgs units unless otherwise stated.

\section{Physical scenario}\label{phys}

The physical scenario on which the simulations are based is the following: We adopted a HMMQ with a circular orbit. This was done for simplicity, although our results should approximately apply to moderately eccentric systems as well. To explore a system with parameters similar to those of Cyg~X-1 and Cyg~X-3, a circular orbit is also a good approximation. For this exploration to be qualitatively applicable to these and other similar potential sources, we adopted an orbital separation distance of $a=10^{12}$~cm and a period of $P_{orb}=1$~day, as these values are in between those of Cyg~X-3 and Cyg~X-1 \citep[periodwise right in the middle in logarithmic scale; see, e.g.,][respectively]{zdz18,mil21}. Such an orbital separation can be derived for that period from a binary consisting of a star and a CO with masses of, for instance, $M_\star=30$~M$_\odot$ and $M_{\rm CO}=10$~M$_\odot$, respectively.

The calculations assumed that the flows evolve without significantly radiating their internal energy away, so an adiabatic equation of state (EoS) describes the gas. This approximation is valid as long as thermal and non-thermal cooling does not significantly affect the flow internal energy: this approximation might not apply to Cyg~X-3, as its stellar wind is very dense, nor it would apply to jets whose pressure is dominated by non-thermal particles that radiate fast their energy. On the other hand, the advantage of an adiabatic EoS is that it allows the results to apply to all sources that fulfil the relation $P_{orb}\propto a$, which given the third law of Kepler, $P_{orb}\propto a^{3/2}/(M_{\star}+M_{\rm CO})^{1/2}$, is actually fulfilled when $P_{orb}/a^{3/2}\propto 1/(M_{\star}+M_{\rm CO})^{1/2}$.  
On the other hand, it is worth noting that at least at an heuristic level, departures by a factor of a few in this relation, even under non-negligible radiative cooling \citep[see, e.g.,][]{cha21}, may still be qualitatively described by our results. 

The study by \cite{per12}, which did not include orbital motion, showed that the presence of clumps in the winds should probably lead to enhanced jet instability. For simplicity we assumed that the stellar wind was smooth, which makes our simulations more conservative from this side. 

Due to the high computational cost of the simulations, we decided to consider a case in which significant (but not strong) jet-wind-orbit interaction was expected. For this, we fixed the intercepted wind-to-jet momentum rate
ratio to a value appropriate for that case, derived using analytical estimates from \cite{bos16} (see also \citealt{zdz15}). First, one can characterize the jet density as:
\begin{equation}
    \rho_j\approx \frac{L_j}{\pi r_j^2\gamma_j(\gamma_j-1)\varv_jc^2}=\frac{\rho_w }{ \chi_j}\frac{\varv_w^2}{\varv_j^2 \gamma_j^2 }\frac{ \theta_jr_s^2 }{\pi r_j^2 }\,,
    \label{eq:rhoj}
\end{equation}
where $\gamma_j$ and $L_j$ are the jet Lorentz factor and power (removing the rest mass energy), $\rho_j$ and $\rho_w$ are the jet and stellar wind densities, $r_j$ and $\theta_j$ are the jet radius and half opening angle, and $r_s$ is the distance to the star.
The parameter $\chi_j$ is the approximate ratio of jet momentum rate ($\dot P_j$) to wind momentum rate ($\dot P_w$) intercepted by the jet \citep[before the orbit influence becomes important; see][]{bos16}, and can be obtained from: 
\begin{equation}
    \chi_j = \frac{\theta_j \dot{P}_w}{4\pi \dot{P}_j} =\frac{\theta_j}{4\pi}\frac{\gamma_j-1}{\gamma_j \beta_j} \frac{\dot{M}_w \varv_w c}{L_{j}}\,,
    \label{eq:chij}
\end{equation}
where $\beta_j=\varv_j/c$, and $\theta_j/4\pi$ is the solid angle fraction with respect to $4\pi$ of the jet as seen from the star. The larger the wind momentum rate captured by the jet, the more the jet is going to bend away from the star. The parameter $\chi_j$ plays the key role in the derivation of an analytical estimate of the deflection angle of the jet with respect to $\hat{z}$ and away from the star, in the CO-star-$\hat z$-plane:
\begin{equation}
    \Phi = \frac{\pi^2 \chi_j}{ 2\pi\chi_j + 4\chi_j^{1/2} +  \pi^2},
    \label{eq:def}
\end{equation}
derived assuming that all the intercepted wind momentum was added to that of the jet. For $\chi_j = 0.3$, Eq.~\eqref{eq:def} yields $\Phi = 0.21$~radians ($\approx 12^\circ$), which is larger than
$\theta_j$ ($\approx 6^\circ$) and thus fulfills the condition for significant orbital effects \citep[see][]{bos16}. 

For the chosen values of $\chi_j $, $\theta_j$ and $\varv_w$, one obtains the relation $L_{j}\approx 10^{37}
\dot{M}_{w,-6}$~erg~s$^{-1}$ for $\gamma_j\gtrsim 2$, and $L_{j}\approx 6\times 10^{35}
\dot{M}_{w,-6}\,\beta_{j,-1}$~erg~s$^{-1}$ in the weakly relativistic jet case {($\dot M_w$ units are
M$_\odot$~yr$^{-1}$)}. This relation between $L_j$ and $\dot M_w$ remains the same even if one modifies $P_{orb}$ and $a$ while keeping $P_{orb}/a$ constant. The fiducial parameter values used in the simulations are shown in Table~\ref{tab:model}.

\section{Simulations}
\label{sim}

To probe two different regimes of jet velocity and the resulting jet-wind-orbit interaction, we ran simulations for a highly relativistic and a weakly relativistic jet. Unfortunately, due to the high computational cost limitations, we had to reduce the weakly relativistic jet simulation resolution to a half, that is, cells are twice larger in that case with respect to the case of the relativistic jet. With these lower jet speed and resolution, we could spare enough time to at least probe the orbital evolution of the interacting structure longer than for the relativistic jet. 

The mentioned differences between the relativistic and the weakly relativistic jet simulations make them not directly comparable, but the obtained qualitative features, like jet (partial) disruption and helical structure formation, should be robust enough against resolution changes despite numerical diffusion. The reason is that numerical diffusion makes in fact the results more conservative when evaluating the degree of flow instability, as a lower resolution tends to stabilize the flow because jet-wind numerical mixing slows the jet flow down, and kinetic energy conversion into heat reduces its Mach number. On the other hand, numerical diffusion will lead to a more extended jet in the low resolution case as jet and wind mix faster, but the core jet should still trace the actual physical behavior. Regarding the helical structure, numerical effects can hardly excite its formation but rather smooth its geometry out. Thus, if both jet disruption and helical structure manifest in the simulated jets, in particular in the weakly relativistic case, this result will probably stand when higher resolutions are employed.

\subsection{Code, EoS and computational grid}\label{codecg}

The simulations were performed in 3D Cartesian coordinates using the {\it PLUTO} code\footnote{Link:
http://plutocode.ph.unito.it/index.html} \citep{mbm07}, which is a modular Godunov-type code entirely written in C and intended
mainly for astrophysical applications and high Mach number flows in multiple spatial dimensions. Energy, and not entropy, is the
independent variable, and an Eulerian scheme is adopted when solving the equations of relativistic hydrodynamics \footnote{See
sections 6.1 and 6.3 of the PLUTO user guide: http://plutocode.ph.unito.it/userguide.pdf}. Special relativity is enough to describe
the problem at hand, as the distances from the CO are much larger than its gravitational radius. Spatial parabolic interpolation, a 3rd order Runge-Kutta approximation in time, and an HLLC Riemann solver were used \citep[e.g.,][]{2005JCoPh.203..344L}. The duration of the time steps is determined by the adopted CFL number of 0.2, to keep the stability of the simulations. We note that adaptive mesh refinement (AMR) was not employed because, due to
the proneness of the simulated flow to develop instabilities, AMR would tend to fill large regions of the computational grid with
higher resolution zones. The high computational cost associated to these zones would make very difficult to choose the optimal initial
resolution to perform the simulation for long enough. A static non-uniform smooth grid  allows otherwise for an adequate control of
the resolution in the whole computational grid during a whole run. This choice {(i.e., static mesh refinement)} also avoids the artificial resolution jumps that
AMR may trigger \citep[e.g.,][]{2014A&A...565A..65B,bbp15}. 

The simulated flow was approximated by the TAUB EoS for an ideal gas \citep[which is that of an ideal gas with one temperature that ranges from non-relativistic to relativistic values,][see also section 7.4 in the PLUTO user guide]{1978AnRFM..10..301T}, and the grid size was taken to be $z \in [0, 15]$, and $x \mbox{ and } y \in [-15, 15]$, in units of the orbital separation distance $a$. We used a non-uniform cell geometry in the computational grid to have an adequate resolution in the central and extended jet regions. To probe different jet velocity regimes, we performed two kinds of simulations: a relativistic high-resolution simulation with total number of cells in each direction: $N_z = 256$, and $N_x = N_y  = 768$; and a weakly relativistic simulation with half the number of cells in each direction. As shown in Table~\ref{tab:grid}\footnote{More details of the description of the grid setup can be found in section~4.1 of the PLUTO user guide; http://plutocode.ph.unito.it/userguide.pdf.}, the grid was divided in three regions, a central one with $N_c$ uniform cells, and two extended, non-uniform cell regions to the left and to the right, with $N_l$ and $N_r$ cells, respectively. At injection, the tracer value was $-1$ for the wind flow, and 1 for the jet flow. 

\begin{table*}
\caption{Parameters of the grid ("left" extended grid, central grid, "right" extended grid; see text for details -Sect.~\ref{codecg}-)}
\begin{tabular}{lccccccc}
\hline
\hline
  Coordinates         &  Left~($\times a$)    &  $N_l$   &   Left-center~($\times a$)  &   $N_c$ &   Right-center~($\times a$) & $N_r$ & Right~($\times a$)  \\
\hline
  High Resolution         &      &     &    &    &   &  &   \\
\hline
&&&&&\\[-5pt]
 $ x  $     & \quad $-15$ & \quad $192$ & \quad $-1.5$ &\quad 384 & \quad $1.5$ &\quad 192 & \qquad $15$ \\
$ y  $     & \quad $-15$ & \quad $192$ & \quad $-1.5$ &\quad 384 & \quad $1.5$ &\quad 192 & \qquad $15$ \\
 $ z  $   &  &    & \quad $0$ & \quad $64$ & \quad $0.5$ &\quad 192 & \qquad $15$  \\
\hline
  Low Resolution         &      &     &    &    &   &  &   \\
\hline
 $ x  $     & \quad $-15$ & \quad $96$ & \quad $-1.5$ &\quad 192 & \quad $1.5$ &\quad 96 & \qquad $15$ \\
$ y  $     & \quad $-15$ & \quad $96$ & \quad $-1.5$ &\quad 192 & \quad $1.5$ &\quad 96 & \qquad $15$ \\
 $ z  $   &  &    & \quad $0$ & \quad $32$ & \quad $0.5$ &\quad 96 & \qquad $15$  \\
\hline
&&&&&\\[-5pt]
\end{tabular}
\label{tab:grid}
\end{table*}

\subsection{Setup}
\label{sec:setup}

At the beginning of the simulation, the computational grid is filled by a spherically symmetric, supersonic stellar wind with
radial speed $\varv_w=2400$~km~s$^{-1}$. A wind azimuthal velocity component, due to stellar rotation and angular momentum
conservation, was neglected because it was $\sim 10$\% of the CO orbital velocity, and we considered that its effect on the
numerical solutions would be small when accounting for other uncertainties and simplifications of the problem. The radius of the star is $0.3\,a$, whereas the radius of the jet injection inlet, that is, the jet base radius, is $0.045\,a$, that is,
$\approx 11$ cells across the jet base ($\approx 6$ cells in the low resolution simulation). The star initial position is
$(0,0.25\,a,0)$. The jet is injected from the CO location at $(0,-0.75\,a,0)$ in the direction $\hat z$, thus perpendicularly to the
$xy$- (orbital) plane, with half-opening angle $\theta_j = 0.1$, and a Lorentz factor $\gamma_j=10$ in the relativistic jet case,
and a velocity $\varv_j = 0.1\,c$ in the weakly relativistic jet case. {The wind injection takes place at the stellar surface defined above, and the jet inlet is a boundary condition in the plane $z=0$. In each injection cell density and pressure are the same, but the velocity corresponds to that of a spherically symmetric flow for the wind, and a radial conical one with half-opening angle $\theta_j$ for the jet.}  As mentioned, the jet inlet moves on the orbital plane
following a counter-clockwise circular path with $P_{orb}=1$~day (i.e., an angular velocity $\Omega_{orb}\approx 7.3\times
10^{-5}$~s$^{-1}$) with a constant distance from the star $a=10^{12}$~cm. The wind and the jet were assumed to be cold, with wind
Mach number $M_w=6.2$, and jet Mach number $M_j = \gamma_j \varv_j/ \gamma_{s,j} c_{s,j} = 85$ (where {\it s} refers to the sound speed).
These Mach numbers are high enough to satisfy the following conditions: 1) neither the wind nor the jet present heat-driven acceleration, nor expansion in the case of the jet; and 2) the flows cross the grid boundaries at supersonic speeds, precluding bouncing waves to form there, hence preventing the associated numerical artifacts. The sketch of the setup, similar to that in \cite{bos16}, can be found in Fig.~\ref{f1}. 

\begin{table}
\caption{Fiducial parameter values used in the models. Velocity is in $c$ units (see text for details -Sect.~\ref{sec:setup}-).}
\begin{tabular}{lccc}
\hline
\hline
  Parameters         &  High Res.    & Both &   Low Res.  \\
\hline
 $ v_w$  [$c$]   &  & $0.008$ &  \\
 $\dot{M}_w$ [$M_\odot$~yr$^{-1}$] &  & $10^{-6}$ &  \\
 $L_j   $ [erg~s$^{-1}$] & $10^{37}$  & & $6\times 10^{35}$ \\
 $ v_j \Gamma_j  $    & 10 & & 0.1 \\
 $P_{orb} [day] $     & & 1 &  \\
$\theta_j  $     & &$0.1$ &   \\
 $ M_w $   &  &6.2  &      \\
 $ M_j  $     &  & 85 &  \\
 $\chi_j$     &  & 0.3 &  \\
\hline
\\[-5pt]
\end{tabular}
\label{tab:model}
\end{table}

We analyzed the numerical solutions of the simulations when the CO is at $(-0.45\,a,0.6\,a,0)$ and $(0.6\,a,-0.45\,a,0)$ for the
relativistic and the weakly relativistic (low resolution) jet case, respectively. These locations correspond to a bit more than
$1/2$ and 1 orbit, or orbital phases $\approx 0.6$ and 0.15, taking the simulation start at phase 0. The simulation running times
were long enough to allow the jet-wind interaction structure to reach a quasi-steady state in the orbit rotating frame. Thanks to the increased simulation speed, the evolution of the structure could be probed for a longer time in the weakly relativistic jet case.
The number of time steps done for the relativistic simulation was $\sim 10^6$, and for the weakly relativistic one was $\sim 2\times10^5$. The interacting flows reach a quasi-steady state, even
when affected by shocks and hydrodynamical instabilities, because they have a much higher velocity than the orbital speed, and
their grid crossing time is $\sim 10^{12}$~cm~$/10^8$~cm~$s^{-1}=10^4$~s, significantly shorter than $P_{orb}$ ($\sim 10^5$~s).

\section{Results}\label{res}

Regarding the jet deflection angle away from the star, $\Phi$, the simulations presented here yield a good agreement between the simulated jet
and the analytical estimate of 0.21~radians derived above. For the relativistic jet, on scales of $\sim a$, we obtained
$\Phi\approx  0.23$~radians, and $\Phi\approx 0.22$~radians for the weakly relativistic jet, with the latter simulation, as
mentioned below, somewhat affected by numerical diffusion. The simulations show that the impact of the wind heats the jet,
which tends to increase its expansion rate. Thus, as will be seen, $\Phi$ increases beyond the analytical estimate for $z>a$ in
the two studied cases.

In addition to the wind push on the jet away from the star, there is also a lateral push produced by Coriolis forces against the orbit sense, which grows with cylindrical radius, $\omega_c=\sqrt{x^2+y^2}$, and eventually leads to strong deflection of the flow against orbital motion at a distance $\omega_c\sim \mbox{few}\times y_{orb}$ \citep{bos16}, with:
\begin{equation}
    y_{orb} = 2\times 10^{12} \frac{A_{orb}}{\chi_j^{1/2}}\frac{\varv_{w,8.5}}{\Omega_{orb,-4}} \mbox{ cm}\,,
    \label{eq:yorb}
\end{equation}
{which is the characteristic jet deflection distance on the orbital plane,} and $A_{orb}$ is a constant $\sim 1$. Substituting in Eq.~\eqref{eq:yorb} the simulation parameters one gets $y_{orb} = 3.8A_{orb}\,a$, which is close to the values obtained in the simulations: $y_{orb}\approx 3.5\,a$ ($A_{orb}\approx 0.9$) and $3.8\,a$ ($A_{orb}\approx 1.1$) for the relativistic and the weakly relativistic jet, respectively. We present now in what follows the simulation results in more detail:

Figure~\ref{fig:trsl} provides with colored tracer maps and velocity vector distributions for the relativistic jet,
showing vertical 2D sections of the grid with different orientations in azimuth. The sections focus on the jet on small scales up to several $a$ (top panel), at intermediate
heights of several $a$ (middle panel), and towards the top of the grid, at $z\gtrsim 10\,a$ (bottom panel). The same information is
displayed in Fig.~\ref{fig:trsllrNR} but for the weakly relativistic jet (low resolution). These maps show the evolution of the jet with $z$ and its disruption, which manifests through color inhomogeneities in the maps, being more apparent in the relativistic jet case.

Figure~\ref{fig:tr} shows the jet tracer isosurfaces in 3D, determined by the locations of cells for which the jet tracer is 0.5 (green) and 0 (brown), for the relativistic jet at phase $\approx 0.6$. We remind that the pure wind tracer value is $-1$. The isosurfaces are shown for two different perspectives, as indicated by the Cartesian diagrams at the bottom left of each panel, illustrating the maximum deviation of the jet away from the star due to the direct impact of the wind (top panel), and the jet deviation against the orbit sense (bottom panel). The same information is displayed in Fig.~\ref{fig:trlrNR} but for the weakly relativistic jet (low resolution) at phase $\approx 0.15$. 

Figure~\ref{fig:analsim} shows a comparison between the trajectories of the relativistic (two left panels) and the weakly relativistic jet (two right panels) for two models: {a simple semi-analytical model with jet radius $\omega_k\propto z^k$ (with $k=0.5$, 0.75, 1, 1.25 and 1.5)} in which all the intercepted wind momentum is transferred to the jet, and the hydrodynamical solution, in which the $xy$-plane mean jet location for different $z$ values is obtained by averaging the relevant quantity over that plane, weighting with the jet tracer. We set the tracer to $>0.3$ to account for the grid cells mostly filled by jet material. The jet trajectory is represented by $\omega_c$ and $\Phi$ for different $z$-values. As seen from these plots, overall the numerical trajectory is relatively similar to the conical jet model ($k=1$), although the detailed hydrodynamical solution predicts stronger and more irregular jet deflection caused by faster shocked jet lateral expansion due to heating and instability growth. This effect may be illustrated by a more similar behavior in both approaches, in some restricted regions, when $k>1$. In the weakly relativistic jet case, numerical diffusion affects the jet tracer averaged quantities. It does so by {apparently} increasing the inclination of the jet, particularly at its base, away from the star (see the bottom right plot in Fig.~\ref{fig:analsim}). We note however that the core of the jet, in green (tracer value 0.5) in Fig.~\ref{fig:trlrNR} shows that the weakly relativistic jet is in fact less affected at its base than the relativistic one, and this is so despite the external layers of the former being more affected by numerical diffusion on those scales (Fig.~\ref{fig:analsim}). The fact that the weakly relativistic jet is overall more stable at the jet base (despite its apparent faster deflection) may be explained by the lower resolution, but this is also an expected outcome of the much larger mass rate and a lower energy per particle, which slow down instability growth.

Colored density maps and velocity vector distributions for the relativistic jet for different $z$ values are shown in Fig.~\ref{fig:rhoVxy}. The panels, from left to right and from top to bottom, correspond to 2D cuts parallel to the $z=0$ at heights $z = 0.1\,a$, $2\,a$, $4\,a$, $6\,a$, $8\,a$, $10\,a$, $12\,a$, and $14\,a$. The same is displayed in Fig.~\ref{fig:rhoVxylrNR} but for the weakly relativistic jet (low resolution). These maps allow the appreciation of the evolution of the jet disruption process with $z$.

Figure~\ref{fig:S} shows colored maps similar to those in Fig.~\ref{fig:rhoVxy} but for $S=P/\rho^{4/3}$ (without velocity information), which can be a measure of the flow disorder (or entropy, thereby the $S$ symbol\footnote{The physical gas entropy proper can be calculated as $s=c_v \log(P/\rho^\gamma) = c_v \log(S)$, where $c_v$ is the thermal capacity at constant volume.}) due to shocks and instability growth. For jet and wind material that did not interact, $S$ should be constant for each flow, but the strong jet-wind-orbit interaction leads to a growingly inhomogeneous distribution of $S$ with height. The same is shown in Fig.~\ref{fig:SlrNR} but for the weakly relativistic jet (low resolution). In both figures, but more prominently in the relativistic jet case, the jet and the wind become more disordered for larger $z$. The low resolution of the weakly relativistic jet yields overall a smoother evolution of the jet-wind interacting structure, although the jet is less powerful than in the relativistic jet case. In any case, the weakly relativistic jet shows clear signs of disruption at high $z$-values after one orbit (and also after half orbit, not shown here).

Figure~\ref{fig:HRRzoom} presents zoomed-in colored maps, in the CO-star-$\hat z$-plane and for the relativistic jet, of the density (top panel), pressure (middle panel), and tracer (bottom panel), together with velocity vector distributions. The same is shown in Fig.~\ref{fig:LRNRzoom} for the weakly relativistic jet (low resolution). These figures allow for a closer inspection to what happens on scales of $\sim a$, and can be more easily compared to previous simulations with smooth stellar winds for similar scales \citep[e.g.,][although some are not quasi-steady state]{per08,per10,yoo15,zdz15}. In particular, the asymmetric recollimation shock, and the jet flow deflection away from the star, also present in those previous simulations, are very apparent in these figures.

Finally, Fig.~\ref{fig:proflog} provides with logarithmic representations of the jet specific internal energy, density, Mach number, and the $z$-component of the 4-velocity, with $z$, for the relativistic (left panel) and the weakly relativistic jet (right panel). Similarly to Fig.~\ref{fig:analsim}, the shown quantities were computed for different $z$ averaging over the $xy$-plane weighting with the jet tracer (taking tracer $>0.3$). The figures provide with an estimate of the typical jet properties for different heights. There are regions in which internal energy turns into kinetic energy and viceversa, with the corresponding changes in specific internal energy, Mach number and 4-velocity $z$-component, and to a lesser extent in density. These changes appear around $z\lesssim a$ (direct jet-wind interaction), and for $z\gtrsim 3\,a$ (Coriolis force effect), particularly for the relativistic jet, but also for the weakly relativistic one. {As suggested above when studying wind-induced jet bending away from the star, the weakly relativistic jet is overall less affected by the direct jet-wind interaction around $z\lesssim a$, and most of the impact on the flow takes place at the Coriolis force interaction distance $\omega_c$. As mentioned, putting aside resolution effects, this larger stability is expected due to the much higher mass rate and lower energy per particle of this jet. This would explain for instance the flat curve of the $z$-component of the 4-velocity but at the farthest regions, where kinetic energy dissipation can be clearly identified.}
   
\section{Summary and discussion}\label{disc} 
 
To summarize, our results clearly show that the interplay of the jet with the stellar wind, and orbital motion, can easily lead to
jet disruption already at $z\sim 10\,a$ in compact HMMQ. In the context of the investigated scenario, the jet-wind and jet-wind-orbit
interactions are likely major factors determining the jet evolution on scales both smaller and larger than the binary size. This possibility was already proposed in \cite{per12} and \cite{bos13}, and further developed in \cite{bos16}, but our simulations represent a qualitative step forward in the
characterization of the jet-wind-orbit interaction structure, at least in HMMQ with orbits in the range of orbits represented by Cyg~X-1 and Cyg~X-3. 

{Let us try to analyze the main differences between Cyg~X-1 and Cyg~X-3 at a quasi-qualitative level. These differences would derive from $y_{orb}$, which is $\propto P_{orb}\chi_j^{-1/2}$ and characterizes the jet-wind-orbit interaction, and $\sim P_{orb}\varv_j$, which determines the step of the spiral (neglecting KH instabilities; see below). Assuming that $\chi_j$ and $\varv_j$ are the same in both sources, changing $P_{orb}$ from 5.6~days \citep[Cyg~X-1;][]{bro99} to 4.8~hours \citep[Cyg~X-3;][]{sin02} leads to a stronger orbit-effect and a shorter spiral step in Cyg~X-3 by a factor $\approx 28$. As jets in both sources are expected to be relativistic, $\varv_j$ should be close to $c$, $\chi_j$ then becomes the major source of uncertainty when comparing these sources, although the dependence of $y_{orb}$ is weaker with $\chi_j$ than with respect to $P_{orb}$. However, as \cite{bos16} already discussed \citep[see also][]{zdz15}, the knowledge of the relevant parameters is still too uncertain to properly assess to which extent the jet-wind-orbit interaction can determine the fate of the jet in Cyg~X-1 and Cyg~X-3.} In the upper (lower) end of the possible value ranges of the jet (wind) momentum rate for these sources, the jet
could propagate quite freely through the stellar wind while the CO and the star orbit each other. On the other hand, in other plausible regions of the parameter space, the jet may be completely destroyed and mixed with the surrounding medium already on scales of a few $a$.  The latter scenario is certainly not the usual state in these sources as
collimated structures are commonly seen in radio on scales $\gg a$ in Cyg~X-1 and Cyg~X-3 \citep{mar00,mar01,sti01,mio01,mil04,fen06}. However, presence of collimated structures on large scales does not necessarily imply lack of significant jet-wind-orbit interaction (see below). In any case, discarding fully destructive cases, there is still a broad range of scenarios with significant jet-wind-orbit interaction that cannot be ruled out. A specific analysis of these two HMMQ under the light of the results presented here is out of the scope of this work, but some explanations are given in what follows to provide perspective when interpreting the simulations under the light of observations. These explanations are followed by some caveats, and a final remark on the radiative consequences of the studied processes.

Our simulations, despite probing quite different jet velocities and resolution levels, point to a very unstable nature
of the structure resulting from the jet-wind-orbit interaction in HMMQ. Although this was already suggested, our numerical
calculations strongly endorse this possibility. Our results, not being an extensive exploration but just two
characteristic examples, do not settle the issue for other parameter choices. Nevertheless, the simulations probe
fiducial cases and their outcomes suggest that the jet is unlikely to leave unscathed the region close to the binary,
say $z\lesssim 10\,a$. Even for smaller $\chi_j$-values and longer $P_{orb}$-values than those explored here, we propose that jet-wind-orbit interaction could induce instability growth, which could
produce precession-like patterns in the jets at larger scale. Even if the jets are so powerful that significant mixing
and disruption does not occur on scales $\sim 1-10\,a$, and $\Phi$ is initially rather small, the growth of
Kelvin-Helmholtz (KH) instabilities is expected \citep[see also, e.g.,][in the context of extragalactic jets]{per19}. In fact, the jet-wind
interaction alone can already perturb the jet \citep{per08,per10}, and stellar wind clumping, not considered in the
present simulations, should also render the jet more prone to disruption \citep{per12}. In addition to KH instabilities, Rayleigh-Taylor and Richtmyer-Meshkov instabilities could also develop at the dynamical jet-wind contact discontinuity, particularly due to the Coriolis force, as in the similar case studied by \cite{bbp15} with a pulsar wind instead of a jet. These additional instabilities would produce non-linear perturbations that would couple to the KH instability, enhancing the disruptive effects of all these processes\footnote{The mangetic field can play an important role enhancing or inhibiting instability growth (L\'opez-Miralles et al., in prep.).}.
We note that jet
precession-like features caused by KH instabilities may or may not follow the orbital period, as these instabilities
could span a broad range of wavelengths. Therefore, in addition to the cases explored here, $\chi_j$-values even smaller
than $\theta_j$ \citep[see][for a discussion on the
$\chi_j-\theta_j$-relation]{bos16} could also lead to distorted jets on scales $\gg a$. Interestingly, this has been observed in radio in Cyg~X-3, but not yet in Cyg~X-1
\citep[see,e.g.][for related radio observations of Cyg~X-3]{mio01,mil04,tud07}. Interestingly, regardless of the $\chi_j$-value, if
the jet gets disrupted on scales $\lesssim 10\,a$, the strong pressure drop outwards of the embedding medium can make the jet-wind mixed flow turn into a collimated structure (see \citealt{bos16}; see also \citealt{mil19} for the case of the HMMQ SS~433, a different but related situation -see Sect.~\ref{intro}-). In the case of a weaker wind-orbit effect on the jet, the jet inertia, and the density drop of the medium, may allow the jet to keep some coherence despite KH instability growth, at
least until interacting with the surrounding medium on much larger scales (see, e.g., \citealt{mar96,gal05,rfg07,sel15}
for Cyg~X-1 radio observations, and \citealt{bor09,bos11b,yoo11} for simulations, of the jet-medium interaction). To
properly study the evolution of orbit-affected jets at $z\gg a$ in HMMQ, devoted simulations are planned.

The simulations done did not take into account that the stellar wind formation can be inhibited in the
hemisphere facing the CO, which happens when there is a strong CO accretion X-ray luminosity that ionizes the
accelerating stellar wind close to the star \citep[see, e.g.,][for Cygs~X-1 and Cygs~X-3, respectively]{mey20,vil21}. As discussed
in \cite{mol19}, the wind moving in that hemisphere should be slowed down when forming close to the star and then, beamed towards
the CO due to its strong gravitational pull \citep[e.g.][]{elm19}, acquiring a higher velocity once within the CO potential well.
These effects can perturb the momentum rate of the wind at least up to a certain jet height, and significantly affect the overall
influence of the wind and the orbit on the jet at scales $\lesssim a$. This is precisely the region where the jet gets most of the
stellar wind push, and thus the dependence of $\Phi$ and $\omega_c$ with $z$ may be modified (see Fig.~\ref{fig:analsim}). However,
although these effects should affect the jet-wind-orbit interaction to some extent, the complicated wind structure expected around
the CO caused by these effects should likely affect the jet in a complex manner as well, perhaps enhancing the growth of
instabilities and jet heating and expansion. To properly study this issue, detailed and specific 3D simulations of jet-wind interaction
including stellar wind inhibition and accretion are needed.

There is an additional possibility that may be also considered in future investigations: the role of accretion-powered
winds screening the jet from the impact of the stellar wind. However, if an accretion wind can indeed screen the jet
from the stellar wind, it may also interfere in a non-trivial manner with the accretion process. Nevertheless,
neglecting this collateral consequence, one must still note the following: Assuming an accretion-powered wind with mass
rate $\dot M_{w,accr}$ similar to that of the Bondi accretion rate \citep{hoy39,bon44,2012MNRAS.421.1351B}, expected to
be $\sim (R_{B}/2a)^2\dot M_w\sim 0.001\dot M_w$ (where $R_B$ is the CO Bondi radius), and a rather high accretion
wind velocity of $\varv_{w,accr}\sim 10\varv_w$, one can derive the standoff distance from the CO of the two-wind
contact discontinuity: $R_{St}\sim \eta^{1/2}a/(1+\eta^{1/2})\approx 0.1\,a$, where $\eta\sim \dot
M_{w,accr}v_{w,accr}/\dot M_wv_{w}\sim 0.01$. This estimate is strictly valid for isotropic winds. A focused stellar wind could
increase $\dot M_{w,accr}$, but balance this effect by increasing the stellar wind ram pressure, so focusing may not modify $R_{St}$
significantly. Another possibility would be Roche-Lobe overflow through the Lagrangian~1 point, which may strongly
enhance $\dot M_{w,accr}$, but this does not seem to be the dominant accretion mechanism in Cyg~X-1 and Cyg~X-3.
Although all these questions need further investigation, it seems to us that under realistic conditions the jet should
significantly interact with the stellar wind for $z\gtrsim 0.1\,a$, making our simulation results at least approximately
valid. It is worth mentioning that wind-wind interactions might have interesting dynamical and radiative consequences
\citep[see, e.g.,][]{kol18,sot21}. 

The setup discussed in this work assumes that the jet is normal to the orbital plane. In general, but particularly if wind accretion is rather chaotic, the jet may be pointing in a different direction, which could also change with time. We will assume in what follows that $\chi_j$ is neither much smaller or larger than assumed in the simulations, and that the jet keeps its initial orientation stable long enough to reach a quasi-steady state at the relevant scales. Under these assumptions, we identify four extreme cases for a jet that is not normal to the orbital plane: the jet points roughly towards the star (1) or away from the star (2); or the jet is normal to the star direction but is directed as the orbital motion (3), or against it (4). In case 1 the jet should be more affected by the stellar wind initially, getting more deflected and possibly more easily disrupted. In case 2, the jet may be less affected initially, but from the beginning it would be more inclined away from the star, becoming more sensitive to Coriolis forces farther from the binary. In cases 3 and 4, the situation may be somewhat similar, as the jet would face more directly the effect of the Coriolis force in case 3 from the beginning, whereas in case 4 the jet would be less affected initially, but the eventual higher inclination would lead to a stronger  lateral push of the wind further away from the binary. Although devoted simulations should be carried out, it seems plausible that a jet normal to the orbital plane may be in fact a more stable configuration than those of cases 1, 2, 3 and 4.

Finally, particularly regarding non-thermal radiation processes, the jet-wind-orbit interaction should have an influence on the
multiwavelength non-thermal emission of the jets of HMMQ. Our simulations hint at a very strong effect of the stellar wind and the
orbit on the jet of HMMQ. Thus, including this effect, as done for instance in \cite{mol18} and \cite{mol19} using semi-analytical dynamical models, seems indeed
necessary to better understand the non-thermal behavior of these sources.

\section*{Acknowledgements}
V.B-R. acknowledges the support by the Spanish  Ministerio de Ciencia e Innovaci\'{o}n (MICINN) under grant PID2019-105510GB-C31 and through the ``Center of Excellence Mar\'{i}a de Maeztu 2020-2023'' award to the ICCUB (CEX2019-000918-M), and by the Catalan DEC grant 2017 SGR 643. V.B-R. is Correspondent Researcher of CONICET, Argentina, at the IAR.
The simulations were carried out on the CFCA~XC30 cluster of the National Astronomical Observatory of Japan (NAOJ). 
We thank the {\it PLUTO} team for the opportunity to use the {\it PLUTO} code. 
The visualization of the results of the numerical simulations were performed in the VisIt package \citep{HPV:VisIt}.

\section*{Data availability}
Data available on request.


\bsp	
\label{lastpage}
\end{document}